\documentclass[final,11pt]{elsarticle}

\usepackage[version=3]{mhchem} 

\usepackage{amssymb}

\usepackage[margin=2.5cm]{geometry}
\usepackage{algorithm}
\usepackage{algpseudocode}

\newcounter{algsubstate}
\renewcommand{\thealgsubstate}{\alph{algsubstate}}
\newenvironment{algsubstates}
  {\setcounter{algsubstate}{0}%
   \renewcommand{\State}{%
     \stepcounter{algsubstate}%
     \Statex {\footnotesize\thealgsubstate:}\space}}
  {}

\usepackage{amsmath,bm}

\usepackage{physics}
\usepackage{subcaption,siunitx,booktabs}
\usepackage{graphicx,epstopdf}
\usepackage{graphics}
\usepackage{multirow}
\usepackage{fancyhdr, ifpdf}
\usepackage{amsxtra}
\usepackage{stmaryrd}
\usepackage{mathrsfs}
\usepackage{psfrag}
\usepackage{epsfig}
\usepackage{rotating}
\usepackage{setspace}
\usepackage{float}
\usepackage{amsfonts}
\usepackage{amsbsy}
\usepackage{amscd}
\usepackage{amsthm}
\usepackage{color}
\usepackage{tabularx}
\usepackage{caption}
\usepackage{subcaption}
\usepackage{sidecap}
\usepackage{dcolumn}
\usepackage{esint}
\usepackage{mathalfa}
\usepackage{cprotect}
\usepackage{ulem}
\usepackage[explicit]{titlesec}

\newcommand{\mat}[1]{{\mathbf #1}}
\newcommand{\del}{\nabla}

\newcommand{\bPsi}{\boldsymbol{\Psi}}

\newcommand{\bvPsi}{\boldsymbol{\varPsi}}
\newcommand{\bvu}{\boldsymbol{u}}

\newcommand{\bPsitildein}{\boldsymbol{\widetilde{\Psi}_{\bf in}}}
\newcommand{\bPsitildef}{\boldsymbol{\widetilde{\Psi}_{\bf f}}}
\newcommand{\bPsitildefDagger}{\boldsymbol{\widetilde{\Psi}_{\bf f}^{\dagger}}}

\newcommand{\bPsitildeRfr}{{\boldsymbol{\widetilde{\Psi}}}^{\bf R}_{\bf fr}}
\newcommand{\bPsitildeo}{{\boldsymbol{\widetilde{\Psi}}}^{\bf o}}

\newcommand{\btau}{\boldsymbol{\tau}}

\newcommand{\bff}{\boldsymbol{f}}

\newcommand{\bk}{\boldsymbol{k}}

\newcommand{\bq}{\boldsymbol{q}}

\newcommand{\bv}{\boldsymbol{v}}
\newcommand{\bx}{\boldsymbol{\textbf{x}}}
\newcommand{\by}{\boldsymbol{\textbf{y}}}

\newcommand{\bDfr}{\boldsymbol{\textbf{D}}_{\textbf{fr}}}
\newcommand{\bE}{\boldsymbol{\textbf{E}}}

\newcommand{\bH}{\boldsymbol{\textbf{H}}}
\newcommand{\bI}{\boldsymbol{I}}
\newcommand{\bJ}{\boldsymbol{J}}
\newcommand{\bK}{\boldsymbol{\textbf{K}}}

\newcommand{\epsilonbar}{\bar{\epsilon}}
\newcommand{\fbar}{\bar{f}}
\newcommand{\bM}{\boldsymbol{\textbf{M}}}
\newcommand{\bL}{\boldsymbol{\textbf{L}}}
\newcommand{\bLDagger}{\boldsymbol{\textbf{L}^{\dagger}}}
\newcommand{\bLinv}{\boldsymbol{\textbf{L}^{-1}}}
\newcommand{\bLinvDagger}{\boldsymbol{{\textbf{L}^{-1}}^{\dagger}}}

\newcommand{\bP}{\boldsymbol{\textbf{P}}}

\newcommand{\bQfr}{\boldsymbol{\textbf{Q}}_{\textbf{fr}}}
\newcommand{\bQPrimefr}{\boldsymbol{\textbf{Q}^{\prime}_{\textbf{fr}}}}
\newcommand{\bR}{\boldsymbol{\textbf{R}}}
\newcommand{\bS}{\boldsymbol{\textbf{S}}}

\newcommand{\bU}{\boldsymbol{U}}
\newcommand{\bUbar}{\boldsymbol{\bar{U}}}
\newcommand{\bV}{\boldsymbol{\textbf{V}}}

\newcommand{\bxi}{\boldsymbol{\xi}}

\newcommand{\varphibar}{\bar{\varphi}}

\newcommand{\ubar}{\bar{u}}

\newcommand{\bsm}{b_{\rm{sm}}}
\newcommand{\bsmi}[1]{b_{\rm{sm}}^{#1}}

\newcommand{\bdeltai}[1]{b_{\delta}^{#1}}
\newcommand{\rhobar}{\bar{\rho}}

\newcommand{\fintd}{\fint \displaylimits}

\newcommand{\dx}{\,d\bx}
\newcommand{\dy}{\,d\by}

\newcommand{\dk}{\,d\bk}
\newcommand{\btH}{\boldsymbol{\widetilde{\textbf{H}}}}
\newcommand{\bbarH}{\boldsymbol{\bar{\textbf{H}}}}
\newcommand{\bhatH}{\boldsymbol{\hat{\textbf{H}}}}
\newcommand{\btHnl}{\boldsymbol{\widetilde{\textbf{H}}^{\textbf{nl}}}}
\newcommand{\btHloc}{\boldsymbol{\widetilde{\textbf{H}}^{\textbf{loc}}}}
\newcommand{\bHnl}{\boldsymbol{{\textbf{H}}^{\textbf{nl}}}}
\newcommand{\bHloc}{\boldsymbol{{\textbf{H}}^{\textbf{loc}}}}
\newcommand{\bHnlk}{\boldsymbol{{\textbf{H}}^{\textbf{nl},\bk}}}
\newcommand{\bHlock}{\boldsymbol{{\textbf{H}}^{\textbf{loc},\bk}}}
\newcommand{\btHlocci}{\boldsymbol{\widetilde{\textbf{H}}}^{\textbf{loc}^{c_i}}}

\newcommand{\bfX}{\boldsymbol{\textbf{X}}}
\newcommand{\bfXb}{\boldsymbol{{\textbf{X}}_{\textbf{b}}}}
\newcommand{\bfYb}{\boldsymbol{{\textbf{Y}}_{\textbf{b}}}}

\newcommand{\Rthree}{ {\mathbb{R}^{3}} }

\newcommand{\ordercomplexity}{\mathcal{O}}

\newcommand{\chieps}{\,\btau^{\varepsilon}}
\newcommand{\bdir}{\boldsymbol{\Upsilon}}
\newcommand{\dir}{\Upsilon}

\newcommand{\atzerob}{\bigg|_{\varepsilon = 0}}

\newcommand{\vselfdelta}[1]{V^{#1}_{\delta}}
\newcommand{\vself}[1]{V^{#1}_{\rm{sm}}}
\newcommand{\vselfeps}[1]{V^{#1}_{{\rm{sm}}_{\varepsilon}}}

\setcitestyle{citesep={,}}

\definecolor{hellgruen}{rgb}{0.2,0.7,0.2}

\newcommand{\QE}{\texttt{QE}~}

\newcommand{\DFTFE}{\texttt{DFT-FE}~}
\newcommand{\DFTFENoSpace}{\texttt{DFT-FE}}
\newcommand{\DFTFEver}{\texttt{DFT-FE 1.0}~}
\newcommand{\DFTFEverprev}{\texttt{DFT-FE 0.6}~}
\newcommand{\DFTFEverNoSpace}{\texttt{DFT-FE 1.0}}
\newcommand{\DFTFEverprevNoSpace}{\texttt{DFT-FE 0.6}}

\newcommand{\nwchem}{\texttt{NWChem}~}
\newcommand{\nwchemNoSpace}{\texttt{NWChem}}

\makeatletter
\newcommand*{\rom}[1]{\expandafter\@slowromancap\romannumeral #1@}
\makeatother

\titleformat{\paragraph}[runin]
  {\normalfont\normalsize\bfseries}{}{11pt}{\theparagraph\hspace*{1em}#1:}
\titleformat{name=\paragraph,numberless}[runin]
  {\normalfont\normalsize\bfseries}{}{11pt}{#1:}
\bibpunct{[}{]}{;}{n}{}{}

\newcounter{bla}

\journal{Computer Physics Communications}

\begin{document}

\begin{frontmatter}



\title{DFT-FE 1.0\,: A massively parallel hybrid CPU-GPU density functional theory code using finite-element discretization}



\author[a]{Sambit Das\fnref{fn1}}
\author[b]{Phani Motamarri\fnref{fn1}}
\author[d]{Vishal Subramanian}
\author[c]{David M. Rogers}
\author[a,d]{Vikram Gavini\corref{author}}
\fntext[fn1]{Sambit Das and Phani Motamarri contributed equally to this work.}
\cortext[author] {Corresponding author.\\\indent \textit{E-mail address:} vikramg@umich.edu}
\address[a]{Department of Mechanical Engineering, University of Michigan, Ann Arbor, MI 48109, USA}
\address[b]{Department of Computational and Data Sciences, Indian Institute of Science, Bangalore, India}
\address[c]{National Center for Computational Sciences, Oak Ridge National Laboratory, Oak Ridge, TN 37831, USA}
\address[d]{Department of Materials Science \& Engineering, University of Michigan, Ann Arbor, MI 48109, USA}

\begin{abstract}
We present \DFTFEverNoSpace, building on \DFTFEverprev [Comput. Phys. Commun. 246, 106853 (2020)], to conduct fast and accurate large-scale density functional theory (DFT) calculations (reaching $\sim 100,000$ electrons) on both many-core CPU and hybrid CPU-GPU computing architectures. This work involves improvements in the real-space formulation---via an improved treatment of the electrostatic interactions that substantially enhances the computational efficiency---as well high-performance computing aspects, including the GPU acceleration of all the key compute kernels in \DFTFENoSpace. We demonstrate the accuracy by comparing the ground-state energies, ionic forces and cell stresses on a wide-range of benchmark systems against those obtained from widely used DFT codes. Further, we demonstrate the numerical efficiency of our implementation, which yields $\sim 20 \times$ CPU-GPU speed-up by using GPU acceleration on hybrid CPU-GPU nodes. Notably, owing to the parallel-scaling of the GPU implementation, we obtain wall-times of $80-140$ seconds for full ground-state calculations, with stringent accuracy, on benchmark systems containing $\sim 6,000-15,000$ electrons.

\end{abstract}
 

\end{frontmatter}
 


\noindent{\bf{Program summary}}
\bigskip

\begin{small}
\noindent
{\it Program Title:}        DFT-FE (https://github.com/dftfeDevelopers/dftfe)                           \\
{\it Journal Reference:}                                      \\
{\it Catalogue identifier:}                                   \\
{\it Licensing provisions:}  LGPL v3                          \\
{\it Programming language:}  C/C++                                 \\
{\it Computer:} Any system with C/C++ compiler, MPI library and GPU support with CUDA (optional).     \\
{\it Operating system:}  Linux                                      \\
{\it RAM:} Ranges from few GBs for small problem sizes to around 50,000 GB for a system with 114,674 electrons.                                              \\
{\it Number of processors used:} Range from 64 to 192,000 MPI tasks, 24 to 22,800 GPUs (demonstrated).   
{\it Classification:}  1.0                                       \\
{\it External routines/libraries:} p4est (http://www.p4est.org/), deal.II (https://www.dealii.org/), \\ BLAS (http://www.netlib.org/blas/), LAPACK (http://www.netlib.org/lapack/), \\ELPA (https://elpa.mpcdf.mpg.de/), ScaLAPACK (http://www.netlib.org/scalapack/), \\Spglib (https://atztogo.github.io/spglib/), ALGLIB (http://www.alglib.net/), \\LIBXC (http://www.tddft.org/programs/libxc/), PETSc (https://www.mcs.anl.gov/petsc), \\ SLEPc (http://slepc.upv.es), \\ NCCL (optional-https://github.com/NVIDIA/nccl). \\ \\
{\it Nature of problem:} Density functional theory calculations.\\
   \\
{\it Solution method:} We employ a local real-space variational formulation of Kohn-Sham density functional theory that is applicable for both pseudopotential and all-electron calculations on periodic, semi-periodic and non-periodic geometries. Higher-order adaptive spectral finite-element basis is used to discretize the Kohn-Sham equations. Chebyshev polynomial filtered subspace iteration procedure (ChFSI) is employed to solve the nonlinear Kohn-Sham eigenvalue problem self-consistently. ChFSI in DFT-FE employs Cholesky factorization based orthonormalization, and spectrum splitting based Rayleigh-Ritz procedure in conjunction with mixed precision arithmetic. Configurational force approach is used to compute ionic forces and periodic cell stresses for geometry optimization.
 \\
   \\
{\it Restrictions:} Exchange correlation functionals are restricted to Local Density Approximation (LDA) and Generalized Gradient Approximation (GGA), with and without spin. The pseudopotentials available are optimized norm conserving Vanderbilt (ONCV) pseudopotentials and Troullier--Martins (TM) pseudopotentials. Calculations are non-relativistic.\\
   \\
{\it Unusual features:} DFT-FE handles all-electron and pseudopotential calculations in the same  framework, while  accommodating  periodic,  non-periodic  and  semi-periodic  boundary conditions.\\
   \\
{\it Running time:} This is dependent on problem type and computational resources used. Timing results for benchmark systems are provided in the paper.\\
   \\


\end{small}

\clearpage
\section{Introduction}
\label{sec:intro}
Electronic structure calculations based on Kohn-Sham density functional theory (DFT) have played a very important role in providing key insights into a wide variety of materials properties, including mechanical, electronic, magnetic and optical properties of materials. A significant fraction of the world's computational resources are being utilized for these calculations, and these have provided predictive capability for material modeling, especially qualitative trends, leading to accelerated materials design and discovery. There are about hundred electronic structure codes available, with the plane-wave (PW)  basis~\cite{VASP,qe2009,gonze2002,CASTEP2005,exciting2014} and the atomic-orbital type basis~\cite{Pople,jensen2002,cp2k2014,blum2009,nwchem} being the most commonly employed basis sets. Though these basis sets enjoy many advantages, the inherent limitations of PW basis sets of restricting the simulations domains to periodic geometries and boundary conditions, the lack of systematic convergence for AO basis sets, compounded by the limited parallel scalability for both PW and AO basis sets have resulted in the development of systematically improvable and scalable real-space discretization techniques like finite-difference~\cite{parsec2006,rescu2016,sparc2017a,octopus2015,gpaw}, finite-elements~\cite{tsuchida1995,tsuchida1996,pask1999,pask2005,bylaska,suryanarayana2010,motamarri2013,SCHAUER2013644,zhou2014,denis2016,kanungo2017,kanungo2019real,motamarri2020,das2019}, and other reduced order basis techniques~\cite{dgdft2015,motamarri2015,xu2018,lin2021a,lin2021b}. We note that finite-element (FE) basis employed in this work is a relatively new entrant to real-space methods for DFT and offers several key advantages in comparison to the other widely discretization schemes. In particular, FE basis provides systematic convergence for any materials system~\cite{motamarri2013}, can accommodate periodic, non-periodic and semi-periodic boundary conditions, and can offer excellent parallel scalability owing to the locality of the FE basis functions~\cite{motamarri2020}. Furthermore, the FE basis sets are amenable to adaptive spatial resolution, not only allowing the possibility of large-scale all-electron calculations~\cite{kanungo2017,rufus2020fast} by combining with enrichment functions, but also mixed all-electron and pseudopotential calculations~\cite{ghosh2021}.

A variety of application areas demand large-scale DFT calculations involving tens of thousands of electrons, which include \textit{ab initio} modeling of dislocation cores~\cite{rodney2017,Arias2000,Trinkle2005,Trinkle2008,Clouet2009,shin2011,shin2013,iyer2015,radhakrishnan2016,das2017} in metals and semiconductors, ion diffusivity and migration barrier predictions in solid-state electrolytes~\cite{leung2020,morgan2021,wang2018}, large-scale bio-molecular simulations~\cite{cole2016,motamarridna2020,takao2018,skylaris2013}, spin-spin interactions in defects in solids for applications in quantum computing~\cite{ghosh2019,Gali2019,ghosh2021}, and many others. Accurate simulations of these large-scale materials systems with DFT usually involve geometry optimizations, or nudged-elastic band calculations, or long time-scale ab-initio molecular dynamics simulations, all of which require a large number of DFT ground-state calculations. Conducting systematically convergent DFT calculations on such large-scale systems has been challenging owing to the cubic scaling computational complexity of DFT calculations with system size and the limited parallel scalability of state-of-the-art DFT codes. To this end, accurate computational methods that are robust, computationally efficient and scalable on evolving heterogeneous parallel computing architectures are crucial to enable large-scale {\it ab-intio} simulations on realistic materials systems involving many tens of thousands of electrons.

Paving the way for large-scale DFT calculations, numerous efforts have focused on reducing the cubic scaling computational complexity~\cite{godecker99,bowler2012,skylaris2005,fattebert2006,ls3df2008,motamarri2014} for both metallic and insulating systems over the past few decades. Alternate approaches to enable large-scale DFT calculations include developing efficient methodologies to reduce the computational pre-factor associated with the cubic scaling computational complexity, and massively scalable numerical implementations taking advantage of the evolving computing architectures. Our recent work on \DFTFEverprev ~\cite{motamarri2020} is along this direction and presents an accurate, efficient and massively parallel finite-element framework for large-scale DFT calculations on many core CPU architectures. The numerical implementation in the work employed computationally efficient strategies which were instrumental in reducing data movement costs and latency on many core CPU architectures. To this end, \DFTFEverprev exhibited close to quadratic scaling till $30,000 - 40,000$ electrons, thereby delaying the onset of cubic scaling complexity. Additionally, excellent parallel scaling performance has been demonstrated up to 192,000 MPI tasks, with significant speedups of up to $10\times$ observed in computational times in comparison to widely used plane-wave codes for material systems with more than 10,000 electrons. The capability of \DFTFEverprev to simulate very large systems reaching up to $\sim 60,000$ electrons has been demonstrated, which remains computationally prohibitive using plane-wave codes. However, despite the demonstrated performance, many inefficiencies remain in \DFTFEverprevNoSpace, which are addressed as a part of this work. In particular, \DFTFEverprev restricts the nuclear charges to be coincident with corner finite-element nodes in the finite-element mesh. This, in turn, requires refined FE discretizations, even in the case of pseudopotential calculations yielding smooth solutions. In addition to increased computational cost, this approach would incur additional overheads during geometry optimization and molecular dynamics simulations due to the requirement of frequent remeshing during the course of geometry updates. Furthermore, the implementation procedures in \DFTFEverprev does not take advantage of the recent advances in heterogeneous computing architectures, and can execute only on massively parallel many-core CPU architectures.

We present here \DFTFEverNoSpace, a massively parallel hybrid CPU-GPU DFT code using finite-element discretization for large-scale density functional theory calculations. The current effort builds on \DFTFEverprev ~\cite{motamarri2020}, addresses the aforementioned limitations of \DFTFEverprevNoSpace, and presents a computationally efficient implementation of \DFTFE on hybrid CPU-GPU architectures. In particular, electrostatic energy and the relevant configurational force expressions (for geometry relaxation) associated with the Kohn-Sham DFT functional are reformulated in terms of smeared nuclear charges, allowing the flexibility for the atoms to float in pseudopotential calculations, in contrast to the use of point charges coincident with finite-element nodes in \DFTFEverprevNoSpace. This enables the use of coarser finite-element meshes with higher-order FE interpolating polynomial up to degree 7, thereby reducing the number of basis functions required to achieve the desired accuracy---by a factor of $\sim3\times$ for the benchmark systems studied in this work---which, in turn, translates to a similar improvement in the computational efficiency. Furthermore, \DFTFEver focuses on efficient GPU porting of key computational kernels in Chebyshev filtered subspace iteration procedure (ChFES)~\cite{motamarri2020,das2019}---the solver employed in \DFTFE for the solution of the Kohn-Sham equations---by utilizing the massive fine grained parallelism of GPUs in both double and complex double datatypes. We note that a good GPU porting efficiency is strongly coupled to the ratio of compute to the data movement associated with the underlying numerical implementation, which in turn is dependent on the algorithms and the basis sets employed. This aspect was also demonstrated in our prior work ~\cite{das2019}, which described a proof of concept GPU implementation for DFT calculations on hybrid CPU-GPU architectures employing the finite-element basis. To this end, our GPU implementation strategies in the current work hinges on exploiting the following HPC-centric ideas: (i) finite-element level matrix-multi vector products in the Chebyshev filtering step of ChFES, (ii) mixed precision arithmetic based processor-boundary communication in the assembly of domain-decomposed finite-element level vectors, (iii) mixed precision based floating point operations and communications in the computation of overlap matrix and projected Hamiltonian, (iv) employing GPU direct communication, and overlapping compute and data movement -via- asynchronous programming paradigms in the Chebyshev filtering step, as well as the computation of overlap matrix and projected Hamiltonian. 

The accuracy and performance of the aforementioned methodology improvements and implementation innovations in \DFTFEver are studied using a wide a range of benchmark systems. To this end, the accuracy is validated on representative benchmark systems using state-of-the art codes like Quantum Espresso (\QE) and \nwchem for pseudopotential and all-electron DFT calculations, respectively. \DFTFEver provided excellent agreement on ground-state energies, ionic forces and stresses with \QE and \nwchem. Furthermore, the aforementioned implementation advances on hybrid CPU-GPU architectures resulted in high throughput performances leading to $\sim 16-20\times$ CPU-GPU speedups for a wide range of material system sizes, ranging from $\sim 4,000-30,000$ electrons. We note that the speedups obtained in the present work are found to be much higher than the reported CPU-GPU speedups from state-of-the-art DFT codes where a GPU implementation is available~\cite{bigdftgpu2009,bigdftgpu2011,vaspgpu2011, espressogpu2012,octopusgpu2013,espressogpu2018}.  Furthermore, excellent parallel scalability has been demonstrated using \DFTFEver on hybrid CPU-GPU architectures over a range of system sizes. In particular, the strong scaling aspect of the current implementation has been instrumental in reducing the wall-time to around $\sim 1-2$ minutes for system sizes of $\sim 6,000-14,000$ electrons for the ground-state DFT calculation. The performance of \DFTFEver was also demonstrated on a very large system comprising of 114,674 electrons, where the ground-state calculation took around $\sim54$ minutes on 3600 GPUs (600 nodes of the Summit supercomputer). 

The remainder of the paper is organized as follows. Section~\ref{sec:ksdft} starts with a discussion on the governing equations in DFT, and presents the local real space formulation implemented in \DFTFEver with an improved electrostatics treatment by employing smeared floating charges to describe the ions. Subsequently, the configurational forces---used to evaluate ionic forces and cell stresses for geometry optimization---in the framework of smeared floating charges is presented, followed by the finite-element discretization of the Kohn-Sham DFT problem. Section~\ref{sec:numerics} describes the numerical methodologies employed in \DFTFEver, and Section ~\ref{sec:arch} discusses the various aspects of the numerical implementation on hybrid CPU-GPU architectures. Section~\ref{sec:results} presents the results on benchmark systems, demonstrating the accuracy, computational efficiency and parallel scaling performance of \DFTFEver on hybrid CPU-GPU architectures. We finally conclude along with an outlook in Section~\ref{sec:conclusions}.

\section{Real-space Kohn-Sham DFT formulation and finite-element discretization }
\label{sec:ksdft}
\subsection{Kohn-Sham DFT governing equations with improved electrostatics treatment}
For a materials system with $N_{at}$ nuclei and $N_e$ electrons, the ground-state properties in Kohn-Sham density functional theory are given by solving the $N$ lowest eigenvalues of the following non-linear eigenvalue problem~\cite{kohn65}:
\begin{equation}\label{eq:contEigen}
\left(-\frac{1}{2} \laplacian + V_{\text{eff}}(\rho,\bR)\right) {\psi}_{i} = \epsilon_{i} {\psi}_{i},\;\;\;\;\; i = 1,2,\cdots N\;\;\;\text{with}\;\;\;N > \frac{N_e}{2}\,,
\end{equation}
where $\bR = \{\bR_1,\,\bR_2,\,\cdots \bR_{N_{at}}\}$ denotes the positions of the $N_{at}$ nuclei, $\rho$ denotes the electron density, $V_{\text{eff}}(\rho,\bR)$ denotes the effective single electron Kohn-Sham potential, and $\epsilon_{i}$ and $\psi_{i}$ denote the eigenvalues and eigenfunctions of the Kohn-Sham Hamiltonian, respectively. We discuss below the formulation in the context of spin-independent Kohn-Sham Hamiltonian. However, extension to spin-dependent case is straightforward~\cite{rmartin}, and is implemented in \DFTFE. Further, we discuss the formulation in an infinite periodic crystal, and obtain the formulation for the non-periodic system as a special case of the periodic formulation with appropriate changes to the boundary conditions. In the periodic case where $V_{\text{eff}}(\rho,\bR)$ is periodic on the unit cell, instead of solving the Kohn-Sham problem on a large periodic supercell, it is computationally efficient to invoke the Bloch theorem~\cite{rmartin,mermin}, which entails solving the reduced Kohn-Sham equations on a much smaller periodic unit cell with periodic boundary conditions on the electronic fields. In particular, the Kohn-Sham eigenfunctions $\psi_{n,\bk}(\bx)$ using the Bloch theorem are expressed as
\begin{equation}\label{bloch}
\psi_{n,\bk}(\bx) = e^{i\bk\cdotp\bx} u_{n,\bk}(\bx) \,,
\end{equation}
where $i = \sqrt{-1}$ and  $u_{n,\bk}(\bx)$ is a complex-valued function that is periodic on the unit cell satisfying $u_{n,\bk}(\bx+\bL_r)=u_{n,\bk}(\bx)$ for all lattice vectors $\bL_r$, and $\bk$ denotes a point in the first Brillouin zone of the reciprocal space. Substituting the above in Eq.~\ref{eq:contEigen}, the reduced Kohn-Sham equations on the unit cell are given by
\begin{align}\label{eq:contEigenBZ}
&\left(-\frac{1}{2} \laplacian  -  i \bk \cdot \grad +\frac{1}{2}{\abs{\bk}}^2+ V_{\text{eff}}^{\text{loc}}(\rho,\bR)\right)  {u}_{n,\bk}\notag\\ & \qquad+ e^{-i\bk\cdot\bx}\int\displaylimits_{\Rthree} V_{\text{eff}}^{\text{nl}}(\bx,\by,\bR) e^{i\bk\cdot\by}\,u_{n,k}(\by) \dy = \epsilon_{n,\bk} \,{u}_{n,\bk}, \;\;\;\;\; n = 1,2,\cdots N \,\,; \,\, \forall \,\,\bk\ \in BZ\,,
\end{align}
where $V_{\text{eff}}^{\text{loc}}(\rho,\bR)$ and $V_{\text{eff}}^{\text{nl}}(\bx,\by,\bR)$ are the local and non-local contributions to $V_{\text{eff}}(\rho,\bR)$, with the non-local contribution arising from the consideration of pseudopotentials. We note that the governing equations for the non-periodic case are recovered from Eq.~\ref{eq:contEigenBZ} by replacing the periodic geometry with a non-periodic simulation domain along with appropriate boundary conditions, and considering only the $\Gamma$-point ($\bk={\bf 0}$). In the above, the electron density $\rho$ is given in terms of the orbital occupancy function  $f_{n,\bk}=f(\epsilon_{n,\bk},\mu) \in \left[0,1\right]$  and the canonical wavefunctions as
\begin{equation}\label{eq:elecdens}
\rho(\bx) =2\sum_{n=1}^N \;\fintd_{BZ}\, f_{n,\bk}\,u^{*}_{n,\bk}(\bx)\,u_{n,\bk}(\bx) \dk \,\,,
\end{equation}
where $\fintd_{BZ}$ denotes the volume average of the integral over the Brillouin zone corresponding to the periodic unit cell $\Omega_p$, and $\mu$ represents the Fermi-energy, determined by the constraint on the total number of electrons in the system ($N_e$) given by
\begin{equation}\label{cons}
\int\displaylimits_{\Omega_p} \rho(\bx) \dx = 2\sum_{n=1}^N \;\fintd_{BZ}\, f_{n,\bk} \dk = N_e\,.
\end{equation}
To avoid numerical instabilities, particularly in metallic systems with large number of eigenstates around the Fermi energy, we use a smooth Fermi-Dirac distribution~\cite{VASP,godecker99} for $f$ given by
\begin{equation}\label{fermidirac1}
f(\epsilon,\mu) = \frac{1}{1 + \exp\left(\frac{\epsilon - \mu}{\sigma} \right)}\,,
\end{equation}
where $\sigma = k_{B} T$ denotes the regularization parameter with $k_B$ denoting the Boltzmann constant and $T$ representing the finite electronic temperature. We note that $f(\epsilon_i,\mu)$ is denoted as $f_i$ in the remainder of the manuscript.

The effective single electron potential, $V_{\text{eff}}(\rho,\bR)$, is composed of:
\begin{equation}\label{eq:veff}
 V_{\text{eff}}(\rho,\bR) = V_{\text{xc}}(\rho) + V_{\text{el}}(\rho,\bR) = \frac{\delta E_{\text{xc}}(\rho)}{\delta \rho} + \frac{\delta E_{\text{el}}(\rho,\bR)}{\delta \rho}\,
\end{equation}
where $V_{\text{xc}}(\rho)$ is the functional derivative of the exchange-correlation energy functional $E_{\text{xc}}(\rho)$, which accounts for the many-body quantum-mechanical interactions between electrons~\cite{XCReview2005}. We adopt the generalized gradient approximation (GGA)~\cite{rmartin,gga1} for the exchange correlation functional description throughout the manuscript, which is given by the following semi-local form:
\begin{equation}\label{exc}
E_{\text{xc}}(\rho) = \int\displaylimits_{\Omega_p} \varepsilon_{xc}(\rho,\grad \rho) \dx.
\end{equation}
We remark that \DFTFE also implements other local and semi-local functional forms (LDA, LSDA, SGGA).

The term $V_{\text{el}}(\rho,\bR)$, in the effective potential (Eq.~\ref{eq:veff}), is the variational derivative of the total classical electrostatic interaction energy inside the periodic unit cell, which is composed of
\begin{equation}
E_{\text{el}}(\rho,\bR) = E_{\text{H}}(\rho) + E_{\text{ext}}(\rho,\bR) + E_{\text{zz}}(\bR)\,,
\end{equation}
where, $E_{\text{H}}$, $E_{\text{ext}}$ and $E_{\text{zz}}$ denote the electrostatic interaction energy between electrons (Hartree energy), interaction energy between nuclei and electrons, and repulsive energy between nuclei, respectively. These are given by
\begin{align}\label{eq:ElecEngy}
E_{\text{H}} =  \frac{1}{2}\int\displaylimits_{\Omega_p} \,\int\displaylimits_{\Rthree} \frac{\rho(\bx)\rho(\by)}{|\bx - \by|} \,\dy\,\dx ,\;\; 
& E_{\text{ext}} =  -\sum_{a} \sum_{r}\int\displaylimits_{\Omega_p} \rho(\bx)  \frac{Z_a}{\abs{\bx-(\bR_a+\bL_r)}}\dx, \notag\\ E_{\text{zz}}= & \frac{1}{2}\sum_{\substack{a,a^{\prime} \neq a}} \sum_r \frac{Z_a Z_{a^{\prime}}}{\abs{\bR_a-(\bR_{a^{\prime}}+\bL_r)}},\;\;
\end{align}
where $Z_a$ denotes the charge on the $a^{th}$ nucleus in the periodic unit cell, and the summation over $r$ extends over all the lattice sites in the infinite periodic crystal. Further, all integrals involving $\bx$ in Eqs.~\ref{eq:ElecEngy} are over the periodic domain represented by the unit cell, whereas the integrals involving $\by$ are over $\Rthree$. Henceforth, unless otherwise specified, we will adopt this convention. In the case of pseudopotential treatment, where only the valence electronic wavefunctions are computed in an effective potential generated by the the nucleus and the core electrons, $E_{\text{ext}}$ in Eq.~\ref{eq:ElecEngy} is defined as
\begin{align}\label{eq:pspengtotal}
    E_{\text{ext}}=\sum_{a} \sum_r \rho(\bx) \int\displaylimits_{\Omega_p}  V^{a}_{\text{psp,loc}}(\abs{\bx-(\bR_a+\bL_r)}) \dx  + E_{\text{psp,nl}}\,.
\end{align}
In the above, $V_{\text{psp,loc}}^a$ denotes the local part of the pseudopotential operator, and $E_{\text{psp,nl}}$  is the non-local pseudopotential energy given by
\begin{align}\label{eq:pspengnl}
 E_{\text{psp,nl}}=\left(2 \sum_{n=1}^{N} \fintd_{BZ} \: \int \displaylimits_{\Omega_p}\: \int\displaylimits_{\Rthree} f_{n,\bk}\; u^{*}_{n,\bk}(\bx)\;e^{-i\bk\cdot\bx}\; \sum_{a} \sum_r V_{\text{psp,nl}}^a(\bx,\by,\bR_a+\bL_r)\;e^{i\bk\cdot\by}\;u_{n,\bk}(\by) \dy \dx \dk\right)\,,
\end{align}
where $V_{\text{eff}}^{\text{nl}}(\bx,\by,\bR):=\sum_{a} \sum_r V_{\text{psp,nl}}^a(\bx,\by,\bR_a+\bL_r)$ denotes the non-local part of the pseudopotential operator.
In \DFTFE, we employ norm-conserving pseudopotentials, and the action of $V_{\text{eff}}^{\text{nl}}(\bx,\by,\bR)$ on a wavefunction is given by the following separable operator form~\cite{rmartin,kb82}:
\begin{align}\label{eq:nlps}
 e^{-i\bk\cdot\bx}\int\displaylimits_{\Rthree} V_{\text{eff}}^{\text{nl}}(\bx,\by,\bR) e^{i\bk\cdot\by}\,u_{n,k}(\by) \dy:= & e^{-i\bk\cdot\bx}\int\displaylimits_{\Rthree} \sum_{a} \sum_r V_{\text{psp,nl}}^a(\bx,\by,\bR_a+\bL_r) e^{i\bk\cdot\by}\,u_{n,k}(\by) \dy \notag \\
 &= \sum_{a} \sum_{lpm} \sum_{r} e^{-i\bk\cdot(\bx - \bL_r)} C_{lpm}^{a,n\bk}\,h^{a}_{lp}\,\,\chi_{lpm}^{a}(\bx,\bR_a+\bL_r)\,,\notag\\
\text{with}\;\;\;\;\;\;\;C_{lpm}^{a,n\bk} =  \int\displaylimits_{\Omega_p} \sum_{r^{\prime}} \chi_{lpm}^{a}(\by,\bR_a+\bL_{r^{\prime}}) & e^{i\bk\cdot(\by - \bL_{r^{\prime}})}  u_{n,\bk}(\by)\dy,  \;\;\;\;
\frac{1}{h^{a}_{lp}} = \ip{\xi_{lm}^a}{\chi_{lpm}^a} \,,
\end{align}
where $l$ denotes the azimuthal quantum number, $m$ denotes the magnetic quantum number, $\ket{\chi_{lpm}}$ denotes the pseudopotential projector with $p$ denoting the index corresponding to the projector component for a given $l$,  $h_{lp}$ denotes the pseudopotential constant, and $\ket{\xi_{lm}^a}$ denotes the single atom pseudo-wavefunction. The above form applies to both Troullier-Martins (TM) pseudopotential~\cite{tm91} and optimized norm conserving Vanderbilt pseudopotential (ONCV) ~\cite{oncv2013}, both of which are implemented in \DFTFE~\cite{motamarri2020}. We note that the all-electron treatment in $E_{\text{ext}}$ is recovered from Eq.~\ref{eq:pspengtotal} by setting $V_{\text{psp,nl}}=0$ and $V^{a}_{\text{psp,loc}}(\abs{\bx-(\bR_a+\bL_r)})=-\frac{Z_a}{\abs{\bx-(\bR_a+\bL_r)}}$\,.

Next, we discuss the reformulation of electrostatics as local variational problems, which is based on our previous works~\cite{motamarri2020,motamarri2018,motamarri2013}, while highlighting the key difference in this work in the efficient treatment of nuclear charges.  First, we briefly discuss the local reformulation using the conventional approach of regularized Dirac distributions, where we begin by defining a nuclear charge distribution $b(\bx,\bR) =  \sum_{a} \bdeltai a (|\bx - \bR_a|)= -\sum_{a}Z_a\tilde{\delta}(|\bx - \bR_a|)$ with 
$\tilde{\delta}(\bx - \bR_a)$ denoting a regularized Dirac distribution centered at $\bR_a$. Using the regularized Dirac distribution and rearranging the electrostatic energy terms into charge neutral groups, the total electrostatic interaction energy in Eqs.~\ref{eq:ElecEngy}--\ref{eq:pspengtotal} is reformulated as
\begin{equation}\label{eq:elecnonlocal}
\begin{split}
  E_{\text{el}} &= \frac{1}{2} \int\displaylimits_{\Omega_p}  \int\displaylimits_{\Rthree}  (\rho(\bx)\, +\, b(\bx,\bR))\, \frac{ (\rho(\by)\, +\, b(\by,\bR))}{|\bx-\by|} \dy \dx- \frac{1}{2} \sum_{a} \int\displaylimits_{\Omega_p} \int\displaylimits_{\Rthree}  \frac{\bdeltai a(|\bx-\bR_a|)\bdeltai a(|\by-\bR_a|)}{|\bx-\by|} \dy \dx  \\&+\sum_{a} \sum_r \int\displaylimits_{\Omega_p} \left(V^{a}_{\text{psp,loc}}(\abs{\bx-(\bR_a+\bL_r)}) -  \int\displaylimits_{\Rthree} \frac{\bdeltai a(\abs{\bx-(\bR_a+\bL_r)})}{|\bx - \by|}\dy\right)\rho(\bx)\dx + E_{\text{psp,nl}} \,.
  \end{split}
\end{equation}
We note that the last two terms are not present in the case of an all-electron treatment. Subsequently, to efficiently compute terms with extended interactions in the  electrostatic interaction energy in Eq.~\ref{eq:elecnonlocal}, we reformulate them by defining auxiliary electrostatic potentials as discussed in  ~\cite{gavini2007,motamarri2013,das2015}, and rewrite the electrostatic energy in Eq.~\ref{eq:elecnonlocal} as the following local form:
\begin{equation}\label{eq:eleclocal}
\begin{split}
  E_{\text{el}} &= \frac{1}{2} \int\displaylimits_{\Omega_p} (\rho(\bx)\, +\, b(\bx,\bR))\, \varphi(\bx,\bR) \dx- \frac{1}{2}\sum_{a}\int\displaylimits_{\Omega_p} \bdeltai a (|\bx-\bR_a|)\vselfdelta a(\bx,\bR_a)\dx \\&+\sum_{a} \sum_r \int\displaylimits_{\Omega_p} \left(V^{a}_{\text{psp,loc}}(\abs{\bx-(\bR_a+\bL_r)}) - \vselfdelta a(\bx,\bR_a+\bL_r)\right)\rho(\bx)\dx + E_{\text{psp,nl}} \,,
  \end{split}
\end{equation}
where $\varphi(\bx,\bR)$ and $\vselfdelta a(\bx,\bR_a)$ are the auxiliary electrostatic potentials corresponding to the total charge distribution $(\rho + b)$ and the $a^{th}$ nuclear charge $\bdeltai a$, respectively, given by:
\begin{equation}\label{eq:pot}
\varphi(\bx,\bR) = \int\displaylimits_{\Rthree} \frac{\rho(\by) + b(\by,\bR)}{|\bx - \by|} \dy\,,
\;\;\;\; 
\vselfdelta a(\bx, \bR_a) =\int\displaylimits_{\Rthree} \frac{\bdeltai a (|\by-\bR_a|)}{|\bx-\by|} \dy
\,. 
\end{equation}
The above potentials can be efficiently computed by taking recourse to the solution of a Poisson problem, as the kernel of the extended interaction corresponds to the Green's function of the Laplace operator. 

The above local formulation of electrostatics has been efficiently implemented in \DFTFEverprev~\cite{motamarri2020} using finite-element (FE) discretization, where the point nuclear charges are enforced to be coincident with the corner nodes of the FE triangulation. The FE discretization provides the regularization of the solution of $\vselfdelta a$ and $\varphi$, and we refer to~\cite{motamarri2013} for a detailed discussion and systematic convergence studies. However, this approach requires a more refined discretization around the nuclei for an accurate solution of the Kohn-Sham ground-state problem, even in the case of pseudopotential calculations with a smooth external potential, $V^{a}_{\text{psp,loc}}(|\bx-\bR_a|)$. To alleviate these issues, in this work, we consider fictitious non-overlapping smeared nuclear charges, $b(\bx,\bR)=\bsm (\bx,\bR) = \sum_r \sum_{a} \bsmi a (\abs{\bx-(\bR_a+\bL_r)})$, in the spirit of ~\cite{Pask2012}. This results in significantly smoother auxiliary electrostatic potentials, which are free of the Coulomb singularities, and are denoted by:
\begin{equation}\label{eq:potsmeared}
\vself a(\bx,\bR_a) =\int\displaylimits_{\Rthree} \frac{\bsmi a (|\by-\bR_a|)}{|\bx-\by|} \dy,\;\;\;\; \varphi(\bx,\bR) = \int\displaylimits_{\Rthree} \frac{\rho(\by) + \bsm(\by,\bR)}{|\bx - \by|} \dy \,,
\end{equation}
where  $\vself a(\bx,\bR_a)$ matches the exact nuclear potential ($-\frac{Z_a}{|\bx-\bR_a|}$) beyond a certain cutoff radius, denoted by $r_c^a$. 
The modified local reformulation of electrostatics incorporating the above auxiliary potentials  is obtained as
\begin{equation}\label{eq:eleclocalsmeared}
\begin{split}
  E_{\text{el}} &= \frac{1}{2} \int\displaylimits_{\Omega_p} (\rho(\bx)\, +\, \bsm(\bx,\bR))\, \varphi(\bx,\bR) \dx- \frac{1}{2}\sum_{r}\sum_{a}\int\displaylimits_{\Omega_p} \bsmi a (\abs{\bx-(\bR_a+\bL_r)})\vself a(\bx,\bR_a+\bL_r)\dx \\&+\sum_{a} \sum_r \int\displaylimits_{\Omega_p} \left(V^{a}_{\text{ext,loc}}(\abs{\bx-(\bR_a+\bL_r)}) - \vself a(\bx,\bR_a+\bL_r)\right)\rho(\bx)\dx + E_{\text{psp,nl}} \,\,,
  \end{split}
\end{equation}
where the all-electron treatment is obtained by setting $V^{a}_{\text{ext,loc}}(|\bx-\bR_a|)=V^{a}_{\text{nuc}}(|\bx-\bR_a|):= -\frac{Z_a}{|\bx-\bR_a|} $, and the pseudopotential treatment is given by setting $V^{a}_{\text{ext,loc}}(|\bx-\bR_a|)=V^{a}_{\text{psp,loc}}(|\bx-\bR_a|) $.
Next, following~\cite{Pask2012}, we define $\bsmi a$ to be
\begin{align}\label{eq:smearedcharge}
\bsmi a (r)= & -Z_a g (r,r_c^a)\,,
\end{align}
where $g (r,r_c^a)$ denotes a smooth radial smeared charge distribution for a unit charge, which is strictly local within $r \le r_c$ and integrates to unity. Further, the value of $r_c^a$ for each atom is chosen to be smaller than half the nearest neighbour distance. Here, we use the following second-derivative continuous form for $g (r,r_c)$~\cite{Pask2012}:
\begin{equation}\label{eq:smearedchargePask}
  g (r,r_c) =
  \begin{cases}
     \frac{-21(r-r_c)^3(6r^2+ 3r r_c +r_c^2)}{5 \pi r_c^8} & r\le r_c\\
    0 &  r > r_c
  \end{cases}
\end{equation}
We remark that a similar smeared nuclear charges strategy has also been adopted in a recent work~\cite{rufus2020fast}, involving enriched finite-elements for solution of all-electron Kohn-Sham DFT equations. Finally, incorporating the functional derivative of the local reformulation of the electrostatic energy with smeared nuclear charges in Eq.~\ref{eq:eleclocalsmeared} into Eq.~\ref{eq:contEigenBZ}, the governing equations of Kohn-Sham DFT for the periodic unit cell are:
\begin{subequations}\label{eq:ksproblem}
\begin{align}
& \left(-\frac{1}{2} \laplacian - i \bk \cdot \grad +\frac{1}{2}{\abs{\bk}}^2+ V_{\text{xc}}(\bx) + \varphi(\bx)   + \sum_{a}\,\sum_r(V^{a}_{\text{ext,loc}}(\abs{\bx-(\bR_a+\bL_r)}) - \vself a (\bx,\bR_a+\bL_r)) \right)u_{n,\bk} \notag\\ & \qquad\qquad+ e^{-i\bk\cdot\bx} \sum_{a}\,\sum_r \int\displaylimits_{\Rthree} V_{\text{psp,nl}}^a (\bx,\by,\bR_a+\bL_r) e^{i\bk\cdot\by}\,u_{n,k}(\by) \dy = \epsilon_{n,\bk} \,{u}_{n,\bk}\qquad\text{on}\;\; \Omega_p, \,\,; \,\, \forall \,\,\bk\ \in BZ\, \label{eq:evp}\\
\;\;\;\;&-\frac{1}{4\,\pi}\laplacian\, \varphi(\bx) =  \rho(\bx) +\bsm(\bx,\bR) \;\;\qquad \text{on}\;\;  \Omega_p\,,\\
& 2\sum_{n}\fintd_{BZ}f(\epsilon_{{n,\bk}},\mu) \dk = N_e\,,\;\;\;\; 
\rho(\bx) = 2\sum_{n} \fintd_{BZ} f(\epsilon_{n,\bk},\mu)|u_{n,\bk}(\bx)|^2 \dk\,,
\end{align}
\end{subequations}
where the action of the non-local pseudopotential operator, $V_{\text{psp,nl}}^a$, is given in Eq.~\ref{eq:nlps}.
We note that the above equations can be easily modified to the non-periodic setting by considering $(\bk={\bf 0}, \bL_r={\bf 0})$ and replacing $\Omega_p$ with $\Omega_0$, which denotes a large non-periodic domain with a vacuum layer surrounding the material system with appropriate Dirichlet boundary conditions on $\varphi$.

\subsection{Configurational forces with smeared nuclear charges}\label{sec:forces}
Evaluation of ionic forces and periodic unit cell stresses in DFT are required for structural optimization and molecular dynamics. In \DFTFENoSpace, we employ the configurational forces approach~\cite{motamarri2018} to evaluate them in an unified and variationally consistent manner. The formulation naturally accounts for the dependence of the basis functions on the ionic positions (Pulay force corrections), in this case the FE basis functions. The configurational forces are evaluated as the generalized directional derivative of the Kohn-Sham free energy functional with respect to the position of the material point $\bx$. The Kohn-Sham free energy for a given position of atoms  is formulated as a local variational problem, whose Euler-Lagrange equations correspond to the Kohn-Sham governing equations (cf. Eq.~\ref{eq:ksproblem}). We refer to~\cite{motamarri2018} for a detailed discussion on the derivation of configurational forces, and to~\cite{motamarri2020} for the configurational force expression used in \DFTFEverprevNoSpace, which includes the expressions involving GGA type exchange correlation functionals. 

The expression for the configurational forces, which we derive below, differs from our previous works in several key aspects of the derivation. First, is the use of non-overlapping smeared nuclear charges, $\bsmi a(\abs{\bx-\bR_a})$, in the local reformulation of the electrostatics in this work (cf. Eq.~\ref{eq:eleclocalsmeared}) instead of the regularized Dirac distributions, $\bdeltai a(\abs{\bx-\bR_a})$, used in our previous works. In our previous formulation~\cite{motamarri2018}, we impose rigid body constraints to the perturbation of the material point  $\bx$ in the compact support of the regularized Dirac distribution nuclear charge in order to preserve the integral constraint $\int \bdeltai a(\abs{\bx-\bR_a}) d\bx=1$. We note that the above regularization and rigid body perturbation constraints are naturally realized in the previous formulation by considering the charge distribution as point charges coincident with the nodes of the FE triangulation. However, in the current formulation, the smeared nuclear charges are independent of the underlying FE discretization. Thus we relax the aforementioned constraints leading to additional terms in the configurational force expression. The second key difference from the previous formulation is in the treatment of the smeared charge self potential, $\vself a(\bx,\bR_a)$, where we no longer assume that this is a radial function dependent only on  $\abs{\bx-\bR_a}$, which may not hold for the FE discretized potential due to discretization errors. In order to compute the relevant response of  $\vself a(\bx,\bR_a)$ with respect to perturbation of the material point $\bx$, we  separately consider the response of perturbations with respect to $\bx$, $\by$ and $\bR_a$ in the non-local expression for $\vself a(\bx,\bR_a)=\int\displaylimits_{\Rthree} \frac{\bsmi a (|\by-\bR_a|)}{|\bx-\by|} \dy$ (cf. Eq.~\ref{eq:potsmeared}.). We develop strategies to evaluate the resulting terms in an accurate and efficient manner. Finally, the other remaining consideration in this work is the use of integration by parts to transfer the gradient operators away from potentially non-smooth fields (e.g. non-local pseudopotential projectors) on to the solution fields (${u}_{n,\bk}(\bx),\varphi(\bx),\vself a(\bx,\bR_a)$), which are relatively smoother in case of pseudopotential calculations. This results in a drastic reduction in the number of quadrature points required to accurately integrate the terms involving the smeared nuclear charges, local part of the pseudopotential, and non-local pseudopotential projectors. 

The starting point for deriving the configurational forces is the local variational problem of the Kohn-Sham free energy $\mathcal{F}_0(\bR)$. Based on prior work~\cite{motamarri2018,motamarri2020}, the local variational problem can be formulated in terms of wavefunctions, fractional occupancies and the electrostatic potentials as:
\begin{equation}\label{saddlepoint}
\mathcal{F}_0(\bR) = \min_{\bff_{\bk}^{\prime} \in [0,1]^{N}}\min_{\bU_{\bk}^{\prime} \in (\mathcal{Y})^{N}} \;\max_{\varphi^{\prime} \in \mathcal{Y}}\; \mathcal{L}(\bff^{\prime},\bU^{\prime},\varphi^{\prime};\bR)\;\;\text{such that}\;\;\int\displaylimits_{\Omega_p} {u^{\prime}_{i, \bk}}^{\ast}u_{j, \bk}^{\prime} \dx = \delta_{ij},\;\;2\sum_{n=1}^{N}\fintd_{BZ} f_{n,\bk}^{\prime} \dk = N_e\,,
\end{equation}
\vspace{-0.25in}
\begin{align*}
\text{where}\;\;\;\mathcal{L}(\bff^{\prime},\bU^{\prime},\varphi^{\prime};\bR) = \widetilde{\mathcal{L}}(\bff^{\prime},\bU^{\prime}) 
+ \min_{\mathcal{V}^{\prime} \in (H^{1}(\Rthree))^{N_a}}\mathcal{L}_{\text{el}}(\bff^{\prime},\bU^{\prime},\varphi^{\prime},\mathcal{V}^{\prime};\bR)\,,
\end{align*}
\vspace{-0.1in}
\begin{equation}
\text{with}\;\;\widetilde{\mathcal{L}}(\bff^{\prime},\bU^{\prime}) = T_{\text{s}}(\bff^{\prime},\bU^{\prime})+ E_{\text{xc}}(\rho) + E_{\text{ent}}(\bff^{\prime})\,,
\end{equation}
where $\mathcal{Y}$ and $H^{1}(\Rthree)$ denote suitable function spaces. In the above, we note that \newline $\bU^{\prime} = \{u_{1, \bk}^{\prime}(\bx),u_{2, \bk}^{\prime}(\bx),u_{3, \bk}^{\prime}(\bx), \cdots, u_{N, \bk}^{\prime}(\bx) \,\, \forall \,\,\bk\ \in BZ\}$, denotes the trial Kohn-Sham wavefunctions, $\bff^{\prime} = \{f_{1, \bk}^{\prime},f_{2, \bk}^{\prime},f_{3, \bk}^{\prime}\cdots f_{N, \bk}^{\prime} \,\, \forall \,\,\bk \in BZ \}$ denotes the vector of trial orbital occupancy factors, and
$\mathcal{V}^{\prime} = \{{V_{\text{sm}}^{\prime}}^{1},{V_{\text{sm}}^{\prime}}^{2},\cdots,{V_{\text{sm}}^{\prime}}^{N_{at}}\}$ denotes the vector containing the trial electrostatic potentials corresponding to all nuclear charges in the periodic unit cell. Here, $T_{\text{s}}(\bff^{\prime},\bPsi^{\prime})$ denotes the kinetic energy of non-interacting electrons and $E_{\text{ent}}(\bff^{\prime})$ denotes the electronic entropy contribution, and the corresponding expressions are given by
\begin{gather}
 T_{\text{s}}(\bff^{\prime},\bU^{\prime}) = 2\,\sum_{n=1}^{N}\;\fintd_{BZ}\,\int\displaylimits_{\Omega_p} f_{n, \bk}^{\prime}\, {u_{n,\bk}^{\prime}}^{\ast}(\bx) \left(-\frac{1}{2} \laplacian -i \bk \cdot \grad + \frac{1}{2} {\abs{\bk}}^2 \right)\,u_{n,\bk}^{\prime}(\bx) \dx \dk \,,\\
 E_{\text{ent}} (\bff^{\prime}) = -2\,\sigma\; \sum_{n=1}^{N} \fintd_{BZ} \left[f_{n, \bk}^{\prime} \ln f_{n, \bk}^{\prime} + (1- f_{n, \bk}^{\prime})\ln(1 - f_{n, \bk}^{\prime})\right]\dk\,.
\end{gather}
The energy functional corresponding to electrostatic energy, $\mathcal{L}_{\text{el}}$ using the local reformulation of the electrostatics with smeared nuclear charges (cf. Eq.~\ref{eq:eleclocalsmeared}) is expressed as:
\begin{equation}
\begin{split}
&\mathcal{L}_{\text{el}}(\bff^{\prime},\bU^{\prime},\varphi^{\prime},\mathcal{V}^{\prime},\bR) = \int\displaylimits_{\Omega_p}\left[-\frac{1}{8\pi} |\grad \varphi^{\prime}(\bx)|^2 + (\rho(\bx) + \bsm(\bx,\bR))\varphi^{\prime}(\bx)\right]\dx \\
&+ \sum_{a} \int\displaylimits_{\Rthree} \left[\frac{1}{8\pi} \abs{\grad {{V^{\prime}_{\text{sm}}}^{a}(\bx, \bR_a)}^2}  -  \bsmi a(\abs{\bx - \bR_a}) {V^{\prime}_{\text{sm}}}^{a}(\bx, \bR_a) \right]\dx\,\\ &+ \sum_{a} \;\sum_r \int \displaylimits_{\Omega_p} \left[V^{a}_{\text{ext,loc}}(\abs{\bx-(\bR_a+\bL_r)}) - \vself a\left(\bx,\bsmi a(|\by-(\bR_a+\bL_r)|)\right)\right]\rho(\bx)\dx \\
 & + 2 \sum_{n=1}^{N} \fintd_{BZ} \: \int \displaylimits_{\Omega_p}\:  f_{n,\bk}^{\prime}\; {u^{\prime}_{n,\bk}}^{\ast}(\bx)\; \sum_{a} \sum_{lpm} \sum_{r} e^{-i\bk \cdot (\bx - \bL_r)}\; {{C^{\prime}}^{a,n\bk}_{lpm}}\,h^{a}_{lp}\,\,\chi_{lpm}^{a}(\bx,\bR_a+\bL_r)\dx\dk\,,
 \end{split}
\end{equation}
where $\vself a$ denotes the electrostatic potential corresponding to the $a^{th}$ nuclear charge.


Let $\chieps : \Omega_p \rightarrow {\Omega_p}^{\prime}$ represent the infinitesimal perturbation of the underlying space that preserves the periodic geometry. The mapping of a material point $\bx$ to a new point $\bx'$, given by
\begin{equation}
    \bx'=\bx+ \epsilon \bdir\,,
\end{equation}
where $\bdir$ denotes the generator of this mapping. Further, the reciprocal space also gets perturbed by the mapping $\kappa^{\prime}={\boldsymbol \kappa}^{\varepsilon}(\bk)$ due to the perturbation of the real-space. Denoting $\mathcal{F}_0(\chieps)$ to be the ground-state free energy in the perturbed space, the configurational force with respect to this perturbation along the generator $\bdir$ is evaluated by computing the G\^ateaux derivative of  $\mathcal{F}_0(\chieps)$ given by
\begin{align}\label{eq:Eshelbyforce}
\frac{d\mathcal{F}_0(\chieps)}{d\varepsilon}\atzerob = &\int\displaylimits_{\Omega_p} \bE:\grad \bdir (\bx) \dx + \int\displaylimits_{\Omega_p} \bE_{\text{ext,corr}}:\grad \bdir (\bx) \dx + \sum_{a} \int\displaylimits_{\Rthree} \bE'^a:\grad \bdir (\bx) \dx \,\notag\\&+\, \text{F}^{\text{sm}}\,+\, \text{F}^{\text{ext,corr}}\,+\, \text{F}^\text{psp,nl} \,+\,  \text{F}^{K}\,,
\end{align}
where $\bE$, $\bE_{\text{ext,corr}}$  and  $\bE'^a$ denote the Eshelby tensors whose expressions in terms of the solutions of the saddle point problem ~\eqref{saddlepoint} on the unperturbed space ($\bx$) are provided below.  If $\bUbar = \{\ubar_{1, \bk},\ubar_{2, \bk},\ubar_{3, \bk} \cdots \ubar_{N, \bk}\}$ denote the Kohn-Sham eigenfunctions corresponding to the lowest $N$ eigenvalues $\bar{\boldsymbol{\epsilon}}=\{\bar{\epsilon}_{1, \bk},\bar{\epsilon}_{2,\bk},\cdots,\bar{\epsilon}_{N,\bk}\}$ with occupancies $\bar{\bff}=\{\fbar_{1, \bk},\fbar_{2, \bk} \cdots \fbar_{N, \bk}\}$, and $\varphibar$ denotes the electrostatic potential---all solutions of the saddle point problem ~\eqref{saddlepoint})---the expressions for the Eshelby tensors: $\bE$, $\bE_{\text{ext,corr}}$, $\bE'^{a}$, and force terms: $\text{F}^{\text{sm}}$, $\text{F}^\text{ext,corr}$, $\text{F}^\text{psp,nl}$ in Eq.~\ref{eq:Eshelbyforce} are given by
\begin{align*}
&\bE = \Biggl(\sum_{n=1}^{N}\;\fintd_{BZ}\,\fbar_{n,\bk}\Bigl(\grad \ubar_{n,\bk}^{*}(\bx)\cdot\grad \ubar_{n,\bk}(\bx) - 2\,i\,\ubar^{*}_{n,\bk}(\bx) \bk \cdot \grad \ubar_{n,\bk}(\bx) + \left(\abs{\bk}^2-2 \bar{\epsilon}_{n,\bk}\right) \,\ubar^{*}_{n,\bk}(\bx)\; \ubar_{n,\bk}(\bx)\Bigr) \dk \notag\\& + \varepsilon_{\text{xc}}(\rhobar,\grad \rhobar)  -\frac{1}{8\pi} |\grad \varphibar(\bx)|^2 +  \rhobar(\bx) \varphibar(\bx)
 \Biggr)\bI \,\notag\\&- \sum_{n=1}^{N}\;\fintd_{BZ}\,\fbar_{n,\bk}\Bigl[\grad \ubar_{n,\bk}^{*}(\bx) \otimes \grad \ubar_{n,\bk}(\bx) 
 + \grad {\ubar_{n,\bk}}(\bx) \otimes \grad \ubar_{n,\bk}^{*}(\bx)-2 \,i\,\ubar_{n,\bk}^{*}(\bx)  \left( \grad {\ubar_{n,\bk}}(\bx) \otimes \bk)\right)\Bigr] \dk \,\notag\\
& - \left.\frac{\partial}{\partial \grad \rho}(\varepsilon_{\text{xc}}(\rho,\grad \rho)) \right|_{\grad \rho=\grad \rhobar} \otimes \grad \rhobar  + \frac{1}{4\pi}\grad \varphibar(\bx) \otimes \grad \varphibar(\bx)\,,\\
&\bE_{\text{ext,corr}} = \Biggl( \sum_a \,\sum_r \Bigl(V_{\text{ext,loc}}^{a}(|\bx-(\bR_a+\bL_r)|) - \vself a\left(\bx,(\bR_a+\bL_r)\right)\Bigr)\rhobar(\bx) \Biggr)\bI\,,\\
&\bE'^{a} = \frac{1}{8\pi} |\grad V_{\text{sm}}^{a}(\bx, \bR_a)|^2 \bI - \frac{1}{4\pi}\grad V_{\text{sm}}^{a}(\bx, \bR_a) \otimes \grad V_{\text{sm}}^{a} (\bx, \bR_a)\,,
\end{align*}
where
\vspace{-0.1in}
\begin{gather}
\rhobar(\bx) = 2\sum_{n=1}^N \;\fintd_{BZ}\, \fbar_{n,\bk}\,\ubar^{*}_{n,\bk}(\bx)\,\ubar_{n,\bk}(\bx) \dk\,,\;\;\;\;\fbar_{n,\bk} = \frac{1}{1 + \exp(\frac{\bar{\epsilon}_{n,\bk} -\mu}{k_B\,T})} \,.\notag
\end{gather}
Further,
\begin{align}
\text{F}^{\text{sm}} = & \sum_{r} \sum_{a}\ \int\displaylimits_{\Omega_p} \bsmi a(\abs{\bx-(\bR_a+\bL_r)}) \grad \varphibar(\bx) \cdot \left(\bdir(\bx) - \bdir((\bR_a+\bL_r)) \right)\dx \notag\\ &- \sum_{a} \int\displaylimits_{\Rthree} \bsmi a(\abs{\bx-\bR_a}) \grad V_{\text{sm}}^{a} (\bx, \bR_a) \cdot \left(\bdir(\bx) - \bdir(\bR_a) \right)\dx\,,\notag
\end{align}
\begin{align}
\text{F}^{\text{ext,corr}} = &\sum_{a}\,\sum_r \int\displaylimits_{\Omega_p} \rhobar(\bx)\Biggl(\grad V^{a}_{\text{ext,loc}}(|\bx-(\bR_a+\bL_r)|)\cdot \left(\bdir(\bx) - \bdir(\bR_a+\bL_r) \right) \notag\\&- \grad_{\bx} \vself a(\bx,\bR_a+\bL_r)\cdot \bdir(\bx) - \left.\frac{\partial \vselfeps a\left(\bx,\bsm(\abs{\chieps(\by)-\chieps(\bR_a+\bL_r)})\right)}{\partial \varepsilon}\right|_{\varepsilon=0} \Biggr)\dx   \,,\notag
\end{align}
and finally, $\text{F}^\text{psp,nl}=\text{F}^\text{nl}\,+\,{\text{F}^\text{nl}}^{*}$, with
\begin{align*}
\text{F}^\text{nl}=-2 \sum_{n=1}^{N} \fintd_{BZ} \: \int \displaylimits_{\Omega_p}\:  & \fbar_{n,\bk}\;  \sum_{a} \sum_{lpm} \sum_{r} e^{-i\bk \cdot (\bx - \bL_r)} \; {\bar{C}}_{lpm}^{a,n\bk}\,h^{a}_{lp}\,\,\chi_{lpm}^{a}(\bx,\bR_a+\bL_r)\,\,\,\notag\\
&\Bigl(\grad \ubar^{*}_{n,\bk}(\bx)\cdot\left(\bdir(\bx) - \bdir(\bR_a+\bL_r)\right)
 -i\bk \cdot \bdir(\bR_a)\Bigr) \dx\dk \,.
\end{align*}
%
In the above, the terms $\text{F}^{\text{sm}}$ and $\text{F}^{\text{ext,corr}}$, which capture the electrostatic contributions related to smeared nuclear charges and external potential correction, are the significant changes compared to previous versions of \DFTFENoSpace~\cite{motamarri2020,motamarri2018}. We refer to ~\ref{app:forces} for details regarding the combined evaluation of
\begin{equation*}
\int\displaylimits_{\Omega_p} \bE_{\text{ext,corr}}:\grad \bdir (\bx) \dx + \text{F}^{\text{ext,corr}}\,,    
\end{equation*}
requiring additional consideration for accurate and efficient evaluation of $\left.\frac{\partial \vselfeps a(\bx,\bsm(\abs{\chieps(\by)-\chieps(\bR_a)})}{\partial \varepsilon}\right|_{\varepsilon=0}$.
Further, we refer to~\cite{motamarri2018} for the detailed expression and derivation of $\text{F}^{K}=\frac{\partial\mathcal{F}_0(\chieps)}{\partial {\boldsymbol \kappa}^{\varepsilon}(\bk)}\frac{\partial {\boldsymbol \kappa}^{\varepsilon}(\bk)}{\partial \varepsilon}\atzerob$. Additionally, we refer to~\cite{das2015,motamarri2018} for details on recasting the integral over $\Rthree$ in the third term in Eq.~\ref{eq:Eshelbyforce}, as sum of integrals over bounded domains and surface integrals on the bounded domains. 

We remark that using Eq.~\ref{eq:Eshelbyforce}, the $i^{\textrm{th}}$ component of the ionic force on any given atom is computed by choosing the compact support of the generator, $\dir_i (\bx)$, to contain only the atom of interest. Along similar lines, the periodic unit cell stress tensor is evaluated by choosing the generator to be an appropriate affine transformation: $\dir_{i} = C_{ij}x_j$, as explained in ~\cite{motamarri2018}. In the context of ionic forces for the pseudopotential case, we employ a further simplification in the evaluation of the configurational forces by taking advantage of the smooth electronic fields to allow  the atom positions to float independent of the underlying space. The ionic forces are computed in this case by setting $\bdir(\bR_a+\bL_r)$ to be an unit vector along the desired force component and setting  $\bdir(\bx)=0$. The ionic force expression thus obtained can be shown to be equivalent to the Hellman-Feynman force~\cite{rmartin}, as would be expected.

\subsection{Discrete Kohn-Sham DFT equations}\label{sec:fem}
We now present the discretization of the Kohn-Sham DFT problem using a finite-element (FE) basis, which is a piecewise polynomial and strictly local basis set~\cite{brenner2002}. In particular, we employ $C^{0}$ continuous Lagrange polynomials generated using Gauss-Lobatto-Legendre (GLL) nodal points, also known as spectral finite-elements (FE). We refer to our prior work~\cite{motamarri2013} for details on the adaptive higher order spectral FE based discretization of the Kohn-Sham equations. We briefly discuss these aspects, and also present those aspects that differ in this work compared to the previous implementation, \DFTFEverprev~\cite{motamarri2020}. We begin by discretization of the electronic fields in the Kohn-Sham problem Eq.~\ref{eq:ksproblem}---the wavefunctions and the electrostatic potentials---in the FE basis given by
\begin{equation}\label{eq:fem}
u^{h}_{n,\bk}(\bx) = \sum_{j=1}^{M} N^{h}_j(\bx) u^{j}_{n,\bk}\,\,,\;\;\;\;\varphi^{h}(\bx) = \sum_{j=1}^{M} N^{h}_j(\bx) \varphi^{j}\,,\;\;\;\;{\vself {a^{h}}}(\bx) = \sum_{j=1}^{M} N^{h}_j(\bx)\vself {a^{j}} \,,
\end{equation}
where $N^{h}_{j}:1\leq j \leq M$ denotes the strictly local Lagrange polynomials generated using the nodes of the FE triangulation, $\mathcal{T}^h$, with the characteristic mesh size denoted by $h$. Further, the order of the interpolating polynomials is denoted by $p$. $u^{h}_{n,\bk}$, $\varphi^{h}$ and $\vself {a^{h}}$ denote the FE discretized fields, with $u^{j}_{n,\bk}$, $\varphi^{j}$ and $\vself {a^{j}}$ denoting the coefficients in the expansion of the $n^{th}$ discretized wavefunction and the electrostatic potentials. The coefficients are also referred to as the nodal values of the discretized fields represented using the FE triangulation. Systematic reduction of the discretization error using the above FE basis is achieved through reducing $h$, or increasing $p$. We refer to~\cite{motamarri2013} for derivation of \textit{a priori} error estimates for the discrete ground-state energy of the Kohn-Sham problem, with respect to $h$ and $p$.

Applying the above FE discretization to the Kohn-Sham eigenvalue problem in Eq.~\ref{eq:ksproblem} results in a generalized hermitian eigenvalue problem (GHEP) given by $\bH^{\bk} \hat{\bvu}_{n,\bk} = \epsilon^{h}_{n,\bk} \bM \hat{\bvu}_{n,\bk}$ where $\bH^{\bk}$  denotes the discrete Hamiltonian matrix, $\bM$ denotes the overlap matrix (or commonly referred to as the mass matrix in finite element literature), and $\epsilon^{h}_{n,\bk}$ denotes the $n^{th}$ eigenvalue corresponding to the discrete eigenvector $\hat{\bvu}_{n,\bk}$. The discrete Hamiltonian matrix is further decomposed as $\bH^{\bk} = \bHlock + \bHnlk$, where 
\begin{align}\label{eq:discreteHam}
\text{H}_{jk}^{\text{loc},\bk} = &\frac{1}{2} \int_{\Omega_p} \grad N^h_{j} (\bx) \cdot \grad N^h_k(\bx) \dx +\int_{\Omega_p} \left(V^{h}_{\text{eff,loc}}(\rho^h,\bR)+ \frac{1}{2} {\abs{\bk}}^2\right)\, N^h_{j}(\bx) N^h_{k}(\bx) \dx\, \notag\\
&+\int_{\Omega_p} \left. \frac{\partial \varepsilon_{\text{xc}}(\rho,\grad \rho)}{\partial \grad \rho} \right|_{\rho=\rho^h}\cdot \left(\grad N^h_{j}(\bx) N^h_{k}(\bx) + N^h_{j}(\bx) \grad N^h_{k}(\bx)\right) \dx 
-\int_{\Omega_p} i \bk \cdot \left( N^h_{j}(\bx) \grad N^h_{k}(\bx)\right) \dx\,.
\end{align}
In the above, $V_{\text{eff,loc}}^{h}(\rho^h,\bR)$ is given by:
\begin{align}\label{eq:vefflocdiscrete}
 V_{\text{eff,loc}}^{h}(\rho^h,\bR) = &  \left.\frac{\partial \varepsilon_{\text{xc}}(\rho,\grad \rho)}{\partial \rho}\right|_{\rho=\rho^h} + \varphi^{h}(\bx) + \sum_a \sum_r \left(V^{a}_{\text{ext,loc}}(|\bx-(\bR_a+\bL_r)|) - {\vself {a^{h}}}(\bx,\bR_a+\bL_r)\right)\,,
\end{align}
where ${\vself {a^{h}}}(\bx,\bR_a+\bL_r)$ is obtained by solving the following discretized Poisson equations:
\begin{align}\label{eq:vselfDiscreteSolve}
\sum_{j=1}^{M}\left[\frac{1}{4\pi}\int_{\Omega_a} \grad N_i^{h}({\bx}). \grad N_j^{h}({\bx})\,{\rm d} {\bf x} \right]V^{a^j}_{\rm{sm}}
=\int_{\Omega_a}\left( \bsmi a(|\bx - \bR_a|) \right)N_i^{h}({\bf x}) \,{\dx} \,, \quad \forall a \,.
\end{align}
Here, $\Omega_a$ denotes a bounded domain enclosing the atom $a$ with Dirchlet boundary conditions set to the exact nuclear potential, $V_{\text{nuc}}^a(|\bx-\bR_a|)$. The distance of the Dirchlet boundary of $\Omega_a$ from $\bR_a$ must be much larger than the smeared charge cut-off, $r_c^a$, for an accurate solution of ${V}^{a^h}_{\rm{sm}}$.

We note that, in the case of all-electron calculations, $V^{a}_{\text{ext,loc}}(|\bx-(\bR_a+\bL_r)|)  = V^{a}_{\text{nuc}}(|\bx-(\bR_a+\bL_r)|) $ and $\bHnlk$ is zero. In the case of pseudopotential calculations, $V^{a}_{\text{ext,loc}}(|\bx-(\bR_a+\bL_r)|)  = V^{a}_{\text{psp,loc}}(|\bx-(\bR_a+\bL_r)|) $ and additionally, $\bHnlk$ is given by
\begin{equation}\label{eq:nonlocalHamDiscrete}
\text{H}_{jk}^{\text{nl},\bk} = 
\sum_{a} \sum_{lpm} {C^{a,\bk}_{lpm,j}}h^{a}_{lp}{C^{{a,\bk}^{\ast}}_{lpm,k}}\,,\;\;\text{where}\;\;\;C^{a,\bk}_{lpm,j} =  \int_{\Omega_p} \sum_r e^{-i\bk\cdot(\bx - \bL_r)} \chi^{a}_{lpm}(\bx,\bR_a+\bL_r) N^h_j(\bx) \dx\,.
\end{equation}

Finally, the matrix elements of the overlap matrix $\bM$ in the GHEP are given by $\text{M}_{jk}=\int_{\Omega_p} N^h_j(\bx) N^h_k(\bx) \dx$. We note that the matrices $\bHlock$ and $\bM$ are sparse as the FE basis functions are strictly local in real space  with a compact support over elements sharing a given FE node. Further, the vectors $C^{a,\bk}_{lpm,j}$ in $\bHnlk$ are also sparse since the projectors $\chi^{a}_{lpm}(\bx,\bR_a+\bL_r)$ have a compact support, thus rendering a sparse structure to the discrete Hamiltonian $\bH^{\bk}$.


We transform the GHEP into a standard Hermitian eigenvalue problem (SHEP), which can be solved more efficiently. Using the fact that there exists an unique positive definite symmetric square root of $\bM$, the GHEP---$\bH^{\bk} \hat{\bvu}_{n,\bk} = \epsilon^{h}_{n,\bk} \bM \hat{\bvu}_{n,\bk}$---is transformed to the following SHEP: 
\begin{equation}
 \btH^{\bk} \widetilde{\bvu}_{n,\bk} =  \epsilon^{h}_{n,\bk} \widetilde{\bvu}_{n,\bk} \label{hep}\,,\quad\quad
 \text{where} \;\;\; \widetilde{\bvu}_{n,\bk} =\bM^{1/2} \hat{\bvu}_{n,\bk}\,,\;\;\;\;
\btH^{\bk} = \bM^{-1/2}\bH^{\bk}\bM^{-1/2}\,.
\end{equation}
However, the evaluation and application of $\bM^{-1/2}$ is computationally expensive. To this end, we employ Gauss-Lobatto-Legendre (GLL) quadrature rules---with quadrature points coincident with the nodes of the spectral FE basis---to evaluate the integrals in the overlap matrix, which renders the overlap matrix diagonal~\cite{motamarri2013}. Furthermore, $\btH^{\bk}$, the Hermitian matrix in the SHEP, retains the sparsity of $\bH^{\bk}$ in the GHEP. Finally,  the discrete Kohn-Sham eigenvalue problem along with the discretized Poisson equation for the total electrostatic potential $\varphi^{h}$  are given by:
\begin{subequations}\label{eq:discreteSolve}
\begin{gather}
  \btH^{\bk} \widetilde{\bvu}_{n,\bk} =  \epsilon^{h}_{n,\bk} \widetilde{\bvu}_{n,\bk}\,\,; \qquad \forall \,\,\bk\ \in BZ\label{eq:kohnShamSolve} \\
\sum_{j=1}^{M}\left[\frac{1}{4\pi}\int_{\Omega_p} \grad N_i^{h}({\bx}). \grad N_j^{h}({\bx})\,{\dx}  \right]\varphi^j
=\int_{\Omega_p}\left(\rho^h({\bx}) + \bsm(\bx,\bR)\right)N_i^{h}({\bf x}) \,{\rm d}{\bf x} \,, \quad \label{eq:phiTotDiscreteSolve}\\
2\sum_{n=1}^{N} \: \fintd_{BZ} \: f(\epsilon^{h}_{n,\bk},\mu) \dk = N_e\,,\;\;\;\;
\rho^{h}(\bx) = 2\sum_{n=1}^{N} \: \fintd_{BZ} \: f(\epsilon^h_{n,\bk},\mu)|{u^h_{n,\bk}}(\bx)|^2 \dk\,,
\end{gather}
\end{subequations}
which are solved self-consistently to obtain the electronic ground-state.  Subsequently, we compute the discrete total ground-state energy  $E^h$  in terms of the discrete ground-state solution fields ($\bar{\epsilon}^h_{n,\bk}, \rhobar^h, \varphibar^{h}$) as follows:
\begin{align}\label{discreteenergy}
E^h = & 2 \sum_{n=1}^{N} \: \fintd_{BZ} \: f(\epsilonbar^h_{n,\bk},\mu) \epsilonbar^h_{n,\bk} \dk\notag\\ & - \int_{\Omega_p}  \left[ \rhobar^h(\bx) \left(\varphi^{h}(\bx)+ \left.\frac{\partial \varepsilon_{\text{xc}}(\rho,\grad \rho)}{\partial \rho}\right|_{\rho=\rho^h}  \right)   +  \left.\frac{\partial \varepsilon_{\text{xc}}(\rho,\grad \rho)}{\partial \grad \rho}\right|_{\grad \rho=\grad \rhobar^h} \cdot \grad \rhobar^h (\bx) \right]\dx \notag\\&+  \int_{\Omega_p} \frac{1}{2} (\rhobar^{h}(\bx) + \bsm(\bx,\bR))\varphibar^{h}(\bx)\dx -\sum_{a} \int_{\Omega_a} \frac{1}{2}  \bsmi a(|\bx - \bR_a|) \vself {a^{h}}(\bx) \dx \,.
\end{align}
We note that the use of smooth smeared nuclear charges, $\bsmi a$  allows the nuclear positions to float independent of the FE triangulation, leading to generation of more efficient FE triangulations with relatively coarser $h$ values and higher values of $p$ ($p$ = 6 or 7). Our earlier implementations with the nuclear positions restricted to corner nodes of FE elements typically used $p$ = 4 or 5 with smaller $h$ values, to maintain an appropriate mesh quality around the atom. In the case of all-electron calculations, we still require the nuclear positions to be located on the corner nodes of FE elements, to capture the cusp in the all-electron wavefunctions.

\section{Numerical methodology}
\label{sec:numerics}
In this section, we discuss the  solution procedure for the spectral finite-element discretized Kohn-Sham equations (cf. Eq.~\eqref{eq:discreteSolve}), implemented in \DFTFE. \DFTFE is built over the deal.II open-source FE library~\cite{dealII90}, and uses its infrastructure for adaptive mesh refinement, FE grid handling, and efficient parallel objects.  \DFTFE accommodates FE basis with adaptive spatial resolution, which is constructed using an octree-based hexahedral mesh generator based on the p4est library~\cite{p4est2011} through deal.II. We refer to \cite{motamarri2020} for a detailed discussion on the adaptive refinement strategies implemented in \DFTFE in the context of both pseudopotenial and all-electron problems.

Starting from an initial guess for Kohn-Sham electron density and eigenfunctions, the discrete nonlinear SHEP is solved self-consistently along with Poisson equations (cf. Eq.~\eqref{eq:discreteSolve}) to compute the Kohn-Sham ground-state solution. In particular, the nonlinear SHEP problem is treated as a fixed point iteration---also referred to as the self-consistent field (SCF) iteration---in terms of electron density: $\rho^h=F(\rho^h)$, where $F(\rho^h)$ involves computing the occupied eigenspace for a given $\rho^h$. We employ Anderson and Broyden electron density mixing schemes~\cite{anderson1965, broyden1965} to accelerate the fixed point iteration. Additionally, we have implemented the Kerker preconditioner~\cite{Kerker1981}, which provides an improved SCF convergence rate that is expected to be independent of system size for homogeneous metallic systems. In each SCF iteration step, we use a computationally efficient and scalable Chebyshev polynomial filtered eigensolver (ChFES)~\cite{saad2006,motamarri2013,motamarri2020} to evaluate the occupied eigenspace of the discrete Kohn-Sham Hamiltonian.  Furthermore, in each SCF iteration step, the total electrostatic potential is computed by solving a Poisson problem, using a Jacobi preconditioned conjugate gradient solver. Algorithm~\ref{alg:scf} lists all the steps in the SCF procedure followed in \DFTFE. We remark that although the formulation in Section~\ref{sec:ksdft} is presented in the context of multiple $k$-points sampling the first BZ zone, going forward we drop the dependence on $k$-points for notational convenience and clarity. All the numerical methods discussed below easily extend to multiple $k$-points.

The most dominant computational step in each SCF iteration (Algorithm~\ref{alg:scf}) is the ChFES procedure, which involves the Chebyshev polynomial filtering (CF) and the Rayleigh-Ritz (RRGEP) steps. CF scales quadratically with system size, and is observed to be the dominant computational cost for small and medium system sizes. However, for large system sizes the cubic-scaling computational cost of RRGEP dominates. The numerical implementation of ChFES in \DFTFEverprev~\cite{motamarri2020} enabled fast and large-scale DFT calculations by using scalable numerical algorithms and reduction of the computational prefactor of the cubic-scaling Rayleigh-Ritz procedure through development of mixed-precision strategies combined with a spectrum splitting approach. The numerical algorithm of ChFES in this work is along similar lines as \DFTFEverprevNoSpace. For the sake of completeness, we will briefly review the numerical methodology of ChFES, and also discuss the modified Rayleigh-Ritz procedure employed in the present work.

\begin{algorithm}
\caption{Self Consistent Field (SCF) iteration in \texttt{DFT-FE}\vspace{2mm}\\ 
\small{$M$: \# FE basis;\quad $N$: \# eigenstates;\quad $(.)^h$: FE discretized field}}
\label{alg:scf}
\small
\begin{algorithmic}[1]
\State Start with an initial guess for input electron density $\rho^h_{\rm in}({\bf x})$, and an initial guess for  $\bm{\widetilde{\Psi}}$.
\State For pseudopotential calculations, compute the FE cell level contributions to $\btHnl$ (cf. Eq.~\ref{eq:nonlocalHamDiscrete}).\vspace{1.5mm}
\State [TEP] Get total electrostatic potential contribution to $V_{\rm eff}^{h}(\rho^h_{\rm in}({\bf x}),{\bf R})$, by solving a discrete Poisson equation. ($\mathcal{O}(M {\rm log}(M))$)

\State [DHM] Compute FE cell level Hamiltonian matrices contributions to $\btHloc$. ($\mathcal{O}(M)$)
\vspace{1.5mm}
\State Perform ChFES procedure: $[\bPsitildeRfr,\bPsitildeo,\bDfr]={\rm ChFES}\left(\bPsitildein,\btH\right)$ 
\begin{algsubstates}
\State [CF] Chebyshev polynomial filtering: $\bPsitildef={\rm CF}\left(\bPsitildein\right)$. (cf. Section~\ref{sec:chebyfiltering}, $\mathcal{O}(MN)$)\vspace{1.5mm}
\State [RRGEP] Rayleigh-Ritz generalized eigenvalue problem in the subspace spanned by $\bPsitildef$: $[\bPsitildeRfr,\bPsitildeo,\bDfr]={\rm RRGEP}\left(\bPsitildef,\btH\right)$. (call Algorithm~\ref{alg:rrgepspectrumsplit}, $\mathcal{O}(MN^2)$ + $\mathcal{O}(N^3)$)
\end{algsubstates}
\vspace{1.5mm}
\State [DC] Compute new output electron density, $\rho^h_{\rm out}({\bf x})$, using $\bPsitildeRfr$,$\bPsitildeo$ and $\bDfr$. \,(call Algorithm~\ref{alg:elec-density}, $\mathcal{O}(MN)$)
\vspace{1.5mm}
\State [DM]  If $\norm{\rho^h_{\rm out}({\bf x}) - \rho^h_{\rm in}({\bf x})} \leq \textrm{tolerance}$, \textit{stop}; Else, compute new $\rho^h_{\rm in}({\bf x})$ using a mixing scheme, and go to step 3.
\end{algorithmic}
\end{algorithm}
\vspace{-0.1in}

\subsection{Chebyshev filtered eigensolver}\label{sec:chebyfiltering}
In order to solve the SHEP problem---\,$\btH \widetilde{\bvPsi}_{i} = \epsilon^{h}_{i} \widetilde{\bvPsi}_{i}$\,---we use the Chebyshev filtering procedure~\cite{saad2006,motamarri2013}, which exploits the fact that the eigenstates of interest correspond to the lowest occupied states (wanted eigenspectrum). The main idea behind the Chebyshev filtering approach is to construct a scaled-and-shifted Hamiltonian ($\bbarH=c_1 \btH +c_2$) such that the desired (wanted) eigenspectrum of the discrete Kohn-Sham Hamiltonian ($\btH$) is mapped to (-$\infty$, -1) and the remainder of the eigenspectrum is mapped to (-1,1). Subsequently, exploiting the property of Chebyshev polynomials, which have a fast growth in (-$\infty$, -1) and (1,$\infty$) but take small values in [-1,1], a subspace that is a close approximation to the desired eigensubspace can be computed. To elaborate, the Chebyshev filtered eigensolver (ChFES) is based on the action of a degree $m$ Chebyshev polynomial filter acting on an input subspace $ \bPsitildein$ given by the following recursive iteration:
\begin{equation}\label{eq:chebyfilter}
\bPsitildef = T_m(\bbarH) \bPsitildein \;\;\; \text{where} \;\;\; T_{m +1}(x) = 2xT_m(x) - T_{m-1}(x)\,, 
\end{equation}
where the choice of $m$ is based on the upper bound of $\btH$, which is related to the smallest mesh size~\cite{motamarri2020}.
The resulting subspace of Chebyshev filtered vectors, $\bPsitildef$, is a good approximation to the desired eigenspace, with the approximation error decreasing systematically with the degree of the polynomial. Further, the Chebyshev filter operations are applied to all wavefunction vectors simultaneously which lends to efficient implementation on modern computing architectures, as will be discussed later in Section~\ref{sec:arch}. Upon computing $\bPsitildef$, the desired eigenstates of $\btH$ required for computing the electron density are obtained by a Rayleigh Ritz procedure (step 5.b in Algorithm~\ref{alg:scf}), which involves projecting the SHEP onto the subspace spanned by $\bPsitildef$ and subsequently solving a much smaller generalized Hermitian eigenvalue problem (GHEP). We next discuss the numerical algorithm of the RRGEP procedure.

\subsection{Rayleigh-Ritz step with spectrum-splitting and mixed precision algorithms, and electron density computation}\label{sec:RR-SSMP}
The cubic scaling RRGEP procedure in our ChFES solver employs two key ideas---spectrum-splitting approach and mixed precision algorithms---to reduce the computational prefactor. First, considering the spectrum-splitting approach, the key idea is that the lowest $N_{\rm oc}$ eigenstates of the Kohn-Sham eigenvalue problem $\btH \widetilde{\bvPsi}_{i} = \epsilon^{h}_{i} \widetilde{\bvPsi}_{i}$ with orbital occupancy $f_i=1$ are not explicitly necessary for the computation of the electron density. This allows us to reformulate the density computation as only requiring the top $N_{\rm{fr}}=N-N_{\rm oc}$ fractionally occupied eigenstates ($\bPsitildeRfr$ of $\bhatH$) and the orthonormalized Chebyshev filtered wavefunctions ($\bPsitildeo$) (cf. Algorithm~\ref{alg:elec-density}).  For typically used Fermi-Dirac smearing temperatures $\sim 500 \,{\rm K}$, $N_{\rm{fr}}$ is 10--15\% of $N$ even for metallic systems, which can provide significant computational prefactor reduction in the RRGEP procedure. We remark that the computational advantages of using spectrum-splitting approach have been demonstrated in  \DFTFEverprev~\cite{motamarri2020}, with the algorithm being adapted from developments in previous works~\cite{rescu2016,amartya2018,motamarri2017}.  The RRGEP procedure implemented in \DFTFEver (cf. Algorithm~\ref{alg:rrgepspectrumsplit}) is composed of the following steps---(i) (RRGEP-OP) Computation of overlap matrix ($\bS=\bPsitildefDagger \bPsitildef $) and projection of the Hamiltonian matrix ($\bhatH = \bPsitildefDagger \btH \bPsitildef $) onto the space spanned by the Chebyshev filtered wavefunctions, $\bPsitildef$, (ii) (RRGEP-D) Solve GHEP: $ \bhatH \bQfr=\bS \bQfr \bDfr$, where $\bQfr$ and $\bDfr$ denote the top $N_{\rm{fr}}$ fractionally occupied eigenstates of the projected Hamiltonian with $\bDfr$ containing the corresponding eigenvalues in ascending order, and (iii) (RRGEP-ST) Transform the fractionally occupied eigenstates to the FE basis ($\bPsitildeRfr=\bPsitildef \bQfr$) as well as compute the orthonormalized Chebyshev filtered wavefunctions ($\bPsitildeo=\bPsitildef \bLinvDagger$). Subsequently, $\bPsitildeRfr$, $\bPsitildeo$, and $\bDfr$ are used to compute the output electron density, $\rho^h_{\rm out}({\bf x})$ at quadrature points inside each FE cell (cf. Algorithm~\ref{alg:elec-density}). In the above, the use of spectrum-splitting approach reduces the computational cost scaling of $\bPsitildeRfr$ in RRGEP-ST-F step to  $\mathcal{O}(MNN_{\rm{fr}})$ from $\mathcal{O}(MN^2)$. Furthermore, in the RRGEP-D step, the resulting SHEP solve (after recasting the GHEP) now entails a partial diagonalization problem, which can also be performed with a reduced prefactor. In \DFTFENoSpace, we perform partial diagonalization using ELPA library's~\cite{elpa2014,elpagpu} direct eigensolver.


\begin{algorithm}
\caption{Generalized eigenvalue problem based Rayleigh-Ritz step with spectrum-splitting procedure: $\left[\bPsitildeRfr,\bPsitildeo, \bDfr\right]={\rm RRGEP}\left(\bPsitildef,\btH\right)$ \vspace{1mm} \\ 
\small{$\bPsitildef$ and $\bPsitildeo$ are $M \times N$ matrices, $\bPsitildeRfr$ is a $M \times N_{\rm{fr}}$ matrix, and $\bDfr$ is a $N_{\rm{fr}} \times N_{\rm{fr}}$ diagonal matrix.} }
\label{alg:rrgepspectrumsplit}
\small
\begin{algorithmic}[1]
    \State [RRGEP-OP] Compute matrices for projection of Kohn-Sham eigenvalue problem onto the subspace spanned by $\bPsitildef$:\vspace{1mm}
    \begin{algsubstates}
       \State [RRGEP-O] Compute overlap matrix: $\bS=\bPsitildefDagger \bPsitildef $. ($\mathcal{O}(MN^2)$)\vspace{1mm}
       \State [RRGEP-P] Compute projected Hamiltonian matrix: $\bhatH = \bPsitildefDagger \btH \bPsitildef $. ($\mathcal{O}(MN^2)$)\vspace{1mm}   
    \end{algsubstates}
    \State [RRGEP-D] Solve for $N_{\rm{fr}}$ highest eigenstates of the generalized hermitian eigenvalue problem (GHEP) : $ \bhatH \bQfr=\bS \bQfr \bDfr$ ($\mathcal{O}(N^3)$)\vspace{1mm}
       \begin{algsubstates}
          \State Perform Cholesky factorization of $\bS=\bL \bLDagger$.
          \State Recast GHEP: $ \bhatH \bQfr=\bS \bQfr \bDfr$ into SHEP: $\left(\bLinv \bhatH \bLinvDagger\right) \bQPrimefr=\bQPrimefr \bDfr$, where $\bQPrimefr=\bLDagger \bQfr$.
          \State Solve the SHEP: $\left(\bLinv \bhatH \bLinvDagger\right) \bQPrimefr=\bQPrimefr \bDfr$ and compute $\bQfr=\bLinvDagger\bQPrimefr$.
       \end{algsubstates}
    \State [RRGEP-ST] Subspace transformation steps:
    \begin{algsubstates}
        \State [RRGEP-ST-F]  Subspace transformation to compute fractionally occupied eigenstates:   $\bPsitildeRfr=\bPsitildef \bQfr$. ($\mathcal{O}(MNN_{\rm{fr}})$)
        \State [RRGEP-ST-O]  Construct orthonormal basis: $\bPsitildeo=\bPsitildef \bLinvDagger$. ($\mathcal{O}(MN^2)$)
    \end{algsubstates}
 \end{algorithmic}
\end{algorithm}

\begin{algorithm}
\caption{Electron density computation $\rho^h_{\rm out}({\bf x})={\rm DC}\left[\bPsitildeRfr,\bPsitildeo, \bDfr\right]$}
\label{alg:elec-density}
\small
\begin{algorithmic}[1]
   \State Compute Fermi-energy ($\mu$) using the constraint:
     \begin{equation*}
       2\left(N_{\rm oc}+\sum_{i=N_{\rm oc}}^{N}f(\epsilon_i^h,\mu)\right)=N_e.
       \end{equation*}    
    \State Scale $\bPsitildeo$ and  $\bPsitildeRfr$: $\bm{{\hat{\Psi}}^{{\rm o}}}= {\bf M}^{-1/2}\bPsitildeo,\,
 \bm{{\hat{\Psi}}_{\rm fr}}^{\bf R}= {\bf M}^{-1/2}\bPsitildeRfr$.
    \State Compute electron density at any point ${\bf x}$ belonging to a FE cell $e$:
       \begin{equation*}
            \rho^h_{\rm out}({\bf x})=2 \,\bm{n^{{e}^{T}}}({\bf x})\left[ {\bm{{\hat{\Psi}}_{\bm e}^{{\rm o}}}}\bm{{\hat{\Psi}}_{e}^{{{\rm o}}^{\dagger}}}
            +\bm{{\hat{\Psi}}_{{\rm fr},e}}^{\bf R}\left(f\left(\bm{{\rm D}_{\rm fr}},\mu \right) -{\bf I_{fr}} \right)\bm{{\hat{\Psi}}_{{\rm fr},e}^{{\bf R}^{\dagger}}}\right]\bm{n^e}({\bf x}),
        \end{equation*}
        where $\bm{n^e}({\bf x})={\left[N^e_1({\bf x}) \,N^e_2({\bf x}) \, \cdots \, N^e_{M_{cell}}({\bf x})\right]}^T$ denotes the FE basis functions associated with the given cell ($M_{cell}$ denotes the number of nodes in the cell).
\end{algorithmic}
\end{algorithm}

To further enable computational prefactor reduction of our RRGEP algorithm, we employ mixed precision arithmetic. Mixed precision computing strategies provide significant reductions in computational and data movement costs on modern computing architectures, and thus are being increasingly adopted in scientific computing. However, due to the stringent accuracy requirements in DFT solvers, there has only been a limited exploration of mixed precision strategies in DFT solvers~\cite{tsuchida1995,motamarri2020,das2019}. In the context of the ChFES procedure, a FP64-FP32 mixed precision algorithm has been developed in \DFTFEverprev~\cite{motamarri2020} for the orthogonalization and Rayleigh-Ritz steps. Similar mixed precision strategies are used in this work in the context of the generalized eigenvalue problem based RRGEP procedure. The key idea exploited here is that the Chebyshev filtered vectors, $\bPsitildef$, adaptively approach the eigenvectors corresponding to the lowest $N$ states of $\btH$. Accordingly, the overlap matrix $\bS=\bPsitildefDagger \bPsitildef $ and the projected  Hamiltonian matrix $\bhatH = \bPsitildefDagger \btH \bPsitildef $  approach diagonal matrices as the SCF approaches convergence. We take advantage of this to use mixed precision strategies for the dominant $\mathcal{O}(MN^2)$ steps in steps 1 and 3 of Algorithm~\ref{alg:rrgepspectrumsplit}. In particular, for the computation of $\bS=\bPsitildefDagger \bPsitildef $ and  $\bPsitildeo=\bPsitildef \bLinvDagger$, the diagonal entries are computed in FP64 while the off-diagonal entries are computed in FP32 as their contribution tends to zero as the SCF approaches convergence. Next, in the case of the projected  Hamiltonian matrix $\bhatH = \bPsitildefDagger \btH \bPsitildef $, we use a mixed strategy which takes advantage of both $\bhatH$ approaching diagonal matrix as well as the spectrum-splitting method in RRGEP. In this approach, we consider the following split of the Chebyshev filtered wavefunctions
\begin{equation}
\bPsitildef=\left[\bm{{\widetilde{\Psi}}^{\rm oc}_{{\rm f}}} \, \bm{{\widetilde{\Psi}}^{\rm fr}_{{\rm f}}} \right],    
\end{equation}
where the columns $\bm{{\widetilde{\Psi}}^{\rm oc}_{{\rm f}}}$ and $\bm{{\widetilde{\Psi}}^{\rm fr}_{{\rm f}}}$ contain the first $N_{\rm oc}$ and the remaining $N_{\rm fr}$ wavefunctions, respectively. Subsequently, we compute  $\bhatH$ in mixed precision as
\begin{gather}\label{eq:partialDiagMixedPrec1}
\left[
\begin{array}{c|c}
{\bf \hat{H}}_{\bf oc-oc} & {\bf \hat{H}}_{\bf oc-fr}  \\
\hline
{\bf \hat{H}}_{\bf fr-oc} & {\bf \hat{H}}_{\bf fr-fr}
\end{array}
\right]
=\left[
\begin{array}{c|c}
{\rm FP32}\left\{\bm{{\widetilde{\Psi}}_{\rm f}^{{\rm oc}^{\dagger}}} {\bf \widetilde{H}} \bm{{\widetilde{\Psi}}_{\rm f}^{{\rm oc}}}\right\} & {\rm FP32}\left\{\bm{{\widetilde{\Psi}}_{\rm f}^{{\rm oc}^{\dagger}}} {\bf \widetilde{H}} \bm{{\widetilde{\Psi}}_{\rm f}^{{\rm fr}}}\right\}\\
\hline
{\rm FP32}\left\{\bm{{\widetilde{\Psi}}_{\rm f}^{{\rm fr}^{\dagger}}} {\bf \widetilde{H}} \bm{{\widetilde{\Psi}}_{\rm f}^{{\rm oc}}}\right\} & {\rm FP64}\left\{\bm{{\widetilde{\Psi}}_{\rm f}^{{\rm fr}^{\dagger}}} {\bf \widetilde{H}} \bm{{\widetilde{\Psi}}_{\rm f}^{{\rm fr}}}\right\}
\end{array}
\right]\,,
\end{gather}
where we exploit the fact that as the SCF approaches convergence, the diagonalization step in RRGEP (step 2 of Algorithm~\ref{alg:rrgepspectrumsplit}) to compute the $N_{\rm{fr}}$ highest eigenstates of the GHEP: $ \bhatH \bQfr=\bS \bQfr \bDfr$, tends to be solely determined by the small ${\bf \hat{H}}_{\bf fr-fr}$ block of $\bhatH$~\cite{motamarri2020}.

The accuracy and robustness of the above mixed precision strategies have been demonstrated in \DFTFEverprev~\cite{motamarri2020} on various pseudopotential benchmark problems. In particular, the number of SCF iterations remains unchanged between double precision and mixed precision simulations, and the error in the energy and ionic forces due to mixed precision is more than two orders of magnitude lower than the discretization errors. In this work, we further demonstrate accuracy and robustness of our mixed precision RRGEP implementation by comparing the energy, ionic forces and cell stresses against \QE for benchmark problems, as will be discussed in Section~\ref{sec:validation}. Finally, the performance benefits of using the above mixed precision strategies on hybrid CPU-GPU architectures are demonstrated in Section~\ref{sec:arch-rr}.


\section{Architecture of the hybrid CPU-GPU implementation}
\label{sec:arch}
\subsection{General implementation strategy}
The central consideration in \DFTFEver implementation strategy for the SCF procedure (cf. Algorithm~\ref{alg:scf}) is on reducing data-movement and data-access costs while exposing fine grained parallelism in the computation on modern many-core CPU and hybrid CPU-GPU architectures. Given the heavy penalty associated with data movements, both intra-node as well as inter-node data transfers, relative to the fast compute performance on hybrid CPU-GPU architectures, such considerations are critical to maximize performance and parallel scalability of the key computational kernels in our SCF procedure, as will be demonstrated below. All the key computational kernels in our Chebyshev filtered eigensolver based SCF procedure, namely CF, RRGEP and DC can be grouped into operations that scale as $\mathcal{O}(MN)$ (CF, DC), $\mathcal{O}(MN^2)$ (RRGEP-OP and RRGEP-ST) and  $\mathcal{O}(N^3)$ (RRGEP-D).

First, in the case of CF, the core computational kernel $\btH \bfX$ (as part of the Chebyshev recursive iteration in Eq.~\ref{eq:chebyfilter}, with $\bfX$ being the intermediate wavefunctions vectors), is a FE discretized sparse-matrix times a dense-matrix (sparse-dense) multiplication. Particularly, in the context of FE discretization, this operation is very amenable to GPU acceleration. We recast the above sparse-dense multiplication as small FE cell level dense matrix times dense matrix (dense-dense) multiplications simultaneously across all FE cells in a given processor. These simultaneous dense-dense multiplications are performed very efficiently by optimized math libraries on either CPUs or GPUs, and combined with our contiguous data access patterns, allows to significantly boost the arithmetic intensity by lowering the data access costs. Section~\ref{sec:arch-cf} discusses more on these aspects and provides the relevant performance benchmarks on hybrid CPU-GPU architectures. Further, efficient construction of these FE cell-level dense matrices ($\btH$) is discussed in Section~\ref{sec:arch-ham} below.

Next, we consider the $\mathcal{O}(MN^2)$ steps in RRGEP-OP and RRGEP-ST that constitute large dense-dense multiplications. In general, such operations are able to achieve high performance on many-core CPU architectures and hybrid CPU-GPU architectures. However, in our case they present a unique challenge due to the extreme aspect ratios ($M/N \gg 1$) involved. To this end, in Section~\ref{sec:arch-rr}, we adopt a scalable approach that boosts the overall performance by reducing the data movement costs significantly. Another important aspect of our implementation is the use of mixed precision strategies and the use of a column-blocked approach to matrix-matrix multiplications in CF, RRGEP-OP and RRGEP-ST. While mixed precision strategies speed up compute as well as reduce data movement costs, the blocked approach to matrix operations provide other benefits---reduction of peak memory, and opportunities to overlap compute and communication. Furthermore, optimal hardware aware data transfer pathways such as GPU direct communication are also used to reduce data movement costs in the RRGEP-OP and RRGEP-ST. Sections~\ref{sec:arch-cf} and~\ref{sec:arch-rr} elaborate on these aspects and demonstrate relevant performance benchmarks on hybrid CPU-GPU architectures. 

Finally, in the case of $\mathcal{O}(N^3)$ RRGEP-D step, we use the scalable direct eigensolver approaches using the state-of-the-art ELPA library~\cite{elpa2014}, which has been efficiently implemented both for many-core and hybrid CPU-GPU architectures~\cite{elpaOpt,elpagpu}. Upon efficient implementation of the above core computational kernels, the bottlenecks in performance shifted to the remaining $\mathcal{O}(MN)$ scaling steps---density computation (DC) and computation of ionic forces (IF)---which are also optimized for both CPUs and GPUs, using similar strategies as outlined above. We also remark that initialization costs (INIT), total electrostatic potential solve (TEP), and electron-density mixing schemes (DM) are the other remaining costs which are not yet ported to GPUs, but optimizations have been performed to ensure they constitute a small portion of the total run-time.

\subsection{FE cell-level Hamiltonian matrix construction}\label{sec:arch-ham}
We describe a computationally efficient procedure for the construction of discretized Hamiltonian matrix at the finite-element cell level focusing on increasing the arithmetic intensity and reducing the data-access costs. This is accomplished by recasting the quadrature integrations arising in Eq.~\ref{eq:discreteHam} in the form of BLAS level 3 matrix-matrix multiplications which can in-turn utilize optimized math kernel libraries in CPUs/GPUs to perform these arithmetic operations. In particular, we focus here on the second and third term of the Eq.~\ref{eq:discreteHam} which is computed for every self-consistent field iteration. Recall from Eq.~\ref{eq:discreteHam}, the contribution to the discretized Hamiltonian matrix for a given FE cell $(e)$ corresponding to  $V^{h}_{\text{eff,loc}}(\rho^{h},\bR)$ can be written as
\begin{equation}\label{eq:quadLocal}
    \int_{\Omega_e} V^{h}_{\text{eff,loc}}(\rho^{h}(\bx),\bR) N^{h}_j (\bx) N^{h}_k (\bx) \dx = \sum_{p=1}^{n_q} w_{p}\;j_{p} V^{h}_{\text{eff,loc}}(\bq_p,\bR) N^{h}_j(\bq_p) N^{h}_k(\bq_p) = \sum_{p=1}^{n_q} P_{Ip}v^{(e)}_{p}
\end{equation}
where $1\leq j,k\leq M_{cell}$ with $M_{cell}$ denoting the number of FE nodes in a given cell (e) and further, $I = (j,k)_{1\leq j,k\leq M_{cell}}$ is the composite index. In the above equation, $n_q$ denotes the number of quadrature points in the given FE cell (e), $\bq_p$ denotes the $p^{th}$ quadrature point while $w_{p}$ denotes the corresponding quadrature weight. $j_{p}$ denotes the determinant of the finite-element Jacobian matrix evaluated at the $p^{th}$ quadrature point corresponding to the mapping from the reference cell in the natural coordinate system to the real-space cell ~\cite{brenner2002}. We also note in Eq. \ref{eq:quadLocal} that $P_{Ip}$, the ${Ip}^{th}$ component of the matrix $\bP$ is given by $N^{h}_i(\bq_p) N^{h}_j(\bq_p)$ while $v_p^{(e)}$, the $p^{th}$ component of the vector $\bv^{(e)} \in \mathbb{R}^{n_q}$ is given by $w_p\, j_p\,V^{h}_{\text{eff,loc}}(\bq_p,\bR)$. Denoting the number of FE cells present in a given MPI task to be $M_c$, we define a matrix $\bV \in \mathbb{R}^{n_q \times M_c}$ with the vector $\bv^{(e)}$ forming the $e^{th}$ column of $\bV$. Subsequently, the quadrature summation in Eq.~\ref{eq:quadLocal} for all the cells can be recast as a single matrix-matrix multiplication operation between the matrices $\bP$ and $\bV$ as:
\begin{equation}\label{eq:gemmLoc}
   \bK^{\text{eff,loc}} = \bP \bV \,.
\end{equation}
We note that for a $p^{th}$ quadrature point, the product $P_{Ip} = N^{h}_i(\bq_p) N^{h}_j(\bq_p)$ is symmetric with respect to $i$ and $j$ for a given composite index $I$, and, consequently, the terms in the product corresponding to $i \ge j$ are only stored in the matrix $\bP$. Hence, the size of the matrix $\bP$ in Eq.~\ref{eq:gemmLoc} is $M_{sym} \times n_q$ where $M_{sym} = M_{cell}(M_{cell} + 1)/2$ 

An additional contribution to the discretized Hamiltonian matrix (third term in Eq.~\ref{eq:discreteHam}) arises in the case of GGA exchange-correlation functional and involves the spatial derivatives of FE basis functions. This can be written in terms of $\sigma(\bx) = |\grad \rho(\bx)|^2$ and the derivatives with respect to reference FE cell coordinate directions [$\bxi = (\xi_1,\xi_2,\xi_3)$] as follows:
\begin{align}\label{eq:quadGGA}
   \nonumber &\left.\int_{\Omega_e} \frac{\partial \varepsilon_{\text{xc}}(\rho,\grad \rho)}{\partial \del \rho}\right|_{\rho=\rho^h} \cdot \left(\del N^h_{j}(\bx) N^h_{k}(\bx) + N^h_{j}(\bx) \del N^h_{k}(\bx)\right) \dx  \\  \nonumber
   &= \left. \int_{\Omega_e} \sum_{r=1}^{3} \sum_{s=1}^{3} 2\; \frac{\partial \varepsilon_{\text{xc}}(\rho,\grad \rho)}{\partial \sigma} \frac{\partial \rho (\bx)}{\partial x_r}\right|_{\rho=\rho^h} \left(J^{-1}_{rs}\frac{\partial N_j(\bx(\bxi))}{\partial \xi_s} N_k(\bx) + N_j(\bx) J^{-1}_{rs} \frac{\partial N_k (\bx(\bxi))}{\partial \xi_s} \right) \dx \\ 
   &= \sum_{p=1}^{n_q} \sum_{r=1}^{3} \sum_{s=1}^{3} 2\, w_p \, j_p \frac{\partial \varepsilon_{\text{xc}}}{\partial \sigma} (\bq_p)\frac{\partial \rho^{h} }{\partial x_r}(\bq_p) J^{-1}_{rs}\left(\frac{\partial N_j}{\partial \xi_s}(\bq_p) N_k(\bq_p) + N_j(\bq_p) \frac{\partial N_k}{\partial \xi_s}(\bq_p) \right) 
\end{align}
where $1\leq j,k\leq M_{cell}$ and $\bJ^{-1}$ denotes the inverse of the finite-element Jacobian matrix. The summation over quadrature points and the spatial directions in Eq.~\ref{eq:quadGGA} for all $M_c$ cells in a given MPI task can be recast as a matrix-matrix multiplication operation as shown below:
\begin{equation}
    \bK^{\text{gga},\text{loc}} = \bP^{g}\bV^{g} \;\;\;\text{with}\;\;\;\bP^{g} \in \mathbb{R}^{M_{sym} \times 3n_q},\;\;\bV^{g}\in \mathbb{R}^{3n_q \times M_c}
\end{equation}
where 
\begin{equation}
  \bP^{g} =   \begin{bmatrix}
      \bP^{q_1} & \bP^{q_2} & \dots & \bP^{q_p} & \dots &  \bP^{q_{n_q}}
    \end{bmatrix};\, 
   {\bV^{g}}^{T} = 
    \begin{bmatrix} 
      \bV^{q_1} &
      \bV^{q_2} &
      \dots&
      \bV^{q_p} &
      \dots &
      \bV^{q_{n_q}}
    \end{bmatrix}
\end{equation}
In the above, $\bP^{q_{p}} \in \mathbb{R}^{M_{sym} \times 3}$ is evaluated for every quadrature point $\bq_p$.
As before, with $I = (j,k)_{1\leq j,k\leq M_{cell};\;j \leq k}$ denoting the composite index and $s = \{1,2,3\}$ denoting the spatial coordinate index, the ${Is}^{th}$ component of matrix $\bP^{q_{p}} \in \mathbb{R}^{m_c \times 3}$ is given by 
\begin{equation}
    P_{Is}^{q_{p}} = \left(\frac{\partial N_j}{\partial \xi_s}(\bq_p) N_k(\bq_p) + N_j(\bq_p) \frac{\partial N_k }{\partial \xi_s} (\bq_p) \right) \,.
\end{equation}
The matrix ${\bV^{q_p}} \in \mathbb{R}^{M_c \times 3}$ evaluated for every quadrature point $\bq_p$. Denoting $e^{th}$ row vector of the matrix ${\bV^{q_p}}$ to be ${\bv^{q_p}}^{(e)}$, the $s^{th}$ component ${\bv^{q_p}}^{(e)}$ is given by
\begin{equation}
    {v^{q_p}_{s}}^{(e)} =  \sum_{r=1}^{3} w_p \, j_p \frac{\partial \varepsilon_{\text{xc}}}{\partial \sigma}(\bq_p)J^{-1}_{rs}\frac{\partial \rho^{h}}{\partial x_r}(\bq_p)\,.
\end{equation}

Similar ideas as discussed above are also used to recast the FE discretized matrix arising in the case of periodic problems involving multiple $k$ point sampling of the Brillouin zone ~\cite{motamarri2020} as BLAS level 3 matrix-matrix multiplications, thereby increasing the arithmetic intensity and reducing the data access costs.

    

%

\subsection{Chebyshev filtering}\label{sec:arch-cf}
\begin{figure}
\centering
\includegraphics[width=0.7\columnwidth]{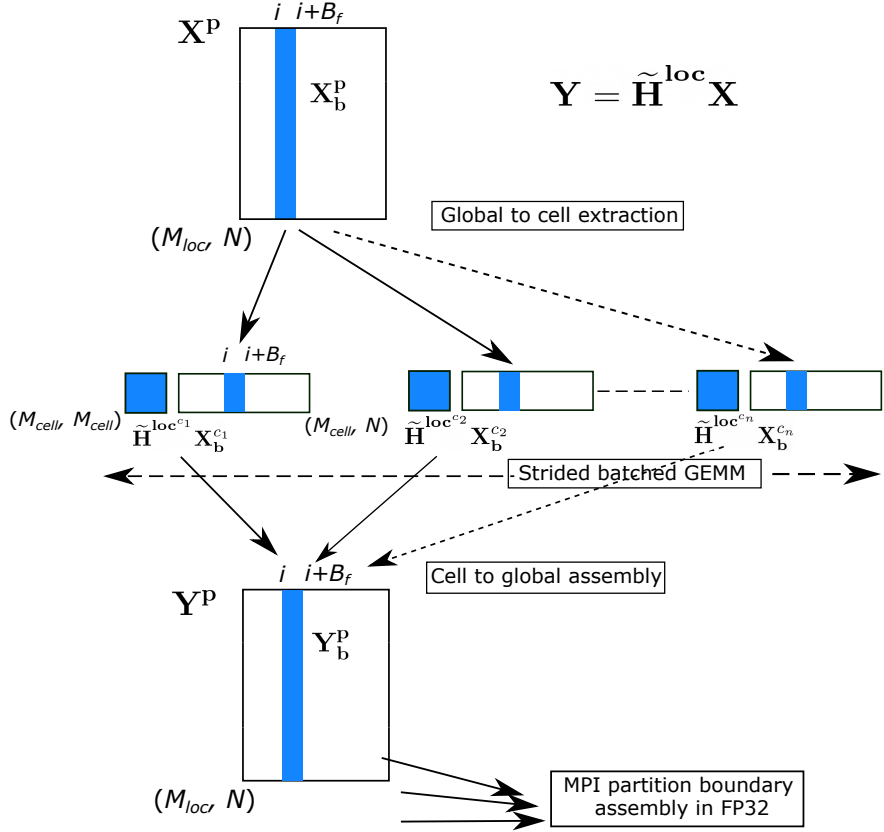}
\caption{\small{Schematic of ${\bf Y}=\btHloc {\bf X}$ computation.  $M_{\rm loc}$ denotes number of DoFs owned locally by a MPI task, and `p' denotes the MPI task Id.}}
\label{fig:HXSchematic} 
\end{figure}

$\btH \bfX $, the core computational kernel in the Chebyshev polynomial filtering (CF)  recursive iteration involves the multiplication of the L\"{o}wdin transformed discrete Hamiltonian matrix, $\btH = \bM^{-1/2}\bH\bM^{-1/2}$, with $\bfX$, which denotes a trial subspace during the course of the recursive iteration. $\btH$ is composed of the local contribution $\btHloc=\bM^{-1/2}\bHloc\bM^{-1/2}$, and additionally in the case of pseudopotential calculations, the non-local contribution $\btHnl=\bM^{-1/2}\bHnl\bM^{-1/2}$, both of which are sparse matrices  owing to the locality of the FE basis and the  pseudopotential projectors (cf. Section~\ref{sec:fem}).  Performing the resulting sparse-dense matrix products using the conventional approaches at the global level is not ideal on modern supercomputing architectures as they incur high global memory access costs~\cite{kronbichler2012}. In \DFTFE, we significantly reduce the memory access costs in the core computational kernel $\btH \bfX$ by employing FE cell level dense matrix operations instead of global sparse matrix approaches. To elaborate, in the $\btHloc \bfX$ operation, we recast it as 
\begin{equation}\label{eq:celldgemm}
{\bf Y}={\rm ASEMB_{FEM}}
\left\{\btHlocci {\bf X}^{c_i}\right\}\,,
\end{equation}
where ${c_i}$ denotes the $i^{\rm th}$ FE cell, $M_{\rm cell}$ denotes the number of DoFs in a FE cell, $\btHlocci$ denotes the $M_{\rm cell} \times M_{\rm cell}$ FE cell level contribution to $\btHloc$, ${\bf X}^{c_i}$ denotes  the $M_{\rm cell} \times N$ cell level wavefunction vectors, and ${\rm ASEMB}$ denotes the assembly operation of vector contributions from all FE cells into ${\bf Y}$. Using the above form, the many small dense-dense matrix multiplications $\left\{\btHlocci {\bf X}^{c_i}\right\}$ are performed very efficiently on both many-core CPU and hybrid CPU-GPU architectures using batched \verb|xGEMM| routines. Particularly on GPUs, the FE cell structure naturally exploits the massive fine grained parallelism available on GPUs by simultaneously performing the dense-dense matrix multiplications for all cells using cuBLAS library's \verb|xGEMMStridedBatched| routine. We further remark that, to reduce the peak memory during the CF procedure, each $\btHloc \bfX$ evaluation is performed using a block approach. To this end, blocks of $B_f$ wavefunction vectors, denoted by $\bfXb$, are filtered sequentially. Figure~\ref{fig:HXSchematic} shows the schematic of the blocked $\btHloc \bfX$ implementation, consisting of three key steps: (i) global $\bfXb$ to cell level ${\bf X}^{c_i}_{\bf b}$ extraction, (ii) \verb|xGEMMStridedBatched|  of $\btHlocci {\bf X}^{c_i}_{\bf b}$, and (iii) the final assembly step into $\bfYb$, which also involves accumulation of contributions from MPI partition boundary DoFs via MPI point-to-point communication calls. 

To efficiently perform the memory bound data-movement operations in the extraction and assembly steps above, we structure the memory layout and  implementation  of data-movement computational kernels (CUDA kernels in case of GPU) such that the multiple wavefunction values are accessed from (written to) the global memory contiguously for each FE degree of freedom. This provides coalesced memory access across GPU threads to maximize the memory bandwidth of reading from the global RAM, as well as improving cache locality on many-core CPU architectures. Such strategies are critical to obtain performance on hybrid CPU-GPU architectures, as will be demonstrated below. The above data-movement aware implementation strategy  also extends to MPI point-to-point communication costs  in the ${\rm ASEMB_{FEM}}$ step.  In particular, we exploit the fact that all wavefunction components of ${\bf Y_b}$ have identical MPI point-to-point communication pattern across the FE domain decomposition partition boundaries. This allows us to perform the MPI communication for all wavefunction components simultaneously, which incurs minimal network latency compared to communicating the wavefunction vectors one by one.

Additionally, in the case of pseudopotential calculations, $\btHnl {\bf X}$ operation (cf. Algorithm~\ref{alg:nonlocalHX}) is performed in every Chebyshev recursive iteration. Our implementation strategy for $\btHnl \bfX$ follows along similar lines as $\btHloc \bfX$. We exploit the sparse structure of $\btHnl$ in real-space  to efficiently compute $\btHnl \bfX$. To this end, we construct FE cell level pseudopotential projector matrices ${\bf C}^{c_i}$ (cf. Algorithm~\ref{alg:nonlocalHX} and Eq.~\ref{eq:nonlocalHamDiscrete}), which denotes a matrix containing the ${C^{a}_{lpm,j}}$ values for all $M_{\textrm{cell}}$ ($j$ index in ${C^{a}_{lpm,j}}$) FE shape functions belonging to the ${c_i}^{\textrm{th}}$ FE cell and for all $N_{\textrm{psp}}^{c_i}=\left\{a,lpm\right\}$ pseudopotential projectors $\chi^{a}_{lpm}(\bx,\bR_a)$ which have a non-trivial compact support over the same FE cell. This renders ${\bf C}^{c_i}$ matrices to be of size $M_{\textrm{cell}} \times N_{\textrm{psp}}^{c_i}$. Subsequently, for the first computational step in Algorithm~\ref{alg:nonlocalHX}, ${\bf P}_{\bf b}^{c_i}={\bf C^{\dagger}}^{c_i} {\bf X}^{c_i}_{\bf b}$, we use \verb|xGEMMStridedBatched| (with zero padding on ${\bf C^{\dagger}}^{c_i}$ due to variable $N_{\textrm{psp}}^{c_i}$) on GPUs to   efficiently perform the FE cell level dense-dense matrix multiplications. Next, we perform the assembly of contributions of the resulting product matrices ${\bf P}_{\bf b}^{c_i}$ from all FE cells within the compact support of the pseudopotential projectors to obtain ${\bf P_b}$ (cf. step 3 in Algorithm~\ref{alg:nonlocalHX}), using MPI point-to-point communication calls. The implementation of remaining steps in $\btHnl {\bf X}$, follow along similar lines as discussed above. We also note that the efficient memory layout structure for coalesced memory access across GPU threads discussed previously in the context of $\btHloc \bfX$ also extends here, where we read/write all the $B_f$ wavefunction values in a given block contiguously for each pseudopotential projector index.


\begin{algorithm}
\caption{FE cell level computation of ${\bf Y_b^{nl}}=\btHnl {\bf X_b}$ based on Eq.~\ref{eq:nonlocalHamDiscrete}}
\label{alg:nonlocalHX}
\begin{algorithmic}[1]
\State Extraction of ${\bf X}^{c_i}_{\bf b}$ from ${\bf X_b}$

\State ${\bf P}_{\bf b}^{c_i}={\bf C^{\dagger}}^{c_i} {\bf X}^{c_i}_{\bf b}$.

\State ${\bf P_b}={\rm ASEMB_{NLP}}\left\{{\bf P}_{\bf b}^{c_i}\right\}$.

\State Scaling of  ${\bf P_b}$ with $h_{lp}^J$

\State ${\bf Y^{nl}_b}^{c_i}={\bf C}^{c_i} {\bf P}^{c_i}_{\bf b}$

\State ${\bf Y^{nl}_b}={\rm ASEMB_{FEM}}\left\{{\bf Y^{nl}_b}^{c_i}\right\}$
\end{algorithmic}
\end{algorithm}


The above implementation innovations lead to a high overall throughput performance for CF in \DFTFEver on hybrid CPU-GPU architectures, as demonstrated in Fig.~\ref{fig:chebyPerfB}. The performance is measured on a BCC Mo benchmark system ($\textrm{Mo}_{\textrm{6x6x6}}$) containing 6,034 electrons (3,600 wavefunctions) and parallelized over 4 Summit GPU Nodes (24 NVIDIA Tesla V100 GPUs), with 72,000 DoFs per GPU. We observe that the performance, measured as the percentage of the theoretical FP64 peak FLOPS corresponding to 24 V100 GPUs, increases with increasing wavefunctions block size, $B_f$, and reaches a value of  $54.4\%$ of the FP64 peak at $B_f=600$. The lower throughput at smaller block sizes is attributed to the increased frequency of strided memory access from the global memory per GPU thread warp, and other overheads resulting from lower utilization of point-to-point MPI communication bandwidth  due to smaller data packet sizes.


\subsubsection{Asynchronous compute and communication, and FP32 boundary communication}
\begin{figure}[t!]
    \centering
    
     \begin{subfigure}[t]{0.5\textwidth}
        \centering
        \includegraphics[scale=0.27]{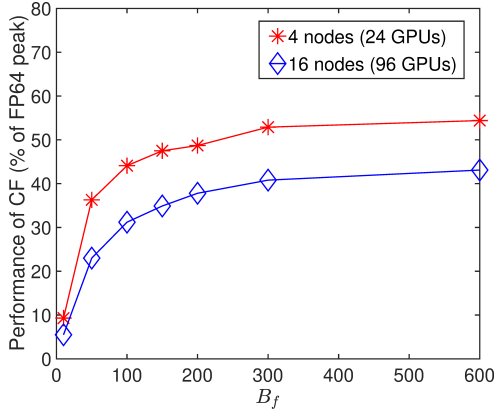}
        \caption{\small{Influence of block size on performance}}
        \label{fig:chebyPerfB}
    \end{subfigure}
    ~
    \begin{subfigure}[t]{0.48\textwidth}
        \centering
        \includegraphics[scale=0.25]{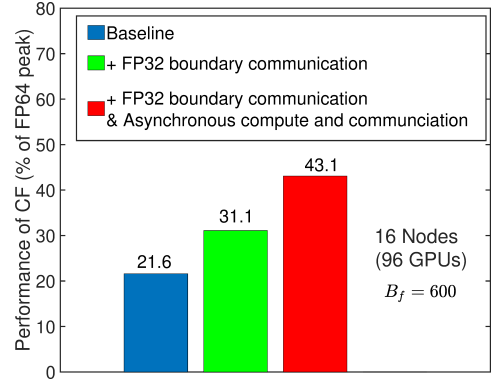}
        \caption{\small{Performance improvements related to communication}}
        \label{fig:chebyPerfA}
    \end{subfigure}
    \caption{\small{Chebyshev filtering (CF) performance profile on multiple Summit nodes, each containing 6 Tesla V100 GPUs. Performance analyzed for various sizes of wavefunction block sizes ($B_f$) and influence of various implementation strategies to reduce communication costs. FP64 peak of a single Tesla V100: 7.8 TFLOPS. Case study: Mo super cell with a mono-vacancy containing 431 atoms (6,034) electrons. FE Mesh DoFs: 1,728,000.}}
    \label{fig:chebyPerf}
\end{figure}

In spite of the MPI communications in both $\btHloc \bfX$ and $\btHnl \bfX$ kernels having a local pattern, the communication cost, particularly in the $\btHloc \bfX$ kernel, constitutes an appreciable portion of the total CF cost when scaling to low degrees of freedom ($<$ 30,000) per MPI task on hybrid CPU-GPU architectures. Hence we have implemented two strategies to further reduce the MPI communication cost.  In the first strategy, we use FP32 datatype instead of FP64 for the FE domain decomposition partition boundary communication (cf. Fig.~\ref{fig:HXSchematic}) in the $\btHloc \bfX$ kernel call. Since the number of FE DoFs on the domain partition boundary are only a small fraction of the total DoFs, it has been observed from our numerical experiments to retain FP64 accuracy in ground-state solutions while reducing the communication cost by $\sim2\times$. In Section~\ref{sec:validation}, we also validate the use of FP32 boundary communication by comparing the accuracy of the \DFTFE ground-state solution against \QE for various pseudopotential, and all-electron benchmark systems against \nwchem. As a second strategy to reduce communication cost, we exploit the aforementioned block approach in CF to overlap the key compute operations---\verb|xGEMMStridedBatched| calls in both the local and non-local HX kernels---of one block of wavefunctions with the MPI communication of the previous block via non-blocking point-to-point MPI communication calls (\verb|MPI_IRecv| and \verb|MPI_ISend|). Figure~\ref{fig:chebyPerfA} demonstrates the performance benefits afforded by the above strategies  on hybrid CPU-GPU architectures. We reconsider the $\textrm{Mo}_{\textrm{6x6x6}}$ benchmark system discussed previously, containing 6,034 electrons, and parallelized over 16 Summit GPU Nodes (96 GPUs) with 18,000 DoFs per GPU. First, by employing the FP32 boundary communication strategy, we obtain 1.4$\times$ improvement in the CF step compared to the baseline performance. Subsequently, by employing the asynchronous compute and communication, we obtain another 1.4$\times$ performance improvement, taking the overall performance improvement to 2$\times$.


\subsection{Rayleigh Ritz procedure}\label{sec:arch-rr}
The Rayleigh-Ritz (RRGEP) step of the ChFES procedure in Algorithm~\ref{alg:scf} involves the computation of 
matrix-matrix multiplications of the form $\mat{A} = \mat{X}^{\dagger}\mat{Y}$ (RRGEP-OP), diagonalization of the projected Hamiltonian matrix (RRGEP-D) and $\mat{X}' = \mat{X}\mat{Z}$ (RRGEP-ST) where $\mat{X}^{\dagger}$ denotes conjugate-transpose of $\mat{X}$. The matrix dimensions of $\mat{X}$, $\mat{Y}$ are  $M \times N$ (RRGEP-OP, RRGEP-ST), while $\mat{X}'$ can be of dimension $M \times N_{\text{fr}}$  (RRGEP-ST-F) or $M \times N$ (RRGEP-ST-O). Furthermore, the dimension of $\mat{A}$ is $N \times N$ (RRGEP-OP) while $\mat{Z}$ can be of dimension $N \times N_{\text{fr}}$ (RRGEP-ST-F) or $N \times N$ (RRGEP-ST-O).

 We remark that the aspect ratios of $\mat{X}$, $\mat{Y}$ and $\mat{X}'$ are very large $M/N \gg 1$, and the computational complexity of these operations scale as $\ordercomplexity{(MN^2)}$. This extreme aspect ratio presents unique challenges to develop a computationally efficient and a scalable implementation for these operations.  Furthermore, scaling \DFTFE calculations to large-size  problems (~ $N > 20,000$) requires strategies to address the $N \times N$ matrix memory footprint of $\mat{A}$ and $\mat{Z}$ owing to limited memory available on each GPU. We now present approaches that address these challenges in the present work.

We first discuss the scalable strategies employed in the matrix-matrix multiplication of the type $\mat{A} = \mat{X}^{\dagger}\mat{Y}$ arising in the computation of the overlap matrix $\mat{S}$ in RRGEP-O and the projected Hamiltonian $\hat{\bH}$ in RRGEP-P. Due to the large aspect ratios of $\mat{X}$ and $\mat{Y}$, the traditional ways of storing these matrices in block-cyclic layout are not well suited as this would make the receiving processors (holding $\mat{A}$) to perform all the computations, leading to a communication bottleneck with a cost scaling as $\ordercomplexity {(N^2\,M)}$. Instead, the optimal data distribution we employ divides $\mat{X}$ and $\mat{Y}$ equally among available processors $P$, into equi-partitioned matrices $\mat{X}_p$ and $\mat{Y}_p$ of size $M_{\text{loc}} \times N$ with $M_{\text{loc}} \approx M/P$. This is achieved via the domain decomposition of the FE mesh. Since the computation of $\mat{X}^{\dagger}\mat{Y}$ is similar to a vector dot product, we compute on GPUs a $N \times N$ matrix $\mat{A_p} = \mat{X_p}^{\dagger}\mat{Y_p}$ associated with each processor locally and then sum these matrices in $\log P$ steps by making use of MPI collectives. This results in a communication cost of $\ordercomplexity{(N^2\log(M))}$, as $P$ is chosen to be proportional to $M$.

Furthermore, for large-scale problems on GPUs, to avoid the huge memory footprint in storing $\mat{A}_p$, and thereby the matrix $\mat{A}$, we employ a blocked approach as shown in Fig.~\ref{blockedDemo}(a) with block size $N\times B_v$. Each block of $\mat{A}$ is computed successively by the accumulation of the local contributions (corresponding block of $\mat{A_p}$) using the \texttt{MPI\_Allreduce} collective. Moreover, we exploit the Hermiticity of $\mat{S}$ and $\hat{\mat{H}}$ to compute only their lower triangular portion. Subsequently, for efficient matrix computations on $\mat{A}$ (Cholesky factorization and subspace diagonalization in RRGEP-D), we distribute it using a 2D block-cyclic grid format. In the case of operations of type $\mat{X}' = \mat{X} \mat{Z}$ arising in RRGEP-ST, we employ the optimal data distribution of $\mat{X}$ as discussed before, and compute $\mat{X}'$ using the local matrix multiplications $\mat{X}'_p = \mat{X}_p\mat{Z}$ ($\mat{X}'_p$ and $\mat{X}_p$ denote the local spatial partitions of the global matrices $\mat{X}'$ and $\mat{X}$, respectively). Further, as shown in Fig.~\ref{blockedDemo}(b), a blocked approach similar to the computation of $\mat{A}$ is employed to handle the memory footprint of $\mat{Z}$. A further exploitation of these blocked approaches is used to develop asynchronous compute-data movement strategies as discussed below.

\subsubsection{Asynchronous compute and communication using mixed precision arithmetic}
\begin{figure}[htbp]
    \centering
    \includegraphics[scale=0.4]{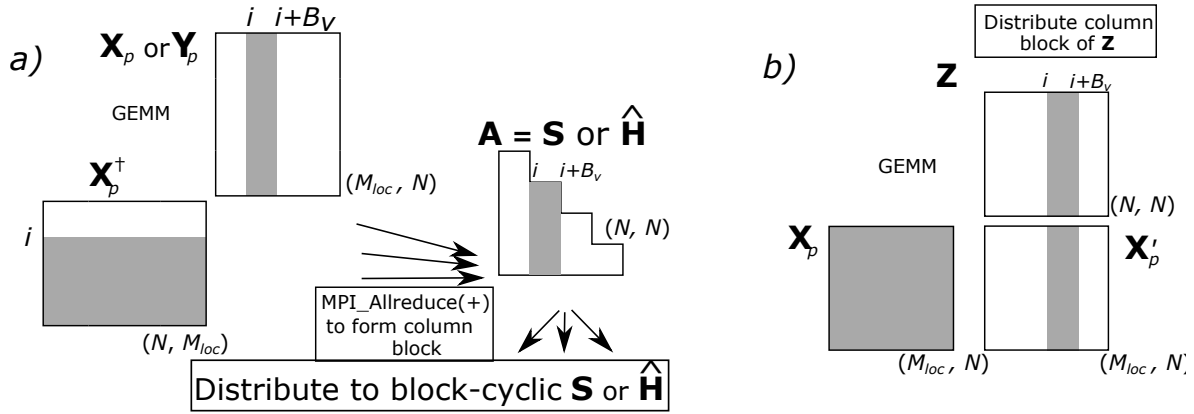}
    \caption{\small{Parallelization strategy for RRGEP steps. (a) Blocked computation of the overlap matrix ($\mat{S}$) and projected Hamiltonian ($\hat{\mat{H}}$) over column blocks. (b) Computation of orthonormal basis in RRGEP-ST with $\mat{Z} = {\mat{L}^{-1}}^{\dagger}$. Blocked approach allows us to asynchronously execute compute operations (\texttt{xGEMM}) of one block with data movement related to a different block (device-to-host/host-to-device data transfers \& MPI calls).}
    }
    \label{blockedDemo}
\end{figure}

\begin{figure}
\centering
\includegraphics[width=0.75\columnwidth]{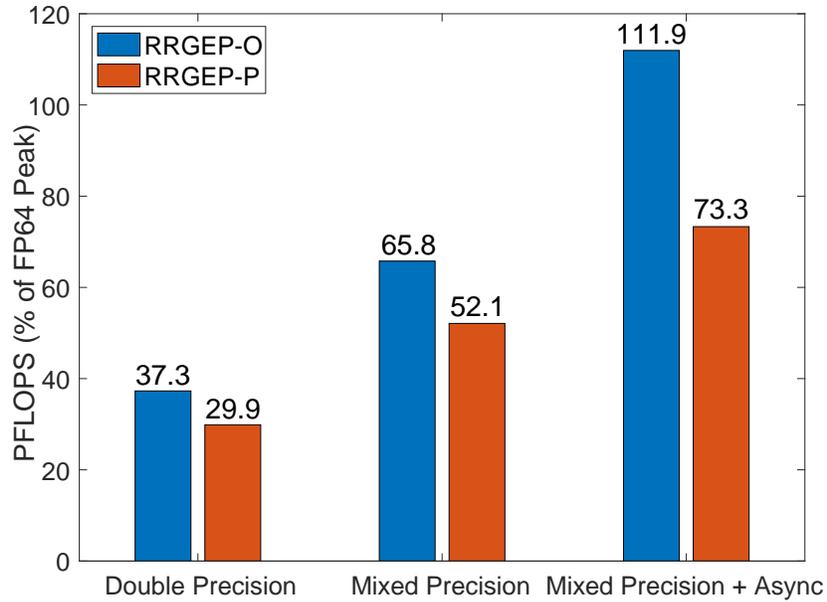}
\caption{\small{Performance profile related to RRGEP-OP step of Algorithm 2 on 160 Summit nodes with 960 V100 GPUs. Case study: Mo super cell with a mono-vacancy containing 4,393 atoms (61,502 electrons). FE Mesh DoFs: 16,194,277}}
\label{fig:RRGEP} 
\end{figure}

 The $\ordercomplexity{(MN^2)}$ operations in Algorithm~\ref{alg:rrgepspectrumsplit} employ the blocked approach as discussed in Sec.~\ref{sec:arch-rr}. In the case of computations of the type $\mat{A} = \mat{X}^{\dagger}\mat{Y}$, we make use of the fact that the GPU computations involving matrix-matrix multiplications $\mat{A}_{p}^{i} = \mat{X}_p^{\dagger}\mat{Y}_p^{i}$ associated with block-$i$ can be executed asynchronously with the data movement related to $\mat{A}_p^{i-1}$, i.e., GPU to CPU copy of $\mat{A}_p^{i-1}$ followed by the \texttt{MPI\_Allreduce} collective operation forming the relevant column block $\mat{A}^{i-1}$. The numerical implementation makes use of two different CUDA streams, one dealing with the compute operations while the other handling the data movement. Additionally, CUDA events are used to record the appropriate compute and data-movement events which allow for inter-stream synchronization of compute and copy operations associated with a given block, while asynchronously executing the data-movement operations related to a previous block. Along similar lines, asynchronous compute and data movement has been implemented for computations of the type $\mat{X}'=\mat{X}\mat{Z}$ (cf. Fig.~\ref{blockedDemo} (b)). Moreover, GPU direct communication through NVIDIA Collective Communications Library (NCCL)~\cite{nccl} is used to accelerate inter-GPU communication associated with the aforementioned collective operations in RRGEP. We observe a factor of $1.5 - 1.8\times$ reduction in computational times for RRGEP-OP and RRGEP-ST-O steps of Algorithm 2 by using NCCL. This performance gain can be attributed to better synchronization among communicating processors during the collective operations achieved -via- NCCL. Furthermore, mixed precision strategies are devised for the RRGEP step in Algorithm~\ref{alg:scf}  as explained in Section 3.  To this end, the diagonal block entries of overlap matrix $\mat{S}$ in RRGEP-O are evaluated in FP64 arithmetic  while the off-diagonal entries are computed in FP32 arithmetic. Similarly in the case of RRGEP-P, the projected Hamiltonian $\hat{\mat{H}}$ has the entries of the lower right diagonal block evaluated in FP64 while the other blocks are evaluated in FP32 arithmetic (cf. section~\ref{sec:RR-SSMP}). These mixed precision strategies reduce the floating point operations cost involved in the matrix-matrix multiplications and further reduce the data movement costs associated with \texttt{MPI\_Allreduce} collective operations involved in the construction of $\mat{S}$ and $\hat{\mat{H}}$.  Figure ~\ref{fig:RRGEP} demonstrates the performance benefits afforded by our mixed precision and asynchronous compute and data-movement strategies on the key kernels involved in RRGEP-O and RRGEP-P of Algorithm 2 in \DFTFEver on hybrid CPU-GPU architectures. The benchmark calculation on Mo$_{13 \times 13 \times 13}$, containing 61,502 electrons, is parallelized over 160 Summit GPU Nodes (960 GPUs) with $\sim17,000$ DoFs per GPU. As evident from the Fig. ~\ref{fig:RRGEP}, mixed precision strategies provide a substantial improvement of $ 1.7-1.8\times$ in comparison to FP64 baseline performance of these compute intensive kernels. Moreover, our asynchronous programming strategies further boosted the performance by an additional $1.4 - 1.7\times$. We also remark that reduction in floating point operation costs and data movement costs is achieved in the case of orthonormal basis construction (RRGEP-ST-O) by computing $\bPsitildeo=\bPsitildef \bLinvDagger$ in mixed precision as $\text{SP}[\mat{\bPsitildef }\mat{\bLinvDagger}_{\text{od}}] + \text{DP}[\mat{\bPsitildef}\mat{\bLinvDagger_d}]$, with $\mat{\bLinvDagger}_{\text{od}}$ and $\mat{\bLinvDagger}_\text{d}$ denoting the off-diagonal and diagonal entries of $\mat{\bLinvDagger}$, respectively. In this case, mixed precision performance in conjunction with asynchronous programming strategies improved the performance of RRGEP-ST-O by $\sim 3.8\times$.

\section{Results and discussion}
\label{sec:results}
In this section, we demonstrate the accuracy, computational efficiency and parallel scalability of \DFTFEver on various benchmark systems on both CPU and hybrid CPU-GPU architectures. The benchmark calculations involve both pseudopotential and all-electron DFT calculations. GGA~\cite{gga1} exchange correlation of the PBE form~\cite{pbe} is employed in all the calculations, and ONCV~\cite{oncv2013} pseudopotentials from the SG15~\cite{oncv2015} and PseudoDojo~\cite{van2018pseudodojo} databases are employed in all the pseudopotential DFT calculations. Further, we use Fermi-Dirac smearing with temperature $T = 500$ $K$. Additionally, for mixing the electron density in the SCF iteration, we use $n$-stage Anderson mixing scheme~\cite{anderson1965} and Kerker preconditioner~\cite{Kerker1981} for non-periodic and bulk metallic benchmark systems, respectively. 

We consider various periodic and non-periodic pseudopotential and all-electron benchmark systems with different sizes to demonstrate the accuracy and performance of \DFTFEverNoSpace. The pseudopotential benchmark systems are validated using plane-wave code Quantum Espresso (\QE)~\cite{qe2009,qe2017}, while all-electron benchmark systems are validated with the \nwchem code~\cite{nwchem}, which uses Gaussian basis. Specifically, we consider the following pseudopotential benchmark systems: (i) body centered cubic (BCC) Mo periodic supercells with a monovacancy, (ii) non-periodic Icosahedron Al nano-particles, (iii) InP-water interface system. In addition, we consider a non-periodic all-electron benchmark system: (iv) Benzamide molecule. For benchmark systems (i), we use norm-conserving ONCV pseudpotential for Mo from the SG15 database~\cite{oncv2015} (version 1.0), and for (ii) and (iii) we use ONCV pseudopotentials from the PseudoDojo database~\cite{van2018pseudodojo}. While performance studies are performed on benchmark systems (i) and (ii), the validation studies are performed on  the smallest system size among the benchmark systems (i) and (ii), and further for the benchmark systems (iii) and (iv). All validation studies are conducted using a $\Gamma$-point calculation. In each of the above benchmark systems, we solve \DFTFE to two levels of discretization errors---medium and high accuracy---and compare against reference values obtained at similar accuracy using \QE/\nwchemNoSpace. The medium accuracy standard corresponds to $\order{10^{-4}}$ Ha/atom in ground-state energy, $\order{ 10^{-4}}$ Ha/Bohr in ionic forces, and $\order{ 10^{-6}}$ Ha/${\rm Bohr}^3$ in cell stress (in periodic benchmark systems). The high accuracy standard corresponds to $\order{10^{-5}}$ Ha/atom in ground-state energy, $\order{10^{-5}}$ Ha/Bohr in ionic forces, and $\order{ 10^{-7}}$ Ha/${\rm Bohr}^3$ in cell stress. In the remainder of this paper, we define chemical accuracy as satisfying the medium accuracy standard.  Further, we note that all simulations using \DFTFE are ran  with mixed precision algorithms enabled in the RRGEP computational step and also employing FP32 communication in the CF computational step (cf.~Sections~\ref{sec:numerics} and~\ref{sec:arch}). We remark that although similar performance and  validation studies were undertaken in previous work~\cite{motamarri2020} (\DFTFEverprevNoSpace) on many-core CPU architectures,  these aspects are revisited here for \DFTFEver due to the modified treatment of electrostatics and configurational forces in the computational methodology and the new hybrid CPU-GPU implementation.

Most of the simulations reported in this work were executed on the Summit supercomputer, which is currently the second fastest supercomputer in the world, with 200.79 PFLOPS FP64 peak. Summit comprises of 4,608 IBM Power System AC922 nodes with two IBM POWER9 processors (42 physical cores) and six NVIDIA Volta V100 GPUs in each node. Each node contains 512 GB of DDR4 memory for use by the POWER9 processors and 16 GB of HBM2 for each V100 GPU. Summit nodes are connected to a dual-rail EDR InfiniBand network providing a node injection bandwidth of 23 GB/s.  On Summit, we have compiled \DFTFE using NVIDIA CUDA/10.1.105, GCC/6.4.0, IBM Spectrum-MPI/10.3.0.0, IBM ESSL/6.1.0, ELPA 2021.05.002, and NVIDIA NCCL 2.10.3. Some of the numerical simulations using CPUs reported in this work were also executed on the Cori supercomputer at the National Energy Research Scientific Computing (NERSC) center. In particular, we used Cori's Phase \rom{2} partition containing 9,688 compute nodes based on  Intel Xeon Phi processors. Each compute node has the following specifications:  single-socket Intel Xeon Phi $7250$ (``Knights Landing'') processor with $68$ physical cores per node @ $1.4$ GHz and $96$ GB of memory per node. Cori uses a Cray Aries with Dragonfly topology for inter-node communication with  $45.0$ TB/s global peak bidirectional bandwidth. The \DFTFE simulations using the Summit GPU nodes reported in this section used all 6 V100 GPUs in each node with multiple MPI tasks (1--3) per GPU via NVIDIA multi-process service (MPS) depending on the system size. \DFTFE simulations using only the CPU cores of the Summit nodes used 42 MPI tasks per node with each MPI task bound to one physical core. Finally, the \DFTFE simulations using the CPU-only Cori supercomputer were run using $32$ MPI tasks per node and $2$ OpenMP threads (for \verb|BLAS| operations).

\subsection{Accuracy validation of \DFTFEver}\label{sec:validation}

In the benchmark systems involving Mo, we consider a periodic supercell constructed from bcc Mo unit cells containing 2 atoms with lattice constant of $5.95$ Bohr. Next, we introduce a monovacancy by removing a Mo atom from the supercell. We consider a supercell size of $6\times6\times6$ Mo unit cells, denoted by $\textrm{Mo}_{\textrm{6x6x6}}$, containing $431$ atoms ($6,034$ electrons). Tables~\ref{tab:pspValidationMed} and~\ref{tab:pspValidationHigh} shows the comparison of ground-state energy, ionic forces and hydrostatic cell stress between \DFTFEver and \QE for medium and high accuracy comparisons respectively, which demonstrates excellent agreement between the two codes for a periodic system. We remark that taking advantage of the modified electrostatics methodology and efficient numerical implementation strategies  in this work, we find the most computationally efficient strategy for pseudopotential DFT calculations is to use relatively higher order finite-elements, ${\rm FE}_{\rm ord}=6/7$ (HEX343SPECTRAL/HEX512SPECTRAL) with coarser element sizes compared to \DFTFEverprev~\cite{motamarri2020}. Importantly, this provides significant computational efficiency improvements in comparison to \DFTFEverprevNoSpace, as will be discussed in Section~\ref{sec:compeff}. 

Next, in the non-periodic benchmark system involving Al, we consider a three-dimensional non-periodic Icosahedron Al nano-particle~\cite{cuNano}. The Icosahedron nano-particles are constructed with nearest neighbour bond length of $6.8$ Bohr and varying the number of shells. We consider a Al nano-particle of size: $\textrm{Al}_{\textrm{3-Shell}}$ containing $147$ atoms ($441$ electrons). For the \DFTFE simulations, we employ non-periodic boundary conditions on a domain with a large vacuum surrounding the Al nano-particle. On the other hand, for the \QE simulations, we choose an artificial periodic domain containing the Al nano-particle. The ground-state energy and ionic forces are converged with respect to the domain size in both \DFTFE and \QE simulations. Tables~\ref{tab:pspValidationMed} and~\ref{tab:pspValidationHigh} shows the comparison between \DFTFE and \QE for the $\textrm{Al}_{\textrm{3-Shell}}$ benchmark system, which demonstrates excellent agreement between the two codes in a non-periodic setting. 

The final pseudopotential benchmark system we consider is the InP-water semiconductor-insulator interface system, which consists of a semiconducting InP slab of $22.178 \times 22.178 \times 44$ Bohr and an insulating water column of $35$ Bohr in the Z direction containing $47$ water molecules. In total, the system contains $277$ atoms ($1,632$ electrons). For both the \DFTFE and \QE simulations, we use periodic boundary conditions on the domain. Tables~\ref{tab:pspValidationMed} and~\ref{tab:pspValidationHigh} shows the comparison between \DFTFEver and \texttt{QE}, which demonstrates excellent agreement in ground-sate energy and ionic forces, similar to Mo and Al benchmark systems. 

\begin{table}[htbp]
\centering
\small
\caption{\label{tab:pspValidation}\small{Validation of \DFTFEver with \QE on pseudopotential benchmark systems. ${\rm FE}_{\rm ord}, h_{\rm min}$ and $h_{\rm max}$ denote the FE polynomial order, minimum element size and maximum element size (Bohr), respectively, employed in \DFTFE. ${ E}_{\rm cut}$ denotes the plane-wave basis energy cut-off (Hartree) used in \QE. ${ E}_{\rm g}$ denotes ground-state energy (Hartree/atom). $\Delta_{\rm max} {f}=\max\limits_{1\le i \le N_{a}} \left\|{\bf f}_i^{\rm DFT-FE} -{\bf f}_i^{\rm QE}\right\|$ (Hartree/Bohr), where ${\bf f}_i$ denotes the force on the $i^{\rm th}$ atom. $\Delta {\sigma_h}=\abs{\sigma_h^{\rm DFT-FE}-\sigma_h^{\rm QE}}$ (Hartree/${\rm Bohr}^3$), where $\sigma_h$ denotes the hydrostatic cell stress.}}
\begin{subtable}[h]{1.0\textwidth}
\centering
\begin{tabular}{|c|c|c|c|c|c|}
\hline
System & DFT-FE & DFT-FE & QE & QE & Difference in forces \\
       & $\left({\rm FE}_{\rm ord}, h_{\rm min},\, h_{\rm max}\right)$ & $E_g$ &  ${E}_{\rm cut}$ & ${ E_g}$  &  \& stress $\left(\Delta_{\rm max} {f},\,\Delta \sigma_h\right)$  \\
\hline
$\textrm{MoVac}_{\textrm{6x6x6}}$ & $7$, $2.1$, $2.1$ & $-68.579265$  & $30$ & $-68.579480$& $\Delta_{\rm max} {f}= 3.4 \times 10^{-4}$\\
                            &               &              &    &             & $\Delta \sigma_h= 6.8 \times 10^{-6}$\\
$\textrm{Al}_{\textrm{3-Shell}}$ & $7$, $2.1$, $8.3$ & $-2.300791$  & $20$ & $-2.300782$& $\Delta_{\rm max} {f}= 1.4 \times 10^{-4}$\\
$\textrm{InP-Water}$ & $6$, $1.0$, $2.0$ & $-21.883224$  & $40$ & $-21.883248$& $\Delta_{\rm max} {f}= 2.5 \times 10^{-4}$\\
\hline
\end{tabular}
		\caption{\small{Medium accuracy level comparisons.}}
		\label{tab:pspValidationMed}
		\vspace{0.3cm}
\end{subtable}

\begin{subtable}[h]{1.0\textwidth}
\centering
\begin{tabular}{|c|c|c|c|c|c|}
\hline
System & DFT-FE & DFT-FE & QE & QE & Difference in forces \\
       & $\left({\rm FE}_{\rm ord}, h_{\rm min},\, h_{\rm max}\right)$ & $E_g$ &  ${E}_{\rm cut}$ & ${ E_g}$  &  \& stress $\left(\Delta_{\rm max} {f},\,\Delta \sigma_h\right)$  \\
\hline
$\textrm{MoVac}_{\textrm{6x6x6}}$ & $7$, $1.6$, $1.6$ & $-68.579561$ & 50 & $-68.579503$ & $\Delta_{\rm max} {f}= 2.9 \times 10^{-5}$\\
                            &               &              &    &             & $\Delta \sigma_h= 4.4 \times 10^{-7}$\\
$\textrm{Al}_{\textrm{3-Shell}}$ & $7$, $1.7$, $6.7$ & -2.300817  & 30 & -2.300789& $\Delta_{\rm max} {f}= 4.0 \times 10^{-5}$\\
$\textrm{InP-Water}$ & $6$, $0.8$, $1.6$ & $-21.883260$  & 70 &$-21.883267$ & $\Delta_{\rm max} {f}=3.8 \times 10^{-5}$\\
\hline
\end{tabular}
		\caption{\small{High accuracy level comparisons.}}
		\label{tab:pspValidationHigh}
		\vspace{0.3cm}
\end{subtable}
\end{table}

We now validate \DFTFEver for all-electron ground-state calculations by considering the Benzamide (${\rm C}_{7}{\rm H}_{7}  {\rm N} {\rm O}$) molecule. In both \nwchem and \DFTFE simulations we conduct non-periodic calculations with a domain size large enough for the boundary effects to be negligible. The appropriate mesh parameters for \DFTFE and choice of Gaussian basis for \nwchem are shown in Table~\ref{tab:aeValidation}. The results demonstrate good agreement between \DFTFE and \nwchem in the ground-state energy.

Overall, from Tables~\ref{tab:pspValidation} and~\ref{tab:aeValidation}, we observe excellent agreement between \DFTFEver and \QE on pseudopotential DFT benchmark systems, and with \nwchem for all-electron benchmark systems.

\begin{table}[htbp]
\centering
\small
\cprotect\caption{\label{tab:aeValidation}\small{Validation of \DFTFEver against \nwchem code on Benzamide molecule, a non-periodic benchmark problem. ${\rm FE}_{\rm ord}, h_{\rm min}$ and $h_{\rm max}$ denote the FE polynomial order, minimum element size and maximum element size (Bohr), respectively, in \DFTFE. ${ E}_{\rm g}$ denotes ground-state energy (Hartree/atom).}}
\begin{tabular}{|c|c|c|c|c|}
\hline
System  &\DFTFE & \DFTFE & \nwchem & \nwchem \\
        & $\left({\rm FE}_{\rm ord}, h_{\rm min},\, h_{\rm max}\right)$ & ${ E_g}$ & Gaussian basis & ${E_g}$  \\
\hline
Benzamide & 5, 0.02, 5.0 &  $-25.040821 $  & pc-2  &  $-25.039621  $ \\
 & &  & pc-3  &  $-25.040755  $ \\
\hline
\end{tabular}
\end{table}




\subsection{Performance of \DFTFEver}\label{sec:performance}
The computational efficiency and parallel scalability of \DFTFEver is discussed in this section on benchmark systems involving periodic Mo supercells and non-periodic Al nanoparticles of various system sizes.  We employ finite-element meshes with discretization errors commensurate with chemical accuracy for all the benchmarks considered here.
\subsubsection{Computational efficiency}\label{sec:compeff}
We first discuss about the number of FE basis functions required and the CPU computational times obtained in the current work (\DFTFEverNoSpace) on varying system sizes of BCC Mo periodic supercells with a monvacancy and compare with that obtained by using \DFTFEverprev ~\cite{motamarri2020}. These comparisons  tabulated in Table~\ref{tab:compv0.6vscompv1.0} show $\sim3\times$ reduction in the number of FE basis functions using \DFTFEverNoSpace, and is accompanied by reduction in CPU computational time by $2-3\times$. This reduction in basis functions is due to the modified treatment of electrostatics using the smeared nuclear charges approach where the atomic positions are no longer constrained to coincide with FE nodes (for pseudopotential calculations), which in turn allowed consideration of coarser FE meshes with higher polynomial orders for a similar accuracy. 
\begin{table}[htbp]
\centering
\small
\caption{\label{tab:cpuTimeComparisonPrevVersion}\small{Number of FE basis functions and computational cost comparison of \DFTFEver (this work) with \DFTFEverprev to achieve chemical accuracy. Computational cost per SCF iteration step  is reported in node-Hrs. Node-hrs for the \DFTFEverprev are obtained from~\cite{motamarri2020}, which used an FE discretization corresponding to $({\rm FE}_{\rm ord}=5, h_{\rm min}=0.74, h_{\rm max}=1.49)$. FE discretization parameters used in DFT-FE 1.0 calculations corresponds to $({\rm FE}_{\rm ord}=7, h_{\rm min}=2.1, h_{\rm max}=2.1)$. All calculations are performed on NERSC Cori KNL nodes.}}
\centering
\begin{tabular}{|c|c|c|c|c|c|}
\hline
 System & Number of atoms & FE basis & Node-Hrs & FE basis  & Node-Hrs \\
 & (Number of electrons)  &   v0.6     &  v0.6      &   v1.0 &      v1.0     \\
\hline
$\textrm{MoVac}_{\textrm{6x6x6}}$ & 431 (6034)      & 5,475,843  & 0.5  & 1,728,000  & 0.2 \\
$\textrm{MoVac}_{\textrm{8x8x8}}$ & 1,023 (14,322)    & 12,942,743 & 4.2  & 4,251,528 & 1.4  \\
$\textrm{MoVac}_{\textrm{10x10x10}}$ & 1,999 (27,986) & 25,229,995 & 17.7 & 7,645,373 & 7.5  \\
\hline
\end{tabular}\label{tab:compv0.6vscompv1.0}
\end{table}

\begin{figure}[htbp]
\includegraphics[scale=0.5]{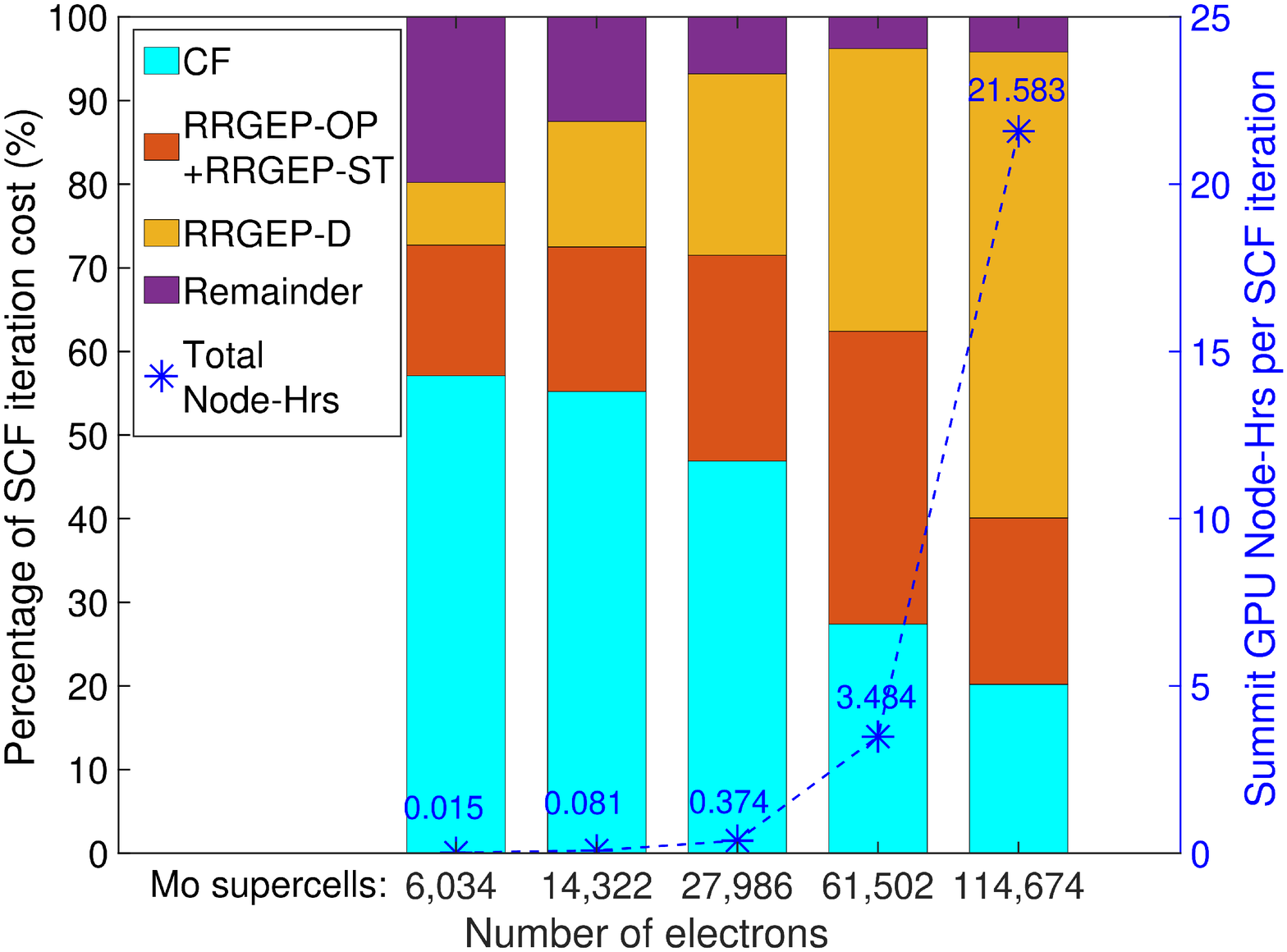}
 \centering
\caption{\small{Bar plot analysis of system size dependence in computational cost breakdown of a single SCF iteration step using \DFTFE on Summit GPU nodes.  Breakdown of single SCF cost into the major computational steps:  CF, (RRGEP-OP+RRGEP-ST), RRGEP-D, and Remainder (DM+TEP+DHM+DC+Overhead). \textbf{Case study}: Pseudopotential DFT calculations on periodic BCC Mo supercells with a monovacancy using $\Gamma$-point (cf. Tables~\ref{tab:cpugpuComparisonBCCMo} and ~\ref{tab:gpuLargeBCCMo} for detailed breakdown and CPU-only/CPU-GPU comparison). }}
\label{fig:breakdownPercentagesMoSupercells}
\end{figure}

We now demonstrate the computational efficiency in terms of CPU-GPU speedups obtained by \DFTFEver using benchmark systems of different sizes on Summit supercomputer, a hybrid CPU-GPU architecture. The CPU-GPU speedups are measured by computing the ratio of the node-hours obtained by running most optimized CPU-only calculation and that obtained by the hybrid CPU-GPU run. To this end, we first consider periodic $\Gamma$-point calculations on  $\textrm{MoVac}_{\textrm{6x6x6}}$ (6,034 electrons), $\textrm{MoVac}_{\textrm{8x8x8}}$ (14,322 electrons) and $\textrm{MoVac}_{\textrm{10x10x10}}$ (27,986 electrons). It is evident from Table ~\ref{tab:cpugpuComparisonBCCMo} and Figure ~\ref{fig:breakdownPercentagesMoSupercells} that the dominant computational cost of a single SCF iteration on GPUs in this regime ($5,000-30,000$ electrons) is the Chebyshev polynomial filtering (CF) ($\sim50\%$ of the single SCF iteration cost) followed by the Rayleigh Ritz step (RRGEP) ($\sim30-40\%$ of the single SCF iteration cost). We note from Table ~\ref{tab:cpugpuComparisonBCCMo} that the CPU-GPU speedups of CF step stay nearly constant for different Mo supercell sizes, and is around $21-23\times$. Recall that the CF step is dominated by dense matrix-matrix multiplications---involving FE discretized cell level Hamiltonian matrices with wavefunction blocks of fixed sizes---and the FP32 processor boundary communication during the assembly operations. Hence, the arithmetic intensity of CF remains more or less independent of system size leading to similar CPU-GPU speedups. In the case of RRGEP step, the computational times for RRGEP-OP and RRGEP-ST show an increasing trend of CPU-GPU speedups with system size ranging from 13$\times$ for $\textrm{MoVac}_{\textrm{6x6x6}}$ to 24$\times$ for $\textrm{MoVac}_{\textrm{10x10x10}}$\,(cf. Table ~\ref{tab:cpugpuComparisonBCCMo}). As discussed in the Section \ref{sec:arch-rr}, the floating point operations in RRGEP-OP and RRGEP-ST scale as $\order{M N^2}$ whereas the communication costs scale as $\order{\log(M) N^2}$. This leads to increased floating point operations per byte of data accessed  (arithmetic intensity) with increasing system size, thus leading to greater CPU-GPU speedups for larger systems. In the case of  $\textrm{MoVac}_{\textrm{6x6x6}}$, we further note that the CPU-GPU speedups for a single SCF iteration is almost the same as the total ground-state run time, and is around $14-15\times$. This was possible as the majority of non-SCF computational steps involving initializations of pseudopotentials and electron density, ionic force computation, multivector Poisson problem solution for self-potential evaluations have all been either significantly optimized, or have been GPU ported wherever amenable. We now examine the CPU-GPU speedups in the case of periodic calculations on $\textrm{MoVac}_{\textrm{6x6x6}}$ supercell with a vacancy employing four irreducible $k$-points for Brillouin zone sampling. This involves the use of complex datatypes to handle complex wavefunctions on GPUs. To this end, we note from the Table \ref{tab:cpugpuComparisonBCCMoKpt} that the CPU-GPU speedups for the dominant computational steps involving CF, RRGEP-OP, RRGEP-ST are higher than $\Gamma$-point calculations employing real data types as reported in Table ~\ref{tab:cpugpuComparisonBCCMo}. The CPU-GPU speedup for a single SCF iteration was found to be $22\times$ in contrast to $15.5\times$ while employing real datatypes (c.f Table ~\ref{tab:cpugpuComparisonBCCMo}). This increase can be attributed to the use of complex datatypes leading to an increased arithmetic intensity ($\sim 2\times$) for various computational steps in comparison to real datatypes. We next present in Table~\ref{tab:cpugpuComparisonAlNP} the CPU-GPU speedups for non-periodic calculations involving icosahedron Al nanoparticles, namely $\textrm{Al}_{\textrm{7Shell}}$ (4245 electrons) and $\textrm{Al}_{\textrm{12Shell}}$ (19575 electrons). Similar to the $\Gamma$-point calculations on BCC Mo supercells, the CPU-GPU speedup of CF step is similar for both systems, while the speedup for RRGEP-OP and RRGEP-ST increases with system size. The CPU-GPU speedup for single SCF iteration is observed to be 17$\times$ for $\textrm{Al}_{\textrm{7Shell}}$ and increased to 20$\times$ in the case of $\textrm{Al}_{\textrm{12Shell}}$. We finally report computational times for two large Mo supercells, namely $\textrm{MoVac}_{\textrm{13x13x13}}$ (61,502 electrons) and $\textrm{MoVac}_{\textrm{16x16x16}}$ (114,674 electrons), in Table ~\ref{tab:gpuLargeBCCMo}. We note that even for such large systems, the ground-state calculations, computed to chemical accuracy, are completed in wall-times of $<1$hr. However, we note that the dominant computational cost for both these system sizes is the diagonalization cost (RRGEP-D) and accounts for around 34\% and 56\% of the total SCF cost for  $\textrm{MoVac}_{\textrm{13x13x13}}$ and  $\textrm{MoVac}_{\textrm{16x16x16}}$ systems, respectively. Computationally efficient and scalable approaches to reduce the diagonalization costs will be a subject of our future investigations. Finally, the studies in this section also throw some light on the computational complexity of our numerical implementation on hybrid CPU-GPU architectures with respect to number of electrons ($N_e$). As can be seen from Figure ~\ref{fig:breakdownPercentagesMoSupercells} and Tables  ~\ref{tab:cpugpuComparisonBCCMo} and ~\ref{tab:gpuLargeBCCMo}, the range of close to quadratic scaling in computational complexity with respect to $N_e$ is observed for much larger systems--- $\order{N_e^{2.28}}$ up to $N_e = 27,986$---and only approach cubic scaling at $N_e = 61,502$ and $N_e=114,674$ electrons.

\begin{table}[htbp]
\centering
\small
\caption{\label{tab:cpugpuComparisonBCCMo}\small{Comparison of computational cost, measured in node-hrs, between CPU-only and hybrid CPU-GPU runs on OLCF Summit nodes using \DFTFE. Breakdown of total computational cost into the following computational steps: a) initialization \& nuclear potential solve (INIT), electron density mixing (DM), total electrostatic potential solve (TEP), discrete Hamiltonian matrix construction (DHM), Chebyshev filtering (CF), Rayleigh-Ritz generalized eigenvalue problem (RRGEP), density computation (DC), and ionic forces (IF). RRGEP is further dividied into the following sub-steps: overlap matrix \& subspace projected Hamiltonian (RRGEP-OP),  subspace diagonalization (RRGEP-D) and subspace transformation to FE basis (RRGEP-ST).  $N_{at}$ denotes total number of atoms, $N_e$ denotes total number of electrons, and $E_g$ denotes the ground-state energy per atom (Hartree). \textbf{Case study}: Pseudopotential DFT calculations on periodic BCC Mo supercells with a monovacancy using $\Gamma$-point. }}
\begin{tabular}{|l|c|c|c|c|c|c|c|c|c|}
\hline
 {\bf Material}                      & \multicolumn{3}{c|} {$\textrm{MoVac}_{\textrm{6x6x6}}$} & \multicolumn{3}{c|} {$\textrm{MoVac}_{\textrm{8x8x8}}$} & \multicolumn{3}{c|} {$\textrm{MoVac}_{\textrm{10x10x10}}$}  \\
 \hline
  $N_{at}$ ($N_e$)                      & \multicolumn{3}{c|} {431 (6,034)} & \multicolumn{3}{c|} {1,023 (14,322)} & \multicolumn{3}{c|} {1,999 (27,986)}  \\
   DoFs                      & \multicolumn{3}{c|} {1,728,000} & \multicolumn{3}{c|} {4,251,528} & \multicolumn{3}{c|} {7,645,373}  \\ 
    $E_{g}$                      & \multicolumn{3}{c|} {-68.57927} & \multicolumn{3}{c|} {-68.57839} & \multicolumn{3}{c|} {-68.57883}  \\
    SCF iterations                     & \multicolumn{3}{c|} {15} & \multicolumn{3}{c|} {16} & \multicolumn{3}{c|} {16}  \\    
\hline
{\bf Resources} & CPU & \multicolumn{2}{c|}{CPU-GPU}  & CPU & \multicolumn{2}{c|}{CPU-GPU}  & CPU & \multicolumn{2}{c|}{CPU-GPU}\\
\hline
 Nodes & 4 & \multicolumn{2}{c|}{4} & 10 & \multicolumn{2}{c|}{14} & 20 &\multicolumn{2}{c|}{40}\\  
 CPU cores & 168 & \multicolumn{2}{c|}{168} & 420 & \multicolumn{2}{c|}{588} & 840 &\multicolumn{2}{c|}{1680}\\  
 GPUs & 0 & \multicolumn{2}{c|}{24} & 0 &\multicolumn{2}{c|}{84} & 0 & \multicolumn{2}{c|}{240}\\ 
 MPI tasks & 168 & \multicolumn{2}{c|}{72} & 420 & \multicolumn{2}{c|}{252} & 840 &\multicolumn{2}{c|}{720}\\  
\hline
{\bf Total run} & Node & Node & Speed  & Node & Node & Speed & Node & Node & Speed \\
{\bf breakdown} & -hrs & -hrs & -up & -hrs & -hrs &-up & -hrs & -hrs &-up\\
\hline
INIT+Others            & 0.053 &  0.069  &  -    & -  & 0.257  &  -&-  & 0.857  & -  \\
Total SCF        & 5.006 & 0.289  & 17.3 &  - & 1.782  & - & -  & 8.300 & - \\
IF               & 0.024 & 0.015   & 1.6  &  - & 0.062  &  -& - & 0.246  &- \\
{\bf Total run}  & 5.082 & 0.372  & {\bf 13.7} &  - & 2.100  & - & - &  9.406 & -\\
\hline
{\bf Single SCF} & Node & Node & Speed  & Node & Node & Speed & Node & Node & Speed \\
{\bf breakdown} & -hrs & -hrs & -up & -hrs & -hrs &-up & -hrs & -hrs &-up\\
\hline
DM+TEP+Others    & 0.0034 & 0.0016  &  - & 0.0136 & 0.0051  & - & 0.0622 &  0.0133 & -\\
DHM              & 0.0061  &0.0008   & 7.6  & 0.0150 & 0.0019  & 7.9 & 0.0278 &  0.0022 & 12.6\\
CF               & 0.1761  & 0.0084  & 20.9   & 1.0123  & 0.0447  & 22.6 & 4.1483  & 0.1756  & 23.6\\
RRGEP-OP         & 0.0192 & 0.0014  &13.7& 0.1814 & 0.0086   & 21.1  &  1.2389 & 0.0511  & 24.2  \\
RRGEP-D          & 0.0007 & 0.0011  & 0.6 & 0.0058  & 0.0121   & 0.5 & 0.0367  & 0.0811  & 0.5  \\
RRGEP-ST         & 0.0120 & 0.0009 & 13.3 & 0.1167 & 0.0054   & 21.6 & 1.010 & 0.0411  &  24.6\\
DC               & 0.0104 & 0.0004 & 26  & 0.0519 &   0.0027 & 19.2 & 0.2211 & 0.0089  & 24.8\\
{\bf Total single SCF}  & 0.2280 &  0.0147 & {\bf 15.5}  & 1.3972  & 0.0809   & {\bf 17.3} & 6.7450  & 0.3744  & {\bf 18.0}\\
\hline
\end{tabular}
\end{table}

\begin{table}[htbp]
\centering
\small
\caption{\label{tab:cpugpuComparisonBCCMoKpt}\small{Comparison of computational cost, measured in node-hrs, between CPU-only and hybrid CPU-GPU runs on OLCF Summit nodes using \DFTFE. \textbf{Case study}: Periodic pseudopotential DFT calculations with 4 irreducible $k$-points on BCC Mo supercell with a monovacancy, which requires complex datatype for the computations. The CPU and hybrid CPU-GPU simulations used both domain decomposition and $k$-point parallelization. }}
\begin{tabular}{|l|c|c|c|}
\hline
 {\bf Material sys.}                      & \multicolumn{3}{c|} {$\textrm{MoVac}_{\textrm{6x6x6}}$---4 irreducible $k$-points}   \\
 \hline
  $N_{at}$ ($N_e$)                      & \multicolumn{3}{c|} {431 (6,034)} \\
   DoFs                      & \multicolumn{3}{c|} {1,728,000}   \\ 
    $E_{g}$ (Ha/atom)                     & \multicolumn{3}{c|} {-68.57850} \\
    SCF iterations                     & \multicolumn{3}{c|} {15}   \\    
\hline
{\bf Compute res.} & CPU & \multicolumn{2}{c|}{CPU-GPU}   \\
\hline
 Nodes & 20 & \multicolumn{2}{c|}{40}  \\  
 CPU cores & 840 & \multicolumn{2}{c|}{1680}  \\  
 GPUs & 0 & \multicolumn{2}{c|}{240}  \\ 
 MPI tasks & 840  & \multicolumn{2}{c|}{240}  \\  
\hline
{\bf Single SCF} & Node-hrs & Node-hrs & Speedup    \\
{\bf breakdown}         &   & &  \\
\hline
DM+TEP+DHM+DC+Others  & 0.352  & 0.042  & 8.4  \\
CF  & 3.396  & 0.117  &   29.0  \\
RRGEP-OP & 0.356 &  0.014 & 25.4   \\
RRGEP-D & 0.027 & 0.015  &  1.8 \\
RRGEP-ST  & 0.226 &  0.009 & 25.7  \\
{\bf Total single SCF}  & {\bf 4.357}  & {\bf 0.197}   & {\bf 22.1}   \\
\hline
\end{tabular}
\end{table}

\begin{table}[htbp]
\centering
\small
\caption{\label{tab:cpugpuComparisonAlNP}\small{Comparison of computational cost, measured in node-hrs, between CPU-only and hybrid CPU-GPU runs on OLCF Summit nodes using \DFTFE. \textbf{Case study}: Pseudopotential DFT calculations on non-periodic Al nanoparticles. }}
\begin{tabular}{|l|c|c|c|c|c|c|}
\hline
 {\bf Material sys.}                      & \multicolumn{3}{c|} {$\textrm{Al}_{\textrm{7Shell}}$} & \multicolumn{3}{c|} {$\textrm{Al}_{\textrm{12Shell}}$}  \\
 \hline
  $N_{at}$ ($N_e$)                      & \multicolumn{3}{c|} {1415 (4,245)} & \multicolumn{3}{c|} {6,525 (19,575)}  \\
   DoFs                      & \multicolumn{3}{c|} {10,272,229} & \multicolumn{3}{c|} {43,034,039}   \\ 
    $E_{g}$ (Ha/atom)                     & \multicolumn{3}{c|} {-2.30939} & \multicolumn{3}{c|} {-2.31226}  \\
    SCF iterations                     & \multicolumn{3}{c|} {54} & \multicolumn{3}{c|} {92}   \\    
\hline
{\bf Compute res.} & CPU & \multicolumn{2}{c|}{CPU-GPU}  & CPU & \multicolumn{2}{c|}{CPU-GPU}  \\
\hline
 Nodes & 20  & \multicolumn{2}{c|}{20} & 100  & \multicolumn{2}{c|}{190} \\  
 CPU cores & 840 & \multicolumn{2}{c|}{840} & 4200 & \multicolumn{2}{c|}{7980} \\  
 GPUs & 0 & \multicolumn{2}{c|}{120} & 0 &\multicolumn{2}{c|}{1140} \\ 
 MPI tasks & 840 & \multicolumn{2}{c|}{360} & 4200 & \multicolumn{2}{c|}{3420} \\  
\hline
{\bf Single SCF} & Node-hrs & Node-hrs & Speedup  & Node-hrs & Node-hrs & Speedup  \\
{\bf breakdown}         &   & &  &  & & \\
\hline
DM+TEP+Others  & 0.012 & 0.016  &  - &0.34  & 0.248  & -  \\
DHM  & 0.031 & 0.005  & 6.2  & 0.182 & 0.021  & 8.1  \\
CF  & 0.72  & 0.025  & 28.8  & 14.5 & 0.567  &  25.6 \\
RRGEP-OP & 0.07 &  0.006 & 11.7 & 3.927 & 0.172  & 22.8 \\
RRGEP-D &0.002  & 0.0027  & 0.74  & 0.074 & 0.140 & 0.42 \\
RRGEP-ST  & 0.054 & 0.0034  & 15.9 & 3.594 &  0.141 &  25.4\\
DC  & 0.213 & 0.0063  & 33.8  & 4.352 & 0.103  & 42.3 \\
{\bf Total single SCF}  & 1.1 & 0.064 & \bf{17.2}  &  26.97 &  1.4 & \bf{19.3} \\
\hline
\end{tabular}
\end{table}

\begin{table}[htbp]
\centering
\small
\caption{\label{tab:gpuLargeBCCMo}\small{Computational cost measured in node-hrs of hybrid CPU-GPU runs involving large system sizes ($>$ 60,000 electrons) on OLCF Summit Nodes using \DFTFE. \textbf{Case study}: Pseudopotential DFT calculations on periodic BCC Mo supercells with a monovacancy.}}
\begin{tabular}{|l|c|c|c|c|}
\hline
 {\bf Material sys.}                      & \multicolumn{2}{c|} {$\textrm{MoVac}_{\textrm{13x13x13}}$} & \multicolumn{2}{c|} {$\textrm{MoVac}_{\textrm{16x16x16}}$}  \\
 \hline
  $N_{at}$ ($N_e$)                      & \multicolumn{2}{c|} {4393 (61,502)} & \multicolumn{2}{c|} {8,191 (114,674)}  \\
   DoFs                      & \multicolumn{2}{c|} {16,194,277} & \multicolumn{2}{c|} {29,503,629}   \\ 
    $E_{g}$ (Ha/atom)                     & \multicolumn{2}{c|} {-68.57884} & \multicolumn{2}{c|} {-68.57890}  \\
    SCF iterations                     & \multicolumn{2}{c|} {18} & \multicolumn{2}{c|} {19}   \\    
\hline
{\bf Compute res.} & \multicolumn{2}{c|}{CPU-GPU}  &  \multicolumn{2}{c|}{CPU-GPU}  \\
\hline
 Nodes   & \multicolumn{2}{c|}{160} &  \multicolumn{2}{c|}{600} \\  
 CPU cores  & \multicolumn{2}{c|}{6720} &  \multicolumn{2}{c|}{25200} \\  
 GPUs  & \multicolumn{2}{c|}{960} & \multicolumn{2}{c|}{3600} \\ 
 MPI tasks   & \multicolumn{2}{c|}{1920} &  \multicolumn{2}{c|}{3600} \\  
\hline
{\bf Total run} & Wall-time & Node-hrs & Wall-time & Node-hrs  \\
{\bf breakdown} & (sec)  &   & (sec)  & \\
\hline
INIT+Others & 117.2 & 5.209  &176.6  & 29.433\\
Total SCF  & 1780 & 79.100 & 3034  & 505.667 \\
IF  & 31.7 & 1.409  &39.5  & 6.583    \\
{\bf Total run}  & 1928.9 & 85.729 & 3250.1   & 541.683  \\
\hline
{\bf Single SCF} & Wall-time & Node-hrs & Wall-time & Node-hrs  \\
{\bf breakdown}         & (sec)  &   & (sec)  & \\
\hline
DM+TEP+DHM+DC+Others  & 3.0  & 0.133  & 5.4  & 0.900    \\
CF  & 21.5 & 0.956 & 26.2  &4.367   \\
RRGEP-OP & 14.2  & 0.631 & 12.4 &  2.067 \\
RRGEP-D & 26.5  & 1.178 & 72.1 &  12.017 \\
RRGEP-ST  & 13.2  & 0.587 & 13.4   & 2.233 \\
{\bf Total single SCF}  & 78.4 & 3.484   & 129.5  & 21.583   \\
\hline
\end{tabular}
\end{table}

\subsubsection{Parallel scalability}
The primary parallelization route in \DFTFE is based on domain decomposition parallelization of the adaptive FE mesh into partitions distributed across different MPI tasks. The domain decomposition parallelization in \DFTFE is implemented through the deal.II finite element library~\cite{dealII90} with p4est~\cite{BangerthBursteddeHeisterEtAl11}. Unlike the all-to-all communication in global basis sets like plane-wave basis, only the FE nodes on the MPI domain partition boundaries are communicated. This has enabled massive parallel scalability of \DFTFE, as has been previously demonstrated on many-core CPU architectures~\cite{motamarri2020}. Notably, for a $\sim$60,000 electrons system, \DFTFE was demonstrated to scale up to 192,000 MPI tasks (6,000 nodes) of the NERSC Cori KNL supercomputer. The parallel scaling advantages of FE basis naturally translates to the hybrid CPU-GPU architectures, although with the additional challenge of compute being significantly faster than communication on these architectures. In this work, our scalable numerical implementation, in conjunction with strategies such as mixed precision algorithms, asynchronous compute and communication, as well as GPU direct communication (cf. Section~\ref{sec:arch}) has enabled excellent parallel scalability on hybrid CPU-GPU architectures, which we demonstrate subsequently.

We now demonstrate the parallel scalability of \DFTFEver on the Summit supercomputer, a hybrid CPU-GPU architecture, using benchmarks systems of different sizes. In particular, we reconsider the BCC Mo periodic benchmark systems containing a monovacancy ($\Gamma$-point calculations)---$\textrm{MoVac}_{\textrm{6x6x6}}$ and  $\textrm{MoVac}_{\textrm{8x8x8}}$ (cf. Section~\ref{sec:validation})---which consist of 6,034 and 14,322 electrons, respectively, and conduct a strong-scaling study. The FE discretization parameters used are commensurate with chemical accuracy in energy per atom and ionic forces (cf. Section~\ref{sec:validation}). 
The speedup in the strong-scaling study is measured relative to the wall-time taken on 12 and 42 GPUs for  $\textrm{MoVac}_{\textrm{6x6x6}}$  and $\textrm{MoVac}_{\textrm{8x8x8}}$, respectively, which are the smallest accessible number of GPUs based on memory requirements. We note that only domain decomposition parallelization is considered in these strong-scaling studies. Although \DFTFE implements a second level of parallelization over wavefunctions (band parallelization), which has been demonstrated to extend parallel scalability on many-core architectures~\cite{motamarri2020}, we find the combined domain decomposition and band parallelization route to only marginally extend the parallel scalability of \DFTFE on hybrid CPU-GPU architectures. 

\begin{figure}[t!]
    \centering
    \begin{subfigure}[t]{0.48\textwidth}
        \centering
        \includegraphics[scale=0.27]{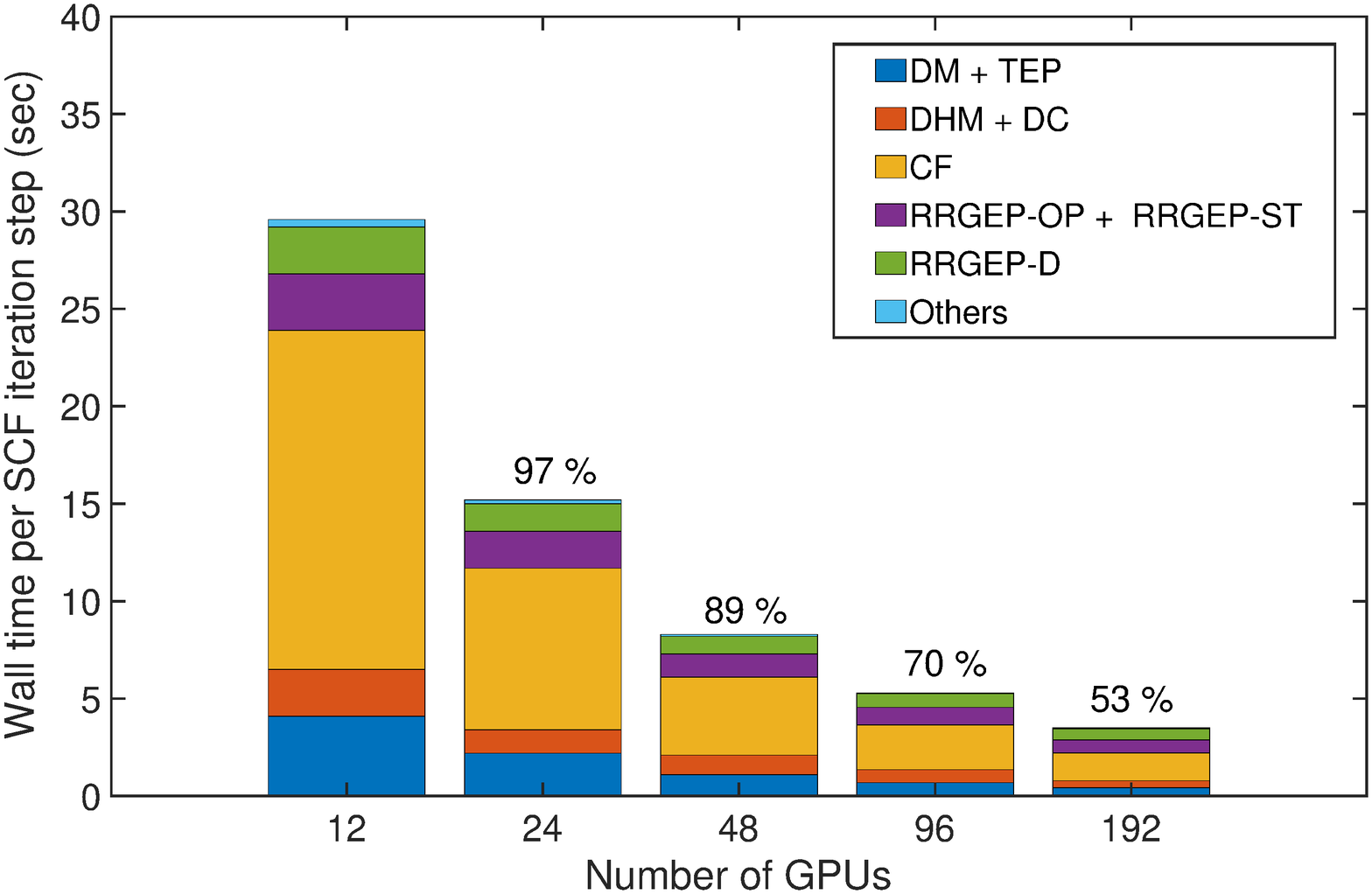}
        \caption{\small{$\textrm{MoVac}_{\textrm{6x6x6}}$ (431 atoms, 6,034 electrons)}}
        \label{fig:scalingStudyMo6x6x6}
    \end{subfigure}
    ~
    \begin{subfigure}[t]{0.5\textwidth}
        \centering
        \includegraphics[scale=0.27]{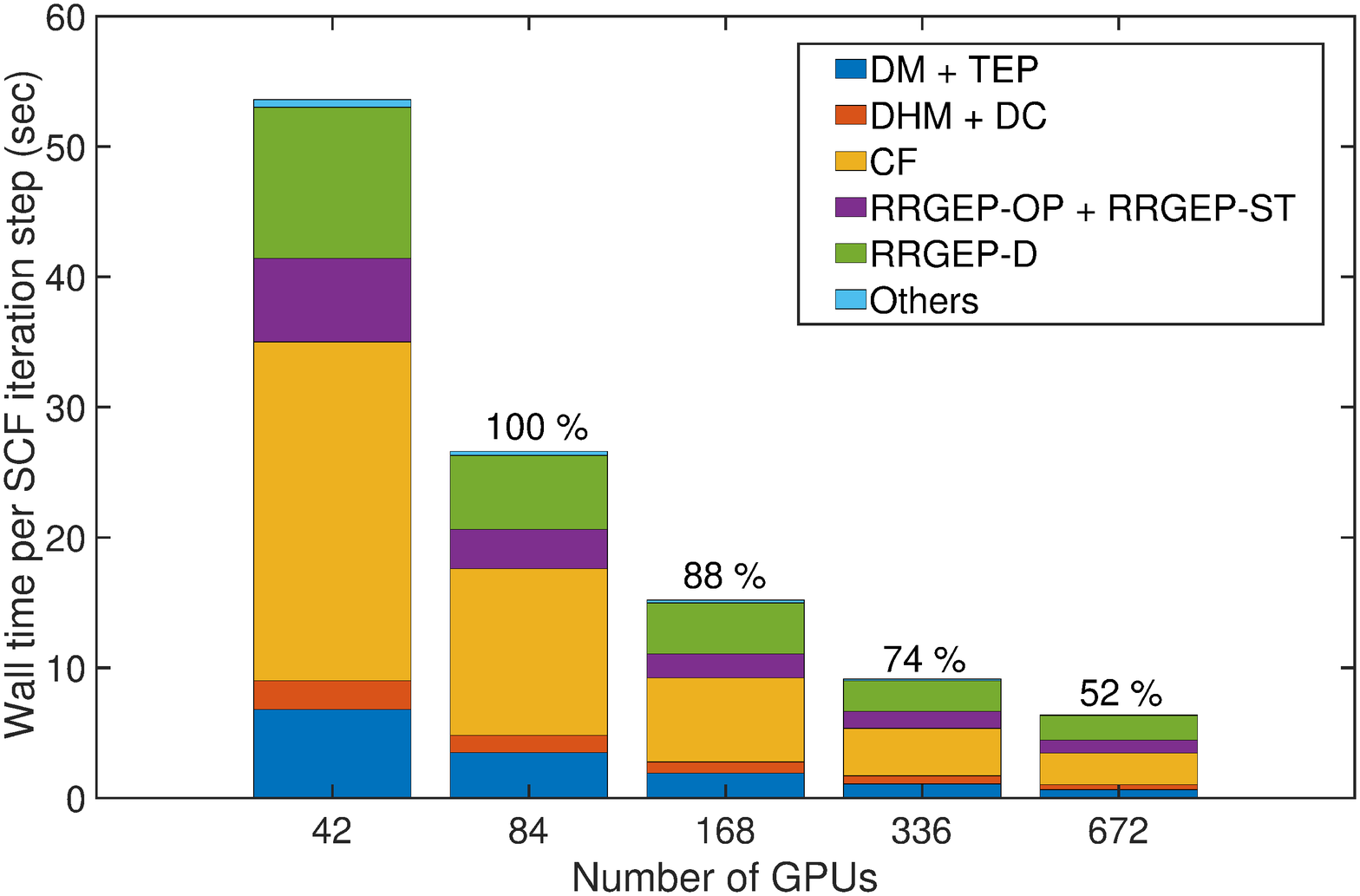}
        \caption{\small{$\textrm{MoVac}_{\textrm{8x8x8}}$ (1,023 atoms, 14,322 electrons)}}
        \label{fig:scalingStudyMo8x8x8}
    \end{subfigure}
    \caption{\small{Strong parallel scaling of wall-time per SCF iteration using \DFTFE on Summit GPU nodes. Pseudopotential DFT case studies: $\textrm{MoVac}_{\textrm{6x6x6}}$, and  $\textrm{MoVac}_{\textrm{8x8x8}}$.  Total DoFs are 1,728,000 and 4,251,528 for $\textrm{MoVac}_{\textrm{6x6x6}}$ and  $\textrm{MoVac}_{\textrm{8x8x8}}$, respectively. Parallel scaling efficiency relative to the run with lowest number of GPUs is indicated on the top of each bar. Breakdown of total wall-time per SCF iteration into the various computational steps:  a) DM+TEP, b) DHM+DC, c) CF, d) RRGEP-OP+RRGEP-ST, e) RRGEP-D and f) Others.}}
    \label{fig:scalingStudyBreakdown}
\end{figure}

We first consider the strong-scaling of $\textrm{MoVac}_{\textrm{6x6x6}}$ containing 6,034 electrons. Figure~\ref{fig:scalingStudyMo6x6x6} demonstrates the parallel scalability of the SCF iteration step and its breakdown into various constituent steps. Using 2 MPI tasks per GPU, via MPS, the domain decomposition parallelization is scaled from 12 GPUs to 192 GPUs. The per SCF wall-time reduced from 29.6 sec at 12 GPUs to 3.5 sec at 192 GPUs, where the average number of FE DoFs per GPU are $\sim9,000$. This amounts to relative speedup of 8.5$\times$ at a parallel scaling efficiency of 53\%. We further note that 70\% parallel scaling efficiency is obtained at 96 GPUs, with an average of $\sim18,000$ FE DoFs per GPU. This excellent parallel scalability is attributed to the scaling of the dominant CF step, where the key computational kernel, involving matrix-vector products of the FE discretized sparse Hamiltonian matrix and the wavefunction vectors, requires only the partition boundary nodes to be communicated. Furthermore, as discussed in Section~\ref{sec:arch-cf}, the boundary communication has been further reduced by using strategies such as single precision communication, overlapping compute and communication, along with optimization of data access overheads. Next, we consider the parallel scalability of the larger system size, $\textrm{MoVac}_{\textrm{8x8x8}}$ containing 14,322 electrons, which is shown in Figure~\ref{fig:scalingStudyMo8x8x8}. Using, 1 MPI task per GPU, the domain decomposition parallelization is scaled from 42 GPUs to 672 GPUs, where the average FE DoFs per GPU are $\sim 6,300$. The per SCF wall-time reduced from 53.6 sec on 42 GPUs to 6.4 sec on 672 GPUs, which amounts to a relative speedup of 8.4$\times$ at a parallel scaling efficiency of 52\%. As evident from the computational cost breakdown, at such system sizes and beyond, the dominant drivers of the scalability are the quadratic scaling CF step and the cubic scaling RRGEP steps. In this regard, the excellent parallel scalability of the RRGEP step is possible due to the implementation advances such as mixed precision algorithms, overlap of compute and communication, and hardware aware communication as discussed in Section~\ref{sec:arch-rr}. Overall, \DFTFE's excellent parallel scalability on hybrid CPU-GPU architectures is primarily a result of the locality of the FE basis, and the scalable numerical implementation of the CF and RRGEP steps that reduces communication costs.

Table~\ref{tab:minWallTimeGPU} reports the minimum wall-times achieved by \DFTFEver on the same benchmark systems, with a parallel scaling efficiency of $\sim40\%$. We observe that the minimum wall-time for the $\textrm{MoVac}_{\textrm{6x6x6}}$ benchmark system containing 6,034 electrons is $\sim80$ sec for the complete calculation including initialization and ionic force computation, with the SCF iteration part of the calculation taking $\sim60$ sec. Notably, for the larger $\textrm{MoVac}_{\textrm{8x8x8}}$ benchmark system containing 14,322 electrons, the wall-time for the complete calculations is $\sim140$ sec on 1,344 GPUs, with the SCF iteration part of the calculation taking $\sim100$ sec. Such low wall-times of $\sim1-2$ minutes for medium to large system sizes at chemical accuracy  are possible both due to the excellent parallel scaling of the SCF iteration procedure discussed earlier, as well as the GPU porting of ionic force computation and optimization of initialization steps. These studies underline the practical capability of \DFTFE to use the pre-exascale supercomputers for conducting fast and accurate DFT simulations of generic materials systems involving structural relaxation, molecular-dynamics, and high-throughput calculations as required in many application areas.


\begin{table}[htbp]
\centering
\small
\caption{\label{tab:minWallTimeGPU}\small{Minimum  wall-time for the total simulation measured in seconds on Summit GPU nodes and its breakdown into (INIT+IF+Others) and Total SCF costs. Case study: Pseudopotential DFT calculations on periodic BCC Mo supercells with a monovacancy. }}
\begin{tabular}{|c|c|c|c|c|c|c|}
\hline
System & Number of atoms        & Number of nodes & INIT & Total & Total\\
       &  (Number of electrons) & (Number of GPUs) &+IF+Others & SCF & run\\
\hline
$\textrm{MoVac}_{\textrm{6x6x6}}$ & 431 (6,034) & 64 (384) & 22.9 & 59.6 & 82.5\\
$\textrm{MoVac}_{\textrm{8x8x8}}$ & 1023 (14,322) & 224 (1,344) & 33.4 & 106.8 & 140.2  \\
\hline
\end{tabular}
\end{table}

\subsection{Molecular dynamics}\label{sec:md}
\begin{figure}[htp]
\includegraphics[scale=0.4]{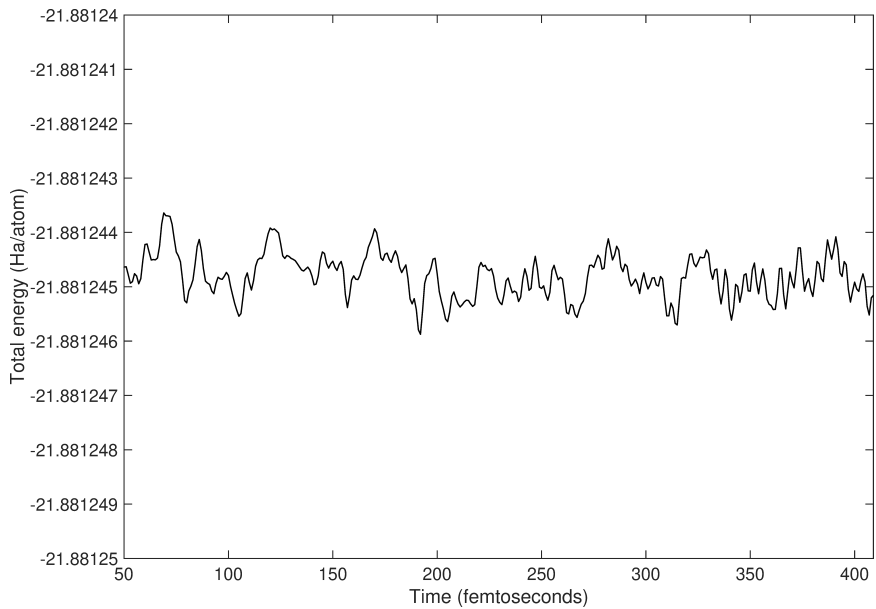}
 \centering
\caption{\small{Variation of total energy during NVE molecular dynamics simulation using \DFTFEver. Case study: InP-Water interface system with 277 atoms (1,632 electrons).}}
\label{fig:nvemd}
\end{figure}

We demonstrate the capability of \DFTFE to conduct Born-Oppenheimer molecular dynamics (MD) calculations~\cite{marx_hutter_2009}. In particular, we consider the InP-water interface system from Section~\ref{sec:validation}, which consists of 277 atoms (1,632 electrons)  and conduct a NVE molecular dynamics simulation. We use an initial ionic temperature of $T = 600$ K and employ a time step of $1.0$ fs using a velocity Verlet time integration algorithm. This simulation is conducted using norm-conserving ONCV pseudopotentials and  GGA exchange correlation. We use the FE discretization parameters from Table~\ref{tab:pspValidation}, which has been demonstrated to achieve chemical accuracy in ground-state energy and ionic forces with respect to \QE. We assign initial velocities using a Maxwell-Boltzmann distribution corresponding to the initial temperature of $600$ K. Figure~\ref{fig:nvemd} shows the variation of the total energy of the system up to $410$ fs. The mean and the standard deviation of the total energy is computed to be $-21.8812448$ Ha/atom and $4 \times 10^{-6}$ Ha/atom, respectively. We also compute the drift in the total energy by evaluating the slope of linear fit, which is found to be $1 \times 10^{-6}$ Ha/atom-ps. The average ionic temperature is around $300$ K.

\section{Conclusions}
\label{sec:conclusions}
In this work, we have developed \DFTFEverNoSpace, which incorporates significant methodology and implementation improvements over the previous version, \DFTFEverprev~\cite{motamarri2020}.  The key methodology improvement in \DFTFEver is the use of a smeared nuclear charges approach in the electrostatics formulation unlike the regularized Dirac distributions used in \DFTFEverprevNoSpace. Importantly, in the case of pseudopotential calculations, the use of smeared nuclear charges allows to decouple the atoms from the FE triangulation nodes. This enables usage of coarser finite-elements with a higher interpolating polynomial order compared to \DFTFEverprev for pseudopotential calculations, leading to a substantial reduction in FE basis functions required to achieve chemical accuracy. The other substantial methodology change in \DFTFEver is the modification of the configurational forces formulation to account for the smeared nuclear charges. Aside from the above formulation improvements, \DFTFEver employs a similar numerical solution procedure as \DFTFEverprev using FE discretization in conjunction with an efficient and scalable Chebyshev filtered eigensolver (ChFES) based SCF iteration procedure. Each SCF iteration involves the following steps: (i) computation of the FE cell level Hamiltonian matrix contributions, (ii)  Chebyshev polynomial filtering (CF) to compute lowest occupied eigensubspace, (iii)  Rayleigh-Ritz (RRGEP) procedure which involves projection of the Kohn-Sham problem onto the filtered subspace and solving the resulting small GHEP problem, and (iv) computation of the electron density (DC). Although the RRGEP procedure implemented in \DFTFEver is slightly different from the orthogonalization and Rayleigh-Ritz approach implemented in \DFTFEverprevNoSpace, the accuracy and robustness of the solver is not affected. 

The major implementation improvement in \DFTFEver is the efficient GPU porting of all the key computational kernels in the SCF iteration, namely CF, RRGEP and DC. Our GPU porting strategy hinges on reducing data-access and communication costs while effectively utilizing the massive fine grained parallelism of GPUs. In the case of the CF step, which involves  sparse-dense matrix products, we achieved a high GPU throughput performance of 54\% of the V100 GPU FP64 peak in a multi-node job with a total of 24 GPUs for a $\sim$6,000 electrons metallic benchmark system. Further, on 96 GPUs, the throughput performance was observed to be 43\% of the FP64 peak. This high GPU throughput performance was achieved by leveraging the FE cell structure to recast the sparse-dense matrix products into a batch of small dense-dense matrix products that have higher arithmetic intensity and naturally exploiting the massive parallelism of GPUs. Further, we efficiently utilize the high memory bandwidth of the GPUs by performing blocked operations over the wavefunctions in the various data movement steps in CF. Additionally, the low boundary communication cost due to the locality of the FE basis and reduction of communication latencies in CF by performing simultaneous point-to-point communications across multiple wavefunctions are also crucial in realizing this performance. We further reduced the communication costs by combining implementation strategies such as asynchronous compute and communication and FP32 boundary communication, the latter being motivated from the small percentage of FE boundary nodes in a given MPI partition. 
Next, in the case of the  $\order{M N^2}$ RRGEP procedure, which is the dominant cost for larger systems, we implemented a scalable approach specifically designed to overcome the communication and peak memory bottlenecks associated with performing parallel dense matrix-matrix multiplications involving matrices with extreme aspect ratios ($M/N \gg 1$). Our implementation reduces the dominant communication cost to $\ordercomplexity{(N^2\log(M))}$ through optimal distribution of compute costs across FE domain decomposition MPI partitions and efficient data movement strategies involving the intermediate $N \times N$ matrices that are stored in a parallel block-cyclic format. Further, we implemented a blocked approach to reduce the peak memory footprint in the intermediate steps. The blocked approach is also exploited to further enhance the performance of the parallel dense matrix-matrix multiplications on GPUs by implementing mixed precision arithmetic, asynchronous compute and communication and GPU-direct communication strategies. To assess the computational efficiency of our RRGEP implementation, we studied the throughput performance of the dominant RRGEP steps on a $\sim$60,000 electrons metallic benchmark system using 960 GPUs. We achieved 30--40\% throughput performance of the FP64 peak, which was significantly boosted to 70--110\% when the mixed precision arithmetic and asynchronous compute and communication strategies are employed. 

In order to assess the accuracy of the improved electrostatics methodology as well as the new GPU implementation, 
we conducted a validation study of \DFTFEver using \QE (pseudopotential calculations) and \nwchem (all-electron calculations). Our pseudopotenial benchmark systems included a periodic bulk metallic system, non-periodic nanoparticle, and a semiconductor-insulator interface system. On all the benchmark systems, we demonstrate excellent agreement between \DFTFEver and \QE on ground-state energies, ionic forces and stresses, when both codes are solved to chemical accuracy. Similarly, in the case of the all-electron validation study, we demonstrate excellent agreement of \DFTFEver with \nwchem on the ground-state energy of a non-periodic aromatic compound.

We demonstrated the computational efficiency and parallel scalability of \DFTFEver for pseudopotential DFT calculations on a subset of the benchmark problems with varying system sizes ($\sim$\,4,000--120,000 electrons), all calculations being solved to chemical accuracy.  First, to assess the computational efficiency afforded for pseudopotential DFT calculations by the improved electrostatics methodology, we compared the CPU computational time per SCF iteration step between \DFTFEverprev and \DFTFEver on periodic BCC Mo benchmark systems of various sizes. These studies performed on a many-core CPU machine indicated $\sim3\times$ reduction both in the CPU time and the number of FE basis functions compared to \DFTFEverprev. Next, we evaluated the performance of \DFTFEver on the Summit hybrid CPU-GPU supercomputer considering both periodic and non-periodic metallic benchmark systems. We find \DFTFEver to achieve $\sim$16--20$\times$ CPU-GPU speedups for system sizes ranging from $\sim$\,4,000--30,000 electrons, where the speedup is measured by comparing the node-hrs per SCF iteration step consumed by a hybrid CPU-GPU run against a CPU-only run on the Summit machine, while  using all the available compute resources on the compute nodes. An analysis of the computational cost breakdown showed that the efficient GPU performance of \DFTFE for both the $\order{M N}$ CF step and the $\order{M N^2}$ RRGEP steps is crucial to realize these CPU-GPU speedups for both medium and large system sizes.  Furthermore, \DFTFEver demonstrated excellent parallel scalability on the hybrid CPU-GPU architectures, with an observed strong scaling of a 14,322 electrons system from 42 to 672 GPUs at a relative speedup of 8.5$\times$.  This enabled very low wall-times for the full ground-state calculation, which includes initialization costs and computation of ionic forces. In particular, \DFTFEver achieved minimum wall-times of $\sim$1--2  minutes for system sizes of $\sim$6,000--14,000 electrons. Finally, we studied the performance of \DFTFEver on large systems with 61,502 and 114,674 electrons, using 960 and 3,600 GPUs respectively. Notably, \DFTFEver wall-times for the full ground-state calculation of the 61,502 and 114,674 electrons systems were 32 and 54 minutes, respectively.

In summary, \DFTFEver combines improved computational efficiency  and the ability to achieve significant acceleration and parallel scalability on hybrid CPU-GPU architectures to enable fast and accurate pseudopotential DFT calculations at larger length-scales than possible heretofore. Furthermore, the high throughput performance and parallel scalability achieved by \DFTFE on pre-exascale hybrid CPU-GPU supercomputers, as demonstrated in this work, makes it well placed to leverage the upcoming exascale supercomputers. In this regard, our future implementation efforts will focus on ensuring performance portability of \DFTFE on a diverse array of accelerators from different vendors requiring different programming languages such as CUDA and HIP. Our ongoing and future efforts will also involve implementation of softer projector augmented-wave pseudopotentials~\cite{bliichl1994projector} in \DFTFE for more efficient pseudopotential DFT calculations, and implementation of enriched finite-element basis~\cite{kanungo2017,rufus2020fast}, which can enable large-scale efficient all-electron DFT calculations. Another important area of focus for further development of \DFTFE is the implementation of advanced exchange-correlation  functionals, where the authors are currently pursuing development of numerical strategies for fast and accurate hybrid exchange-correlation functional calculations. Also related to exchange-correlation functional development is the ongoing effort of implementing systematically convergent inverse DFT approaches~\cite{kanungo2019exact} into \DFTFE to compute exact exchange-correlation potentials from many-body ground-state electron densities~\cite{kanungo2021}, and subsequently improve the exchange-correlation models from this data. Finally, implementation of real-time time-dependent DFT~\cite{kanungo2019real} and its GPU porting is also being pursued.

\section*{Acknowledgements}
We thank B. Kanungo, C.-C. Lin, K. Ramakrishnan, N. Kodali and N. Rufus for independently testing \DFTFEverNoSpace, providing useful feedback, and providing reference data for validating the all-electron calculations reported in this work. We gratefully acknowledge the support from the Department of Energy, Office of Basic Energy Sciences (Award number DE-SC0008637) and the Toyota Research Institute that funded the development of \DFTFEverNoSpace. V.G. gratefully acknowledges the support of Air Force Office of Scientific Research (Grant number FA9550-21-1-0302) that supported the work on improved electrostatics treatment. This research used resources of the Oak Ridge Leadership Computing Facility, which is a DOE Office of Science User Facility supported under Contract DE-AC05-00OR22725. This research used resources of the National Energy Research Scientific Computing Center, a DOE Office of Science User Facility supported by the Office of Science of the U.S. Department of Energy under Contract No. DE-AC02-05CH11231. This work used the Extreme Science and Engineering Discovery Environment (XSEDE), which is supported by National Science Foundation Grant number ACI-1053575. V.G. gratefully acknowledges the support of the Army Research Office through the DURIP grant W911NF1810242, which also provided computational resources for this work. P.M gratefully acknowledges the seed grant from Indian Institute of Science and SERB Startup Research Grant from the Department of Science and Technology India (Grant Number: SRG/2020/002194) for the purchase of a GPU cluster, which also provided computational resources for this work.
\appendix

\section{Efficient evaluation of terms related to the external potential correction in the configurational force expression}
\label{app:forces}
We describe here the procedure for the combined evaluation of the terms $\int\displaylimits_{\Omega_p} \bE_{\text{ext,corr}}:\grad \bdir (\bx) \dx + \text{F}^{\text{ext,corr}}$ in the configurational force expression (cf. Eq.~\ref{eq:Eshelbyforce} in Section~\ref{sec:forces}) implemented in \DFTFEverNoSpace. The notations followed here are consistent with that of Section~\ref{sec:ksdft}. Considering the expression for
\begin{align}\label{eq:a1}
&\int\displaylimits_{\Omega_p} \bE_{\text{ext,corr}}:\grad \bdir (\bx) \dx + \text{F}^{\text{ext,corr}}\,\notag\\
=& \int\displaylimits_{\Omega_p} \Biggl(\sum_a \,\sum_r \Bigl(V_{\text{ext,loc}}^{a}(|\bx-(\bR_a+\bL_r)|) - \vself a\left(\bx,(\bR_a+\bL_r)\right)\Bigr)\rhobar(\bx) \Biggr)\bI :\grad \bdir (\bx) \dx \notag\\&+ \sum_{a}\,\sum_r \int\displaylimits_{\Omega_p} \rhobar(\bx)\Biggl(\grad V^{a}_{\text{ext,loc}}(|\bx-(\bR_a+\bL_r)|)\cdot \left(\bdir(\bx) - \bdir(\bR_a+\bL_r) \right) \notag\\&- \grad \vself a(\bx,\bR_a+\bL_r)\cdot \bdir(\bx) - \left.\frac{\partial \vselfeps a\left(\bx,\bsm(\abs{\chieps(\by)-\chieps(\bR_a+\bL_r)})\right)}{\partial \varepsilon}\right|_{\varepsilon=0} \Biggr) \dx\,,
\end{align}
presents two challenges for its computationally efficient and accurate evaluation. The first challenge is that the direct evaluation of integrals containing $\grad V^{a}_{\text{ext,loc}}(|\bx-(\bR_a+\bL_r)|)$ would be inefficient due to requirement of denser quadrature grids owing to their potentially non-smooth nature. However, employing integration by parts to transfer the gradient on to smoother fields, as done in other places in Section~\ref{sec:forces}, needs to be carefully employed here to ensure that the terms in the final expression have cancellation of the tails of $V^{a}_{\text{ext,loc}}(|\bx-(\bR_a+\bL_r)|)$ and $\vself a(\bx,\bR_a+\bL_r)$ for convergent evaluation of their combined sums over the  periodic lattice sites $r$. The second challenge in evaluation of Eq.~\ref{eq:a1} is the accurate and efficient computation of $\left.\frac{\partial \vselfeps a(\bx,\bsm(\abs{\chieps(\by)-\chieps(\bR_a)})}{\partial \varepsilon}\right|_{\varepsilon=0}$.

First, we discuss our methodology to address the first challenge. To begin with, we take integration by parts of terms involving   $\grad V_{\text{ext,loc}}^{a}(|\bx-(\bR_a+\bL_r)|)$ and $\grad \vself a(\bx,\bR_a+\bL_r)$ to rewrite Eq.~\ref{eq:a1} as follows
\begin{align}\label{eq:a2}
& \int\displaylimits_{\Omega_p} \sum_{a}\,\sum_r \left(\vself a(\bx,\bR_a+\bL_r)-V^{a}_{\text{ext,loc}}(|\bx-(\bR_a+\bL_r)|)\right)\left( \grad \rhobar(\bx) \cdot \bdir(\bx) \right) \dx \notag\\
&+\int\displaylimits_{\Omega_p} \sum_{a}\,\sum_r  V^{a}_{\text{ext,loc}}(|\bx-(\bR_a+\bL_r)|) \left(\grad \rhobar(\bx) \cdot \bdir(\bR_a+\bL_r)  \right) \dx\notag\\
&-\sum_{a}\,\sum_r \int\displaylimits_{\Omega_p} \rhobar(\bx)\Biggl(  \left.\frac{\partial \vselfeps a\left(\bx,\bsm(\abs{\chieps(\by)-\chieps(\bR_a+\bL_r)})\right)}{\partial \varepsilon}\right|_{\varepsilon=0} \Biggr) \dx.
\end{align}
Next, considering the last term in the above, we further facilitate transfer of derivatives by splitting the integration over $\Omega_p$ for each $(a,r)$ pair into two regions inside the unit cell:  (i) $\Omega_{a,r} \,\cap\, \Omega_p$, which is the intersection of a bounded domain $\Omega_{a,r}$ surrounding the atom at $(\bR_a+\bL_r)$ with the unit cell periodic domain  $\Omega_p$, and (ii) $\Omega_p-\Omega_{a,r}$, which is the remainder region inside  $\Omega_p$ excluding the intersection with $\Omega_{a,r}$. Thus, Eq.~\ref{eq:a2} is rewritten as 
\begin{align}\label{eq:a3}
& \int\displaylimits_{\Omega_p} \sum_{a}\,\sum_r \left(\vself a(\bx,\bR_a+\bL_r)-V^{a}_{\text{ext,loc}}(|\bx-(\bR_a+\bL_r)|)\right)\left( \grad \rhobar(\bx) \cdot \bdir(\bx) \right) \dx \notag\\
&+\int\displaylimits_{\Omega_p} \sum_{a}\,\sum_r  V^{a}_{\text{ext,loc}}(|\bx-(\bR_a+\bL_r)|) \left(\grad \rhobar(\bx) \cdot \bdir(\bR_a+\bL_r)  \right) \dx \notag\\
&-\sum_{a}\,\sum_r \int\displaylimits_{\Omega_{a,r} \,\cap\, \Omega_p} \rhobar(\bx)\Biggl(  \left.\frac{\partial \vselfeps a\left(\bx,\bsm(\abs{\chieps(\by)-\chieps(\bR_a+\bL_r)})\right)}{\partial \varepsilon}\right|_{\varepsilon=0} \Biggr) \dx \notag\\
&-\sum_{a}\,\sum_r \int\displaylimits_{\Omega_p-\Omega_{a,r}} \rhobar(\bx)\Biggl(  \left.\frac{\partial \vselfeps a\left(|\bx-\chieps(\bR_a+\bL_r)|)\right)}{\partial \varepsilon}\right|_{\varepsilon=0} \Biggr) \dx.
\end{align}
The choice of $\Omega_{a,r}$ in the above is such that outside $\Omega_{a,r}$, $\vself a(\bx,\bR_a+\bL_r)$ can be assumed to be a smooth radially symmetric function, given by $ \vself a(|\bx-(\bR_a+\bL_r)|$. We use this assumption to rewrite the last term in Eq.~\ref{eq:a3} as
\begin{align}\label{eq:a4}
-&\sum_{a}\,\sum_r \int\displaylimits_{\Omega_p-\Omega_{a,r}} \rhobar(\bx)\Biggl(  \left.\frac{\partial \vselfeps a(|\bx-\chieps(\bR_a+\bL_r)|)}{\partial \varepsilon}\right|_{\varepsilon=0} \Biggr) \dx\notag\\
=&\sum_{a}\,\sum_r \int\displaylimits_{\Omega_p-\Omega_{a,r}} \rhobar(\bx)\Biggl( \grad \vself a(|\bx-(\bR_a+\bL_r)|) \cdot \bdir(\bR_a+\bL_r)  \Biggr)\dx.
\end{align}
Further, noting that form of Eq.~\ref{eq:a4} is now amenable to apply integration by parts, we transform it using integration by parts into a sum of a surface integral term and a volume integral term which subsequently do not have $\grad \vself a(|\bx-(\bR_a+\bL_r)|$. This is desired to ensure cancellation of the asymptotic tail of $ \vself a(|\bx-(\bR_a+\bL_r)|$ with that of $V^{a}_{\text{ext,loc}}(|\bx-(\bR_a+\bL_r)|)$ in Eq.~\ref{eq:a3}. We perform this step as follows
\begin{align}\label{eq:a5}
&\sum_{a}\,\sum_r \int\displaylimits_{\Omega_p-\Omega_{a,r}} \rhobar(\bx)\Biggl( \grad \vself a(|\bx-(\bR_a+\bL_r)|) \cdot \bdir(\bR_a+\bL_r)  \Biggr)\dx\notag\\
=&\sum_{a}\,\sum_r \int\displaylimits_{\partial (\Omega_p-\Omega_{a,r})} \rhobar(\bx)\vself a(|\bx-(\bR_a+\bL_r)|)  \bdir(\bR_a+\bL_r)  \cdot \hat{\boldsymbol{n}}\; d{\bf s} \notag\\&\qquad-\sum_{a}\,\sum_r \int\displaylimits_{\Omega_p-\Omega_{a,r}} \vself a(|\bx-(\bR_a+\bL_r)|)\Biggl( \grad \rhobar(\bx) \cdot \bdir(\bR_a+\bL_r)  \Biggr) \dx\notag\\
=&\sum_{a}\,\sum_r \int\displaylimits_{\partial (\Omega_p-\Omega_{a,r})-\partial \Omega_p} \rhobar(\bx)\vself a(|\bx-(\bR_a+\bL_r)|)  \bdir(\bR_a+\bL_r)  \cdot \hat{\boldsymbol{n}}\; d{\bf s} \notag\\&\qquad-\sum_{a}\,\sum_r \int\displaylimits_{\Omega_p-\Omega_{a,r}} \vself a(|\bx-(\bR_a+\bL_r)|)\Biggl( \grad \rhobar(\bx) \cdot \bdir(\bR_a+\bL_r)  \Biggr) \dx\,
\end{align}
where $\hat{\boldsymbol{n}}$ denotes the outward normal to the surface $\partial (\Omega_p-\Omega_{a,r})$, and the last equality follows from cancellation of contributions on the periodic domain surface, $\partial \Omega_p$.
Substituting Eqs.~\ref{eq:a4}--\ref{eq:a5} in Eq.~\ref{eq:a3} and rearranging the terms based on cancellation of asymptotic tails, we obtain the final expression as
\begin{align}\label{eq:a6}
&\int\displaylimits_{\Omega_p} \bE_{\text{ext,corr}}:\grad \bdir (\bx) \dx + \text{F}^{\text{ext,corr}}\,\notag\\
=& \int\displaylimits_{\Omega_p} \sum_{a}\,\sum_r \left(\vself a(\bx,\bR_a+\bL_r)-V^{a}_{\text{ext,loc}}(|\bx-(\bR_a+\bL_r)|)\right)\left( \grad \rhobar(\bx) \cdot \bdir(\bx) \right) \dx \notag\\
&-\int\displaylimits_{\Omega_p-\Omega_{a,r}} \sum_{a}\,\sum_r \left(\vself a(|\bx-(\bR_a+\bL_r)|)- V^{a}_{\text{ext,loc}}(|\bx-(\bR_a+\bL_r)|) \right) \left(\grad \rhobar(\bx) \cdot \bdir(\bR_a+\bL_r)  \right) \dx \notag\\
&+\sum_{a}\,\sum_r \Biggl[\:\:\int\displaylimits_{\Omega_{a,r} \,\cap\, \Omega_p}  V^{a}_{\text{ext,loc}}(|\bx-(\bR_a+\bL_r)|) \left(\grad \rhobar(\bx) \cdot \bdir(\bR_a+\bL_r)  \right) \dx \notag\\
&- \int\displaylimits_{\Omega_{a,r} \,\cap\, \Omega_p} \rhobar(\bx)\Biggl(  \left.\frac{\partial \vselfeps a\left(\bx,\bsm(\abs{\chieps(\by)-\chieps(\bR_a+\bL_r)})\right)}{\partial \varepsilon}\right|_{\varepsilon=0} \Biggr) \dx \notag\\
&+ \int\displaylimits_{\partial (\Omega_p - \Omega_{a,r})-\partial \Omega_p} \rhobar(\bx)\vself a(|\bx-(\bR_a+\bL_r)|)  \bdir(\bR_a+\bL_r)  \cdot \hat{\boldsymbol{n}}\; d{\bf s}\Biggr]\,.
\end{align}

Next, we discuss our methodology for addressing the second challenge, which is the evaluation of \begin{equation}\label{eq:appvsmperturbcharge}
    \left.\frac{\partial \vselfeps a\left(\bx,\bsm(\abs{\chieps(\by)-\chieps(\bR_a+\bL_r)})\right)}{\partial \varepsilon}\right|_{\varepsilon=0},
\end{equation}
present in the third term in Eq.~\ref{eq:a6}.
We consider the following three different cases: (a) all-electron ionic forces---configurational force is evaluated with respect to a generator perturbing the atom and the underlying space in a compact support around the atom. This is required in all-electron calculations where the ionic positions are coupled to the corner nodes of FE triangulation elements to capture the cusp in the electronic wavefunctions; (b) pseudopotential ionic forces---configurational force is evaluated by perturbing the atoms without perturbing the underlying space i.e. floating nuclear charges with $\bdir(\bx)=0$ and $\bdir(\bR_a+\bL_r)\neq 0$. This approach is used in the case of pseudopotential calculations with smooth electronic fields; and (c) periodic unit-cell stress computation---configurational force is computed with respect to a affine deformation generator ($\dir_{i} = C_{ij}x_j$). First, in the case of all-electron ionic forces (case (a)), we assume $\vself a(\bx,\bR_a+\bL_r)$ to be a radially symmetric field, $\vself a(\abs{\bx-\bR_a+\bL_r})$, which is a reasonable assumption as refined FE triangulations with much smaller element sizes near the ionic positions are employed in the case of all-electron calculations in \DFTFENoSpace. Thus, Eq.~\ref{eq:appvsmperturbcharge} now simplifies to  
\begin{equation}
    \left.\frac{\partial \vselfeps a\left(\abs{\bx-\chieps(\bR_a+\bL_r)}\right)}{\partial \varepsilon}\right|_{\varepsilon=0}=- \grad \vself a(\abs{\bx-(\bR_a+\bL_r)}) \cdot \bdir(\bR_a+\bL_r) \,.
\end{equation}
Second, in the case of pseudopotential ionic forces (case (b)), the assumption of floating nuclear charges ($\bdir(\bx)=0$ and $\bdir(\bR_a+\bL_r)\neq 0$) simplifies Eq.~\ref{eq:appvsmperturbcharge} to
\begin{equation}
    \left.\frac{\partial \vselfeps a\left(\bx,\bsm(\abs{\by-\chieps(\bR_a+\bL_r)})\right)}{\partial \varepsilon}\right|_{\varepsilon=0}=  \left( \;\;\int\displaylimits_{\Rthree}  \frac{- \grad \bsmi a (|\by-(\bR_a+\bL_r)|)}{|\bx-\by|} \dy \right) \cdot \bdir(\bR_a+\bL_r)\,,
\end{equation}
where the integral over the Coulomb kernel is obtained by solutions of Poisson problems to compute the potential associated with the gradient of the smeared charge distribution. The Poisson problems are solved on bounded domains $\Omega_{a,r}$ with Dirichlet boundary conditions set to the gradient of the exact nuclear potential, $\grad V_{\text{nuc}}^a\abs{\bx-(\bR_a+\bL_r)}$. We note that only the Poisson problems corresponding to a finite number of bounded domains $\Omega_{a,r}$ having non-trivial intersection with the periodic unit cell ($\Omega_{a,r} \,\cap\, \Omega_p \neq \emptyset$) need to be considered. Finally, for the case of periodic unit-cell stress computation (case (c)), we employ central finite difference scheme to evaluate Eq.~\ref{eq:appvsmperturbcharge} corresponding to the six unique affine strain components. We remark that a similar approach would be impractical for all-electron ionic forces (case (a)) as it would incur a large cost to evaluate  Eq.~\ref{eq:appvsmperturbcharge} by finite difference for all the different $\chieps$ corresponding to individual force components over all the atoms. 

\clearpage
\bibliographystyle{elsarticle-num}
\bibliography{citations}







\end{document}